\documentclass[11pt,a4paper]{article}
\pdfoutput=1

\usepackage{jcappub}

\title{\boldmath Stability analysis of inflation with an $SU(2)$ gauge field}

\author[a]{Kei-ichi Maeda}
\author[b]{and Kei Yamamoto}

\affiliation[a]{Department of Physics, Waseda University,\\
Shinjuku, Tokyo 169-8555, Japan}
\affiliation[b]{Department of Applied Mathematics and Theoretical Physics, 
University of Cambridge,\\
Wilberforce Road, Cambridge CB3 0WA, United Kingdom}

\emailAdd{maeda@waseda.jp}
\emailAdd{K.Yamamoto@damtp.cam.ac.uk}

\abstract{We study anisotropic cosmologies of a scalar field interacting with
an $SU(2)$ gauge field via a gauge-kinetic coupling. 
We analyze Bianchi class A models, which 
include Bianchi type I, II, VI$_0$, VII$_0$, VIII and IX.
 The linear stability of isotropic inflationary solution with background 
magnetic field is shown, which generalizes the known results for $U(1)$
gauge fields.
We also study anisotropic inflationary solutions, all of which turn out to be
unstable.
Then nonlinear stability for the isotropic inflationary solution
is examined by numerically investigating the dependence
of the late-time behaviour on the initial conditions. We present  a number
 of novel features that may well affect physical predictions and viability of 
the models. First, in the absence of spatial curvature, strong initial 
anisotropy leads to a rapid oscillation of gauge field, thwarting convergence 
to the inflationary attractor. Secondly, the inclusion of spatial curvature 
destabilizes the oscillatory attractor and the global stability 
of the isotropic
inflation with gauge field is restored.  Finally, based on the numerical
evidence combined with the knowledge of the eigenvalues for various 
inflationary solutions, we give a generic lower-bound for the duration 
of transient anisotropic inflation, which is inversely proportional
to the slow-roll parameter.}

\begin{document}
\maketitle

\section{Introduction}
\label{introduction}
%%%%%%%%%%%%%%%%%%%%%%%%%%%%%%%%%%%%%%%%%%%%%%%%%%%%%%%%%%%%%%%%%%%%%%%%%%%%%%%
%%%%%%%%%%%%%%%%%%%%%%%%%%%%%%%%%%%%%%%%%%%%%%%%%%%%%%%%%%%%%%%%%%%%%%%%%%%%%%%
As a phenomenological model of the universe beyond the energy scale of TeV, 
inflation
has had a remarkable success in explaining the homogeneity and flatness of the 
observed
universe and the origin of the almost scale-invariant power spectrum of Cosmic 
Microwave
Background Radiation (CMBR)
\cite{Starobinsky1980,Sato1981,Guth1981,Albrecht1982,Linde1982,Linde1983,
Linde2005,Linde2006,Linde2008,McAllister2008,Lyth2008}. 
These predictions are guaranteed by the slow-roll
 conditions
imposed on the potential of scalar inflaton. Ironically those conditions make 
it difficult for
inflation to find a place in the low-energy effective field theories 
derived from the unified theories of fundamental interactions such as string 
theory where 
typical scalar fields appear to have rather steep potentials
\cite{Townsend2003}. While a specific 
realization in
type IIB super string theory has been proposed in \cite{Kachru2003, Kachru2003a}, 
the universality of such scenarios is largely unknown. 

Recently it has been noticed that interactions between inflaton and gauge fields
 can lift
the slow-roll conditions and make inflation possible for steep scalar potentials \cite{Maleknejad2013}. 
The initial interests in inflaton-gauge interactions rooted from the quest
 to generating anomalous
features during inflation, e.g. statistical anisotropy of CMBR
\cite{Yokoyama2008,Watanabe2009,Bartolo2009a,Bartolo2009,
Dimopoulos2009,Moniz2010,Dulaney2010,Gumrukcuoglu2010,Watanabe2010,
Emami2010,Watanabe2011,Murata2011,Shiraishi2011,Namba2012,Bartolo2012}, primordial 
gravitational 
waves that potentially break the Lyth bound
\cite{Barnaby2011,Anber2012}, large scale seed magnetic fields
\cite{Kanno2009,Barnaby2012,Ferreira2013}, and primordial non-Gaussianity
\cite{Valenzuela-Toledo2009,Valenzuela-Toledo2010,Dimastrogiovanni2010,
Barnaby2011,Karciauskas2011,Valenzuela-Toledo2011,Jain2012,Rodriguez2013,Almeida2013}. Eventually, it has been found that the back 
reaction of the generated gauge fields slows down the inflaton and 
alleviates the necessity of flat potential \cite{Anber2010,Dimopoulos2010a,Dimopoulos2011}. 
By properly treating the background dynamics including both scalar and
vector fields, an exponential type potential and a gauge-kinetic function with 
coupling constants of order unity, which are expected to be found 
in supergravity theories,
have been shown to lead to an acceptable background evolution of the space-time
 \cite{Kanno2010,Hervik2011,Do2011,Do2011a,Ohashi2013}. 
While the inflationary regime is anisotropic for the case of a $U(1)$ 
gauge field, multiple gauge fields generically favor the isotropic triad 
configuration and isotropic inflation with non-vanishing background gauge 
fields is dynamically realized \cite{Yamamoto2012}. Its generalization
to non-Abelian gauge field, in which the presence of multiple vectors is 
automatic, has been considered in \cite{Maeda2012}. 
In contrast to the $U(1)$ gauge fields for 
which both electric and magnetic components are 
equally capable of surviving the accelerated expansion, 
it has been revealed that the Yang-Mills interaction destabilizes the
inflation with non-vanishing electric components
 so that only magnetic components can 
lead to stable accelerated expansion without requiring a flat potential for 
the scalar field. 

A similar mechanism involving a non-Abelian gauge field and Chern-Simons
coupling to a pseudo-scalar field can also accommodate inflation with a steep
scalar potential, where the background contains both electric and magnetic 
components \cite{Maleknejad2011,Maleknejad2012,Adshead2012,Adshead2012a,
Sheikh-Jabbari2012,Martinec2012}.  Its linear stability has been recently 
discussed in \cite{Dimastrogiovanni2012,Adshead2013,Adshead2013a}.

Detailed calculations of power spectrum and bispectrum of curvature 
perturbation generated from quantum fluctuations 
have been carried out for the cases 
involving background $U(1)$ gauge fields \cite{Dulaney2010,Gumrukcuoglu2010,
Watanabe2010,Yamamoto2012,Bartolo2012,Funakoshi2012}.
Although the presence of instability arising
 from gauge
field perturbations that are quantum mechanically generated during inflation 
has been claimed in \cite{Bartolo2012}, physical 
interpretation of the infrared divergence for massless excitations in 
de-Sitter space-time 
is a subtle problem and we will not discuss it here. While the recent data
 from Planck point to
extremely feeble interactions of the inflaton with other fields \cite{PlanckCollaboration2013a}, it should be
 noted that the constraint on the isotropic configuration of gauge fields is 
relatively weak
and the background energy density of gauge fields can be as great as the 
kinetic energy of the inflaton according to \cite{Funakoshi2012}. 
Hence further studies 
for the variants of these models are needed to decide their viability and 
search for potentially observable signatures.

In the present article, we investigate the stability of the isotropic 
inflation with a
background $SU(2)$ gauge field and a gauge-kinetic coupling against 
long-wavelength
perturbations by using the dynamical systems analysis of spatially homogeneous 
cosmologies. 
When the potential and gauge-kinetic function are both of exponential 
type, various isotropic and anisotropic inflationary solutions appear as fixed 
points of the dynamical system.
We first carry out the systematic linear stability 
analysis around these fixed points and find that isotropic inflation 
with background magnetic
field is the only locally stable solution. This is a generalization of the 
result obtained in \cite{Yamamoto2012} to non-Abelian gauge fields
in the presence of spatial curvature. 
In order to figure out the global dynamics, we perform numerical 
calculations with a variety of initial conditions. Despite the local stability of 
the isotropic inflation,  the nonlinear nature of the system leads to
a number of novel features that may well affect physical predictions.
\begin{enumerate}
\item In the absence of spatial curvature, the parameter deciding the 
fate
of the universe is the strength of Yang-Mills interaction with respect to the 
energy
scale of inflation. When the gauge coupling is negligible, where the $SU(2)$ 
gauge
field can be regarded as a triplet of Abelian gauge fields, all 
trajectories eventually 
reach the isotropic inflation. However, the convergence is much slower than 
the single
scalar inflation because of the temporary attraction towards anisotropic 
inflationary 
fixed points,  which typically scales inversely proportional to the 
slow-roll parameter. On the other hand, a stronger  initial Yang-Mills 
coupling leads to the emergence of 
oscillatory attractor states and an inflationary final state is not reached 
for strongly anisotropic initial conditions.
\item The inclusion of spatial curvature restores the stability of the isotropic
magnetic inflation for strongly anisotropic initial conditions by 
destabilizing the oscillatory solutions. 
For a certain range of model parameters, the spatial curvature
may also generate anomalously long periods of transient anisotropic inflation.
\end{enumerate}

In the next section, we derive the dynamical system
with an appropriate choice of variables.
 In section 3, all the inflationary fixed points are listed 
and the eigenvalues of the perturbations around those fixed points
 are computed. Because of the necessity 
to take into account the Yang-Mills interaction of the gauge field,
it turns out that some of the fixed points  
appear not as time-independent  
solutions but as asymptotic ones in our formulation. In section 4 and 5, we 
survey certain ranges of initial conditions which are of greater physical 
interests  for a representative set of model parameters
 in the absence (Bianchi I) and
presence of spatial curvature (Bianchi II, VIII, and IX). 
We also analytically identify the
oscillatory attractors and discuss their properties. 
 Section 6 demonstrates
the robustness of our results against different sets of model parameters.
In Appendix, we summarize the Abelian case as a reference. 
We also show the numerical results 
for Bianchi type VI$_0$ and  VII$_0$  where slightly different prescriptions
for initial data are used.

%%%%%%%%%%%%%%%%%%%%%%%%%%%%%%%%%%%%%%%%%%%%%%%%%%%%%%%%%%%
%%%%%%%%%%%%%%%%%%%%%%%%%%%%%%%%%%%%%%%%%%%%%%%%%%%%%%%%%%%
%%%%%%%%%%%%%%%%%%%%%%%%%%%%%%%%%%%%%%%%%%%%%%%%%%%%%%%%%%%
\section{$SU(2)$ gauge fields in Bianchi cosmologies}
%%%%%%%%%%%%%%%%%%%%%%%%%%%%%%%%%%%%%%%%%%%%%%%%%%%%%%%%%%%
%%%%%%%%%%%%%%%%%%%%%%%%%%%%%%%%%%%%%%%%%%%%%%%%%%%%%%%%%%%
%%%%%%%%%%%%%%%%%%%%%%%%%%%%%%%%%%%%%%%%%%%%%%%%%%%%%%%%%%%
As the cosmological model space-times, we take those admitting a
three-dimensional group of isometry characterised by space-like Killing
vectors $\{ \boldsymbol{\xi }_A  \} , A = 1,2,3$, which divides the space-time
into equivalence classes of the points connected by group elements.  We
focus our attention on so-called Bianchi cosmological models for which the
isometry group acts on each equivalence class simply transitively. Then the
equivalence classes are three-dimensional space-like hypersurfaces which
are identified to be homogeneous spatial slices of the universe. Their internal
geometry is classified into nine types according to the group structure of the
isometry. A group invariant orthonormal spatial triad $\{ \mathbf{e}_A \}$ is 
introduced by
\begin{equation}
\mathbf{g} ( \mathbf{e}_A ,\mathbf{e}_B ) = \delta _{AB} \ , \quad 
\left[ \mathbf{e}_A , \boldsymbol{\xi } \right] = 0 \ . 
\end{equation}
Capital Latins denoting the spatial indices are raised and lowered by
Kronecker's delta whence we will not distinguish between superscripts
and subscripts; the position is chosen for notational convenience. We 
construct an orthonormal frame $\{ \mathbf{e}_{\mu } \} , \mu = 0,1,2,3$ 
by taking $\mathbf{e}_0$ to be the unit normal of the homogeneous 
hypersurface. In the followings, all the tensor components are written
in this frame unless otherwise stated. The geometry of the space-time
is encoded in the commutation functions $\gamma ^{\lambda }_{\ \mu \nu }$, 
which can be defined as
\begin{equation}
d\mathbf{e}^{\lambda } = -\frac{1}{2} \gamma ^{\lambda }_{\ \mu \nu } 
\mathbf{e}^{\mu } \wedge \mathbf{e}^{\nu } \,,
\end{equation}
where $\{ \mathbf{e}^{\mu } \}$ is the dual 1-form basis of $\{ 
\mathbf{e}_{\mu } \} $. 

One can introduce a proper time coordinate $t$ by
\begin{equation}
\mathbf{e}_0 = \frac{\partial }{\partial t} 
\end{equation}
and denote the time derivatives by overdots. The non-zero components of
the commutation functions, which are all functions of only $t$, are written as
 follows;
\begin{align}
\begin{split}
& \gamma ^A_{\ 0B} = -H \delta _{AB} - \sigma _{AB} - \epsilon _{ABC} 
\Omega _C \ , \\
& \gamma ^A_{\ BC} = \epsilon _{BCD} n^{AD} + a_B \delta _{CA} - a_C 
\delta _{BA} \ , 
\end{split}
\end{align}
where $\epsilon _{ABC}$ denotes the three-dimensional Levi-Civita symbol. 
The Hubble
expansion rate $H$ and shear expansion rate $\sigma _{AB}$ are related to 
the extrinsic
curvature of the homogenous spatial slice by
\begin{equation}
\left( \nabla \mathbf{e}_0 \right) _{AB} = H \delta _{AB} + \sigma _{AB} \ . 
\end{equation}
The auxiliary quantity $\Omega _A$ will be eliminated after the gauge freedom
associated with $O(3)$ rotations of spatial triad is used.
The spatial part satisfies Jacobi identities of which the algebraic ones are
\begin{equation}
n^{AB}a_B = 0 \ .
\end{equation}
We sometimes call $n_{AB}$ and $a_A$ spatial curvature variables since
they determine the Ricci tensor of the homogeneous hypersurface as
\begin{equation}
\begin{split}
{}^{(3)}R_{AB}= & 2n_{AC}n_{CB} - n_{CC} n_{AB} - \frac{1}{2} \left( n_{CD} n_{CD} - n_{CC}^2 \right) \delta _{AB} \\
& - \epsilon _{CDA}n_{BC}a_D + \epsilon _{CDB}n_{AC}a_D - 2a_E a_E \delta _{AB} 
\,.
\end{split}
\end{equation}

Let us introduce the Lie algebra of $SU(2)$ gauge group by
\begin{equation}
\left[ T_a , T_b \right] = f_{abc} T_c \ \quad a , b,  \dots  = 1,2,3 .
\end{equation}
One can choose the generators $\{T_a\}$ such that 
\begin{equation}
{\rm tr}\left( T_a , T_b \right) = \delta _{ab} \ , \quad f_{abc} = 
\epsilon _{abc} \ .
\end{equation}
The gauge potential $\mathbf{A}$ is a Lie-algebra-valued 1-form defined by
\begin{equation}
\mathbf{A} = A^a_{\mu } T_a \mathbf{e}^{\mu } 
\,.
\end{equation}
The field strength is computed by the extended exterior calculus as
\begin{align}
\begin{split}
\mathbf{F} = & d\mathbf{A} + g \mathbf{A} \wedge \mathbf{A} \\
=& \frac{1}{2} \left[ \mathbf{e}_{\mu } \left( A^a_{\nu } \right) - 
\mathbf{e}_{\nu }\left( A^a_{\mu } \right) + g\epsilon _{abc}A^b_{\mu } 
A^c_{\nu } \right] T_a \mathbf{e}^{\mu } \wedge \mathbf{e}^{\nu }  + A_{\mu }^a T_a d\mathbf{e}^{\mu } \ ,
\end{split}
\end{align}
where $g$ is the gauge coupling constant. Given the group invariant frame 
$\{ \mathbf{e}_{\mu } \}$, we assume the components of vector potential are
homogeneous; that is 
\begin{equation}
\mathbf{e}_A \left( A^a_{\mu } \right) = 0 \ .
\end{equation}
One can introduce electric and magnetic fields ($ E^a_A, B^a_A$) by
\begin{equation}
\mathbf{F} = E^a_A T_a \mathbf{e}^A \wedge \mathbf{e}^0 + \frac{1}{2} 
\epsilon _{ABC} B^a_C T_a \mathbf{e}^A \wedge \mathbf{e}^B \ , 
\end{equation}
which are written by definition in terms of $A^a_{\mu }$ as
\begin{align}
& E^a_A = -\dot{A}^a_A - H A^a_A - \sigma _{AB} A^a_B + \epsilon _{ABC} A^a_B 
\Omega _C - g\epsilon _{abc}A^b_0 A^c_A \ ,  \label{eq:electric} \\
& B^a_A = \frac{g}{2} \epsilon _{abc}\epsilon _{ABC} A^b_B A^c_C - n_{AB} 
A^a_B \ . \label{eq:magnetic}
\end{align}
The Bianchi identities 
\begin{equation}
d\mathbf{F} + g[ \mathbf{A}\wedge \mathbf{F} ]= 0 
\end{equation}
become 
\begin{align}
\begin{split}
0 =& \dot{B}^a_A + 2HB^a_A - \sigma _{AB}B^a_B + \epsilon _{ABC} 
\Omega _B B^a_C - n_{AB} E^a_B -\epsilon _{ABC} a_B E^a_C \\
& + g\epsilon _{abc} \epsilon _{ABC} A^b_B E^c_C + g\epsilon _{abc} A^b_0 
B^c_A \ ,
\end{split} \label{eq:dynamicalB} \\
0 = & 2a_A B^a_A - g\epsilon _{abc} A^b_A B^c_A \ . \label{eq:constraintB}
\end{align}

The dynamical equations are derived from the Lagrangian
\begin{equation}
\mathcal{L} = \sqrt{-g}\left( \frac{M_{pl}^2}{2} R -\frac{1}{2}
\nabla _{\mu } \varphi \nabla ^{\mu } \varphi - V(\varphi ) -
 \frac{f^2(\varphi ) }{4}F^a_{\mu \nu }F^{a\mu \nu }  \right) \,,
\end{equation}
where $M_{pl}$ is the Planck mass, which we shall set unity.

We restrict ourselves to exponential type potential and gauge-kinetic
 function
\begin{equation}
V(\varphi ) = V_0 e^{-\alpha \varphi } \ , \quad f^2(\varphi ) = 
e^{\lambda \varphi } \ ,
\end{equation}
which can be most commonly seen in supergravity theories. 
$\alpha$ is assumed to be positive without loss of generality.
Although these 
choices
are made for computational convenience, the dynamics during inflation will 
be similar
for a wider class of scalar potentials and gauge-kinetic functions.
The Einstein equations can be found in \cite{Wainwright1997} as
\begin{align}
\begin{split}
& \dot{H} = - H^2 - \frac{1}{3} \sigma _{AB}\sigma ^{AB} -\frac{1}{6} (\rho + 3P ) \ , \\
& \dot{\sigma }_{AB} = -3H \sigma _{AB} + \left( \epsilon _{CDA}\sigma _{BC} + \epsilon _{CDB}\sigma _{AC} \right) \Omega _D 
- {}^{(3)}R_{AB} + \frac{1}{3}{}^{(3)}R \delta _{AB} + \pi _{AB} \ , \\
&  3H^2 - \frac{1}{2} \sigma _{AB}\sigma ^{AB} + \frac{1}{2}{}^{(3)}R 
= \rho  \ , \\
&  3\sigma _{AB}a_B - \epsilon _{ABC} \sigma _{BD}n_{DC}= q_A \ 
\end{split}
\end{align}
along with the irreducible decomposition of energy-momentum tensor
\begin{equation}
T_{00} = \rho \ , \quad T_{0A} = -q_A \ , \quad T_{AB} = P \delta _{AB} + \pi _{AB} \ , \quad \pi _{AA} = 0 \ .
\end{equation} 
 The time evolution of the curvature variables 
$n_{AB}$ and $a_A$
are determined  through the remaining Jacobi identities.
The evolution equation for the scalar field is given by
\begin{equation}
\ddot{\varphi } + 3H \dot{\varphi } -\alpha V(\varphi ) -
\frac{\lambda }{2}f^2(\varphi ) \left( E^2 - B^2 \right) 
=0 \ ,
\end{equation}
where 
\begin{equation}
E^2 = E^a_A E^a_A \ , \quad B^2 = B^a_A B^a_A \ .
\end{equation}
The equations of motion for the gauge field are written as
\begin{equation}
d{}^\ast \mathbf{F} + g [\mathbf{A} \wedge {}^\ast \mathbf{F} ] +f^{-2}(\varphi ) 
df^2(\varphi ) \wedge {}^\ast \mathbf{F} = 0 \ , 
\end{equation}
with the dual field strength given by
\begin{equation}
{}^\ast \mathbf{F} = 
- B_A^a T_a \mathbf{e}^A \wedge \mathbf{e}^0 + \frac{1}{2} 
\epsilon _{ABC} E^a_C T_a \mathbf{e}^A \wedge \mathbf{e}^B \ .
\end{equation}
In terms of the components, they become
\begin{align}
\begin{split}
0 =& \dot{E}^a_A + 2\left( H + \frac{f_{,\varphi }}{f}\dot{\varphi } \right) 
E^a_A - \sigma _{AB} E^a_B + \epsilon _{ABC} \Omega _B E^a_C \\
& + n_{AB}B^a_B + \epsilon _{ABC}a_B B^a_C -g\epsilon _{abc}\epsilon _{ABC} 
A^b_B B^c_C + g\epsilon _{abc}A^b_0 E^c_A \ ,
\end{split} \label{eq:dynamicalE}  \\
0 =& 2a_A E^a_A - g\epsilon _{abc}A^b_A E^c_A \ . \label{eq:constraintE}
\end{align}
The energy-momentum tensor components of the scalar field are 
\begin{align}
\begin{split}
\rho_{\varphi } = & \frac{1}{2} \dot{\varphi }^2  + V(\varphi ) \ , \\
P_{\varphi } =& \frac{1}{2} \dot{\varphi }^2 - V(\varphi ) \ .
\end{split}
\end{align}
 
For the gauge field, 
its energy density, energy flux, pressure and anisotropic stress are given by
\begin{align}
\begin{split}
& \rho_{\rm YM} = \frac{1}{2}\left( E^2 + B^2 \right) \ , 
\quad q_A = \epsilon _{ABC}
 E^a_B B^a_C \ , \\
& P_{\rm YM} = \frac{1}{6} \left( E^2 + B^2 \right) \ , 
\quad \pi _{AB} = E^a_A E^a_B 
+ B^a_A B^a_B - \frac{1}{3} \left( E^2 +B^2 \right) \delta _{AB} \ .
\end{split}
\end{align}
Here we drop the subscript YM for the energy flux and anisotropic stress
because those terms contain only the gauge field contribution.

In this article, we focus our attention on Bianchi class A space-times 
characterized by $a_A =0$, which includes Type I, II, VI$_0$, VII$_0$, VIII and IX.
The $SU(2)$ gauge symmetry can be used to set
\begin{equation}
A^a_0 = 0 \ .
\end{equation}
In Bianchi class A, requiring $q_A = 0$ is equivalent to
\begin{equation}
\sigma _{C[A}n_{B]C} = 0 \ , \label{eq:diagCon}
\end{equation}
which enables us to find a frame in which all the anisotropic variables
become diagonal. This ansatz reduces the number of variables significantly.

We then parametrize the diagonal elements as follows:
\begin{align}
\begin{split}
& \sigma _{AB} = \left(   \begin{array}{ccc}
    -2\sigma _+  & 0 & 0 \\ 
    0 & \sigma _+ + \sqrt{3}\sigma _- & 0 \\ 
    0 & 0 & \sigma _+ - \sqrt{3} \sigma _- \\ 
  \end{array} \right) \ , \\
& n_{AB} = \left(   \begin{array}{ccc}
    n_1  & 0 & 0 \\ 
    0 & n_2 & 0 \\ 
    0 & 0 & n_3 \\ 
  \end{array} \right) \ , \quad A^a_A = \left(  \begin{array}{ccc}
    A_1  & 0 & 0 \\ 
    0 & A_2 & 0 \\ 
    0 & 0 & A_3 \\ 
  \end{array} \right) \ , \\
& E^a_A = \left(   \begin{array}{ccc}
    E_1  & 0 & 0 \\ 
    0 & E_2 & 0 \\ 
    0 & 0 & E_3 \\ 
  \end{array} \right) \ , \quad B^a_A = \left(  \begin{array}{ccc}
    B_1  & 0 & 0 \\ 
    0 & B_2 & 0 \\ 
    0 & 0 & B_3 \\ 
  \end{array} \right) \ .
\end{split}
\end{align}
We introduced the obvious matrix notation for gauge field components. 
Note that the constraint equations for the gauge field (\ref{eq:constraintB})
and (\ref{eq:constraintE}) are automatically satisfied. One can also show
that $\Omega _A = 0$. 

%%%%%%%%%%%%%%%%%%%%%%%%%%%%%%%%%%%%%%%%%%%%%%%%%%%%%%%%%%%
%%%%%%%%%%%%%%%%%%%%%%%%%%%%%%%%%%%%%%%%%%%%%%%%%%%%%%%%%%%
%%%%%%%%%%%%%%%%%%%%%%%%%%%%%%%%%%%%%%%%%%%%%%%%%%%%%%%%%%%
%\subsection{Expansion normalization}
%%%%%%%%%%%%%%%%%%%%%%%%%%%%%%%%%%%%%%%%%%%%%%%%%%%%%%%%%%%
%%%%%%%%%%%%%%%%%%%%%%%%%%%%%%%%%%%%%%%%%%%%%%%%%%%%%%%%%%%
%%%%%%%%%%%%%%%%%%%%%%%%%%%%%%%%%%%%%%%%%%%%%%%%%%%%%%%%%%%
Normalization of the variables with respect to the Hubble expansion rate
$H$ has proven to be fruitful for $U(1)$ gauge fields \cite{Yamamoto2012} 
and the isotropic non-Abelian gauge fields \cite{Maeda2012}. 
In the presence
of spatial curvature, the standard normalization in literature is
\begin{equation}
\Sigma _{\pm } = \frac{\sigma _{\pm }}{H} \ , \quad N_A =\frac{n_A}{H} \ , 
\quad \varpi = \frac{\dot{\varphi }}{H} \ , \quad \Omega _V = \frac{V_0 
e^{-\alpha \varphi }}{3H^2} \ . \label{eq:normal1}
\end{equation}
On the other hand, the analysis in \cite{Maeda2012} suggests the following
normalization for the gauge field:
\begin{equation}
\mathcal{A}_A = \sqrt{\frac{ge^{\lambda \varphi }}{H}}A_A \ , \quad 
\mathcal{E}_A = \frac{e^{\lambda \varphi /2}E_A }{H} \ , \quad \Gamma 
= \sqrt{\frac{g}{e^{\lambda \varphi /2}H}} \ .
\end{equation}
This does not result in polynomial equations because of the appearance of
$\Gamma ^{-1}$. Since the important fixed points lie on the $U(1)$ boundary
of the system for which $\Gamma =0$, this choice of variables is rather 
inconvenient. 
It turns out that modifying the variables $N_A$ by
\begin{equation}
\mathcal{N}_A = \frac{N_A}{\Gamma }  \label{eq:normalN}
\end{equation}
is the most helpful for our purpose. In order to tidy up the equations, 
normalized
magnetic field is derived from (\ref{eq:magnetic}) 
\begin{equation}
\mathcal{B}_1 = \mathcal{A}_2 \mathcal{A}_3 - \mathcal{N}_1 \mathcal{A}_1 \  ,
 \quad \mathcal{B}_2 = \mathcal{A}_3 \mathcal{A}_1 - \mathcal{N}_2 
\mathcal{A}_2 \ , \quad \mathcal{B}_3 = \mathcal{A}_1 \mathcal{A}_2 - 
\mathcal{N}_3 \mathcal{A}_3 \ , \label{eq:magdef}
\end{equation} 
and the relevant density parameters are defined as
\begin{equation}
\Omega _K = \frac{1}{6}\varpi ^2  \ , \quad \Omega _E = \frac{1}{6} 
\left( \mathcal{E}_1^2 + \mathcal{E}_2^2 + \mathcal{E}_3^2 \right) \ , 
\quad \Omega _B = \frac{1}{6} \left( \mathcal{B}_1^2 + \mathcal{B}_2^2 
+ \mathcal{B}_3^2 \right) \ . \label{eq:densityParameters}
\end{equation}
Introducing the e-folding number as the time coordinate by
\begin{equation}
d\tau = H dt \ , 
\end{equation}
whose derivatives are denoted by primes, and defining the deceleration 
parameter
\begin{equation}
q = -1 - \frac{H^{\prime }}{H} \ , 
\end{equation}
we obtain the following dynamical system.

\begin{description}
\item[ Einstein equations] 
\begin{align}
\begin{split}
& 1 = \Omega _K + \Omega _V + \Omega _E + \Omega _B + \Sigma _{+}^2 
+ \Sigma _{-}^2 \\
& \quad \quad + \frac{\Gamma ^2}{12} \left[ \mathcal{N}_1^2 + \mathcal{N}_2^2 
+ \mathcal{N}_3^2 -2\left( \mathcal{N}_1 \mathcal{N}_2 + \mathcal{N}_2 
\mathcal{N}_3 + \mathcal{N}_3 \mathcal{N}_1 \right) \right] \ , 
\end{split}\label{eq:Friedmann}\\
& q = 2 \left( \Sigma _{+}^2 +\Sigma _{-}^2 \right) + 2\Omega _K + \Omega _E
 + \Omega _B - \Omega _V \ , 
\label{deceleration_parameter}\\
\begin{split}
& \Sigma _{+}^{\prime } = (q-2) \Sigma _{+} -\frac{1}{6} \left( 
\mathcal{E}_2^2 + \mathcal{E}_3^2 -2\mathcal{E}_1^2 + \mathcal{B}_2^2 
+ \mathcal{B}_3^2 - 2\mathcal{B}_1^2 \right) \\ 
& \quad \quad  - \frac{\Gamma ^2 }{6}\left[ \left( \mathcal{N}_2 
-\mathcal{N}_3 \right) ^2 - \mathcal{N}_1 \left( 2\mathcal{N}_1 
-\mathcal{N}_2 - \mathcal{N}_3 \right)  \right] \ , 
\end{split} \label{eq:shearPlus}\\
\begin{split}
& \Sigma _{-}^{\prime } = (q-2) \Sigma _{-} - \frac{1}{2\sqrt{3}} 
\left( \mathcal{E}_2^2 - \mathcal{E}_3^2 +\mathcal{B}_2^2 -\mathcal{B}_3^2 
\right) \\
& \quad \quad - \frac{\Gamma ^2 }{2\sqrt{3}}\left( \mathcal{N}_3 
-\mathcal{N}_2 \right) \left( \mathcal{N}_1 -\mathcal{N}_2 -\mathcal{N}_3 
\right) \ , 
\end{split}
\end{align}
\item[ Jacobi identities ]
\begin{align}
\mathcal{N}_1^{\prime } =&\frac{1}{2} \left( q -1 + \frac{\lambda }{2} \varpi 
- 8\Sigma _{+} \right) \mathcal{N}_1 , \label{eq:curv1} \\
\mathcal{N}_2^{\prime } =& \frac{1}{2} \left( q-1 + \frac{\lambda }{2} \varpi 
+4\Sigma _{+} +4\sqrt{3}\Sigma _{-} \right) \mathcal{N}_2 , \\
\mathcal{N}_3^{\prime } =& \frac{1}{2} \left( q-1 +\frac{\lambda }{2} \varpi 
+4\Sigma _{+} -4\sqrt{3}\Sigma _{-} \right) \mathcal{N}_3 . \label{eq:curv3}
\end{align}
\item[ Definition of electric fields ]
\begin{align}
\mathcal{A}_1^{\prime } =& \frac{1}{2} \left( q-1 +\frac{\lambda }{2} \varpi 
+ 4\Sigma _{+} \right) \mathcal{A}_1 - \Gamma \mathcal{E}_1 , \label{eq:a1} \\
\mathcal{A}_2^{\prime } =& \frac{1}{2} \left( q-1 +\frac{\lambda }{2} \varpi 
- 2\Sigma _{+} -2\sqrt{3} \Sigma _{-} \right) \mathcal{A}_2 -\Gamma 
\mathcal{E}_2 , \label{eq:a2} \\
\mathcal{A}_3^{\prime } =& \frac{1}{2} \left( q-1 +\frac{\lambda }{2} \varpi
 -2\Sigma _{+} +2\sqrt{3}\Sigma _{-} \right) \mathcal{A}_3 
-\Gamma \mathcal{E}_3 . \label{eq:a3} 
\end{align}
\item[ Dynamical Yang-Mills equations ]
\begin{align}
\mathcal{E}_1^{\prime } =& \left( q-1 -\frac{\lambda }{2} \varpi -2\Sigma _{+}
 \right) \mathcal{E}_1 -\Gamma \left( \mathcal{N}_1 \mathcal{B}_1 
- \mathcal{A}_2 \mathcal{B}_3 - \mathcal{A}_3 \mathcal{B}_2 \right) \ , 
\label{eq:ele1} \\
\begin{split}
\mathcal{E}_2^{\prime } =& \left( q-1 -\frac{\lambda }{2} \varpi +\Sigma _{+} 
+\sqrt{3}\Sigma _{-} \right) \mathcal{E}_2  -\Gamma \left( \mathcal{N}_2 
\mathcal{B}_2 - \mathcal{A}_3 \mathcal{B}_1 - \mathcal{A}_1 \mathcal{B}_3 
\right) \ , 
\end{split}\\
\begin{split}
\mathcal{E}_3^{\prime } =& \left( q-1 -\frac{\lambda }{2} \varpi +\Sigma _{+} 
-\sqrt{3} \Sigma _{-} \right) \mathcal{E}_3  - \Gamma \left( \mathcal{N}_3 
\mathcal{B}_3 - \mathcal{A}_1 \mathcal{B}_2 - \mathcal{A}_2 \mathcal{B}_1 
\right)  \ .
\end{split} \label{eq:ele3} 
\end{align}
\item[ Scalar field equations ]
\begin{align}
& \varpi ^{\prime } = (q-2) \varpi + 3\alpha \Omega _V  + 3\lambda 
\left( \Omega _E - \Omega _B \right) \ , \label{scalar1} \\
& \Omega _V^{\prime } = \left[ 2 (q+1 ) -\alpha \varpi \right] \Omega _V \ .
 \label{scalar2}
\end{align}
\item[ Evolution of the normalised Yang-Mills coupling ]
\begin{equation}
\Gamma ^{\prime } = \frac{1}{2} \left( q+1 -\frac{\lambda }{2}\varpi \right) 
\Gamma \ . \label{eq:GammaEv}
\end{equation}
\end{description}
Note that (\ref{eq:GammaEv}) was derived essentially from the equation for $H^{\prime}$.
These equations define a 13-dimensional autonomous dynamical system which
is everywhere regular. For most purposes, it is convenient to take 
$\Sigma _{\pm },
\mathcal{N}_A ,\varpi , \mathcal{A}_A , \mathcal{E}_A$ and $\Gamma $ as the 
independent
variables with $\Omega _V $ being given by Friedmann constraint 
(\ref{eq:Friedmann}) and the deceleration parameter $q$ by 
Raychauduri equation (\ref{deceleration_parameter}).  Nevertheless, when
we solve the equations numerically in the later sections, we will use the above
14 dynamical equations (\ref{eq:shearPlus}) - (\ref{eq:GammaEv}) 
and monitor the
constraint (\ref{eq:Friedmann}) for consistency check.
From physical point of view, $\mathcal{B}_A$'s would be preferred over 
$\mathcal{A}_A$'s
since their amplitudes represent the dynamical importance of magnetic fields 
in the
space-time dynamics. While this choice is not practical since solving 
(\ref{eq:magdef})
for $\mathcal{B}_A$ introduces non-polynomial terms, it later proves to be 
helpful
to write down the evolution equations for magnetic field:
\begin{align}
\mathcal{B}_1^{\prime } =& \left( q-1 +\frac{\lambda }{2} \varpi -2\Sigma _{+} 
\right) \mathcal{B}_1 +\Gamma \left( \mathcal{N}_1 \mathcal{E}_1 
- \mathcal{A}_2 \mathcal{E}_3 - \mathcal{A}_3 \mathcal{E}_2 \right) \ , 
\label{eq:mag1} \\
\begin{split}
\mathcal{B}_2^{\prime } =& \left( q-1 +\frac{\lambda }{2} \varpi +\Sigma _{+} 
+\sqrt{3}\Sigma _{-} \right) \mathcal{B}_2  +\Gamma \left( \mathcal{N}_2 
\mathcal{E}_2 - \mathcal{A}_3 \mathcal{E}_1 - \mathcal{A}_1 \mathcal{E}_3 
\right) \ , 
\end{split}\\
\begin{split}
\mathcal{B}_3^{\prime } =& \left( q-1 +\frac{\lambda }{2} \varpi +\Sigma _{+} -
\sqrt{3} \Sigma _{-} \right) \mathcal{B}_3  + \Gamma \left( \mathcal{N}_3 
\mathcal{E}_3 - \mathcal{A}_1 \mathcal{E}_2 - \mathcal{A}_2 \mathcal{E}_1 
\right)  \ .
\end{split} \label{eq:mag3}
\end{align}

%%%%%%%%%%%%%%%%%%%%%%%%%%%%%%%%%%%%%%%%%%%%%%%%%%%%%%%%%%%
%%%%%%%%%%%%%%%%%%%%%%%%%%%%%%%%%%%%%%%%%%%%%%%%%%%%%%%%%%%
%%%%%%%%%%%%%%%%%%%%%%%%%%%%%%%%%%%%%%%%%%%%%%%%%%%%%%%%%%%
\section{Inflationary solutions and their local stability}\label{sec:stability}
%%%%%%%%%%%%%%%%%%%%%%%%%%%%%%%%%%%%%%%%%%%%%%%%%%%%%%%%%%%
%%%%%%%%%%%%%%%%%%%%%%%%%%%%%%%%%%%%%%%%%%%%%%%%%%%%%%%%%%%
%%%%%%%%%%%%%%%%%%%%%%%%%%%%%%%%%%%%%%%%%%%%%%%%%%%%%%%%%%%

First of all, we summarize the known properties of inflationary solutions with
gauge-kinetic coupling, studied in Bianchi type I spacetime with 
 three (or more)
Abelian gauge fields (Abelian Bianchi I) \cite{Yamamoto2012} 
and Freedmann-Lem\^{a}tre-Robertson-Walker spacetime with non-Abelian 
gauge field (non-Abelian FLRW) \cite{Maeda2012}.
\begin{itemize}
\item There are three types of Abelian inflationary fixed points with 
non-vanishing 
electric fields and zero magnetic fields. The configuration of electric fields
is either singlet, dyad or triad. The electromagnetic duality guarantees the 
existence of the counterparts with non-vanishing magnetic fields and zero
electric fields. 
\item In Abelian Bianchi I, the isotropic triad solutions are stable. 
The other
anisotropic ones are unstable against perturbations of gauge field components
perpendicular to the background ones. 
\item In non-Abelian FLRW, the duality is broken due to the Yang-Mills 
interaction
and the isotropic inflation with electric components 
(Isotropic Electric Inflation)
becomes unstable. The inflation with magnetic
triad (Isotropic Magnetic Inflation) remains to be a stable local attractor.
\end{itemize}
It turns out that in the full non-Abelian system under consideration, 
inflationary
fixed points are possible only for $\Gamma =0$. This invariant set has the 
identical
structure as Abelian Bianchi I except that there are additional curvature 
variables $\mathcal{N}_A$ which decouple from the others. Hence one
expects that those fixed points share their eigenvalues with 
Abelian Bianchi I, which
means all but the isotropic magnetic inflation are known to be unstable. 
However,
it should be mentioned that Abelian and non-Abelian systems may in principle 
exhibit different dynamics. If we were to start from an Abelian triplet, namely
 $g=0$,
our normalization in the previous section would not have worked. There is no 
need 
to introduce $\mathcal{N}_A$ as $N_A$ are good variables. It is also noted 
that the magnetic
field components would be given by
\begin{equation}
B^a_A = -n_{AB}A^a_B \ , 
\end{equation}
which means they should all be zero for Bianchi type I (where $n_{AB}=0$).
To include both electric and magnetic fields in Abelian Bianchi I,
 we would have
to add terms dependent on spatial coordinates to the vector potential, or 
working 
on electric and magnetic fields from the beginning without using vector 
potential. 
For this reason, also for clarity and completeness, we present the exhaustive 
list of 
inflationary fixed points with their eigenvalues below despite that there are 
some
overlaps with the contents of \cite{Yamamoto2012}. It will also help to 
understand the existence of Abelian ($\Gamma =0$) fixed points 
that appear as asymptotic 
solutions of non-Abelian system. The dynamics in the Abelian triplet model is
discussed in Appendix A.

%%%%%%%%%%%%%%%%%%%%%%%%%%%%%%%%%%%%%%%%%%%%%%%%%%%%%%%%%%%
%%%%%%%%%%%%%%%%%%%%%%%%%%%%%%%%%%%%%%%%%%%%%%%%%%%%%%%%%%%
%%%%%%%%%%%%%%%%%%%%%%%%%%%%%%%%%%%%%%%%%%%%%%%%%%%%%%%%%%%
\subsection{The conventional power-law inflation without gauge field}\label{sec:scalar}
%%%%%%%%%%%%%%%%%%%%%%%%%%%%%%%%%%%%%%%%%%%%%%%%%%%%%%%%%%%
%%%%%%%%%%%%%%%%%%%%%%%%%%%%%%%%%%%%%%%%%%%%%%%%%%%%%%%%%%%
%%%%%%%%%%%%%%%%%%%%%%%%%%%%%%%%%%%%%%%%%%%%%%%%%%%%%%%%%%%
This is an obvious fixed point obtained by setting
\begin{align}
\begin{split}
& \varpi = \alpha  \ ,\quad q = \frac{\alpha ^2 -2}{2} \ , \quad \Omega _V 
= -\frac{\alpha ^2 -6}{6} \ ,\\
& \Sigma _{\pm } = \mathcal{N}_1 = \mathcal{N}_2 = \mathcal{N}_3 = \Omega _E 
= \Omega _B = \Gamma = 0 \ . \label{eq:fix1}
\end{split}
\end{align}
Assuming the potential energy is positive, this solution is physical for 
$\alpha < \sqrt{6}$ 
and supports an accelerated expansion for $\alpha < \sqrt{2}$. 
Note that the power exponent of the volume expansion, 
which we define by $p=Ht$, 
is given by $p=2/\alpha^2$.

Assuming the perturbed variables are proportional to $\exp(\omega\tau)$,
and denoting the eigenvalues 
by $\omega$ with subscript representing the corresponding eigenmode, we 
obtain the following:
\begin{align}
\begin{split}
& \omega _{\varpi } = -\frac{1}{2} \left( \alpha ^2 + 6\right) \ , \quad 
\omega _{\Sigma _+} = \omega _{\Sigma _-} = \frac{\alpha ^2 -6}{2} \ , \\
& \omega _{\mathcal{N}_1}= \omega _{\mathcal{N}_2} = \omega _{\mathcal{N}_3} 
= \omega _{\mathcal{A}_1} = \omega _{\mathcal{A}_2} = \omega _{\mathcal{A}_3}
 = \frac{\alpha ^2 + \alpha \lambda -4}{4} \ , \\
& \omega _{\mathcal{E}_1} = \omega_{\mathcal{E}_2} = \omega _{\mathcal{E}_3} 
= \frac{\alpha ^2 -\alpha \lambda -4}{2} \ , \quad \omega _{\Gamma } 
= \frac{\alpha ^2 - \alpha \lambda }{4} \ .
\end{split}
\end{align}
It becomes unstable when
\begin{equation}
\alpha \left( \alpha + | \lambda | \right) > 4 \ .
\end{equation}
Note that the instability of $\Gamma $ itself does not leave any observable
signature as long as $\mathcal{N}_A , \mathcal{A}_A , \mathcal{E}_A,
A = 1,2,3$ stay small  (in fact, we even did not have to set it zero
 in (\ref{eq:fix1})). 
This is a simplest way of illustrating the violation
of cosmic no-hair conjecture \cite{Wald1983, Kitada1992, Kitada1993, Maleknejad2012a} 
through inflaton-gauge interactions.

Although this pawer-law single scalar fixed point is not of our interest as 
a model of inflation since it is unstable, it turns out that the 
time-evolution of the 
space-time geometry is well-approximated by this solution when the gauge field 
rapidly oscillates as we will see in the next section.

%%%%%%%%%%%%%%%%%%%%%%%%%%%%%%%%%%%%%%%%%%%%%%%%%%%%%%%%%%%
%%%%%%%%%%%%%%%%%%%%%%%%%%%%%%%%%%%%%%%%%%%%%%%%%%%%%%%%%%%
%%%%%%%%%%%%%%%%%%%%%%%%%%%%%%%%%%%%%%%%%%%%%%%%%%%%%%%%%%%
\subsection{Inflation with the electric components (Electric Inflation)}
%%%%%%%%%%%%%%%%%%%%%%%%%%%%%%%%%%%%%%%%%%%%%%%%%%%%%%%%%%%
%%%%%%%%%%%%%%%%%%%%%%%%%%%%%%%%%%%%%%%%%%%%%%%%%%%%%%%%%%%
%%%%%%%%%%%%%%%%%%%%%%%%%%%%%%%%%%%%%%%%%%%%%%%%%%%%%%%%%%%
Within the invariant set $\Gamma =0$, there is the electro-magnetic 
duality so that each fixed point for 
non-vanishing electric components with 
 $\mathcal{B}_A =0$ has a counterpart (non-vanishing electric components with
 $\mathcal{E}_A =0$). 
It does not mean the stability  is the
same for both, however, since the full state space does not support the 
duality. As we have to use vector potential instead of magnetic field, 
magnetic cases are more complicated than electric ones. Hence, 
we first list the fixed points with electric components
although magnetic ones are physically more important and 
will be studied numerically in later sections.

%%%%%%%%%%%%%%%%%%%%%%%%%%%%%%%%%%%%%%%%%%%%%%%%%%%%%%%%%%%
%%%%%%%%%%%%%%%%%%%%%%%%%%%%%%%%%%%%%%%%%%%%%%%%%%%%%%%%%%%
%%%%%%%%%%%%%%%%%%%%%%%%%%%%%%%%%%%%%%%%%%%%%%%%%%%%%%%%%%%
\subsubsection{Anisotropic electric inflation {\rm (}single component{\rm )}}
%%%%%%%%%%%%%%%%%%%%%%%%%%%%%%%%%%%%%%%%%%%%%%%%%%%%%%%%%%%
%%%%%%%%%%%%%%%%%%%%%%%%%%%%%%%%%%%%%%%%%%%%%%%%%%%%%%%%%%%
%%%%%%%%%%%%%%%%%%%%%%%%%%%%%%%%%%%%%%%%%%%%%%%%%%%%%%%%%%%
Let us assume only one electric component is non-trivial, i.e.,
$\mathcal{E}_1 \neq 0 , \mathcal{E}_2 = \mathcal{E}_3 = 0$.
We obtain the following fixed point: 
\begin{align}
\begin{split}
& \varpi = \frac{12(\alpha - \lambda )}{(\alpha - \lambda )(\alpha -3\lambda ) 
+8 } \ , \quad q = \frac{5\alpha ^2 - 2\alpha \lambda -3\lambda ^2 -8}{
(\alpha - \lambda )(\alpha - 3\lambda )+8} \ , \\
& \Sigma _+ = 2 \frac{\alpha ^2 - \alpha \lambda -4}{(\alpha - \lambda )
(\alpha - 3\lambda ) +8} \ , \quad \Sigma _- = \Omega _B = \Gamma 
= \mathcal{E}_2 =\mathcal{E}_3 = 0 \ , \\
& \Omega _V = -3\frac{(\lambda ^2 - \lambda \alpha +4 )(\alpha ^2 + 2\alpha 
\lambda -3\lambda ^2 -8)}{\left[ (\alpha - \lambda )(\alpha - 3\lambda )
+8 \right] ^2 } \ ,  \\
& \Omega _E  = -3 \frac{ \left( \alpha ^2 - \alpha \lambda -4 \right) 
\left( \alpha ^2 + 2\alpha \lambda -3\lambda ^2 -8 \right)}{\left[ 
\left( \alpha -\lambda \right) \left( \alpha - 3\lambda \right) 
+8 \right] ^2 } \ .
\end{split} \label{eq:singleE}
\end{align}
The power exponent $p$ of the volume expansion is given by 
\begin{align}
p:={1\over q+1}=
{(\alpha-\lambda)(\alpha-3\lambda)+8\over 6\alpha(\alpha-\lambda)}
\,.
\end{align}

We are interested in $\lambda <0$ for which the conventional power-law 
inflation
may become unstable against electric field perturbations. Then the existence 
condition ($\Omega _E , \Omega _V >0$) reduces to
\begin{equation}
\alpha ^2 + 2\alpha \lambda -3\lambda ^2 -8 <0 \ , \quad \alpha ^2 - \alpha 
\lambda -4 >0 \ ,
\end{equation}
which implies $\Sigma _+ >0$. 

Since we could have equally assumed 
$\mathcal{E}_2 \neq 0$
or $\mathcal{E}_3 \neq 0$, there are two equilibrium points 
which 
represent physically identical space-times, which can be obtained by
rotation 
\begin{equation}
\left(   \begin{array}{c}
    \Sigma _+ \\ 
    \Sigma _-  \\ 
  \end{array} \right) = \left(   \begin{array}{cc}
    \cos \phi  & \sin \phi \\ 
    -\sin \phi & \cos \phi  \\ 
  \end{array} \right) \left(   \begin{array}{c}
    2\frac{\alpha ^2 + \alpha \lambda -4}{(\alpha + \lambda )(\alpha 
+ 3\lambda )+8}  \\ 
    0 \\ 
  \end{array} \right) \label{shear_rot} \,, 
\end{equation}
where the rotation angle is $\phi = 2\pi /3 $ or 
 $4\pi /3$.

Upon linearization in terms of perturbed variables, 
most of them are by themselves eigenmodes.
 The only
complication arises from 
\begin{align}
\begin{split}
& \delta \Sigma _+^{\prime } =  \left( q-2+6\Sigma _+^2 \right) \delta 
\Sigma _+ + \Sigma _+ \varpi \delta \varpi + 2 \left( \Sigma _+ + 1\right) 
\delta \Omega _E \ , \\
& \delta \varpi ^{\prime } = \left( q-2+\varpi ^2 -\alpha \varpi \right) 
\delta \varpi + 6\left( \varpi - \alpha \right) \Sigma _+ \delta \Sigma _+  
+ \left[ 2\varpi - 3\left( \alpha - \lambda \right) \right]  \delta \Omega _E 
\ , \\
& \delta \Omega _E^{\prime } = 2\Omega _E \left( 2 \left( 3\Sigma _+ - 1 
\right) \delta \Sigma _+ - \left( \varpi + \frac{\lambda }{2} \right) \delta 
\varpi + 2\delta \Omega _E \right) \ , 
\end{split}
\end{align}
where the symbol $\delta $ signifies the perturbation of the following 
quantity 
while the variables without $\delta $ refer to their values at the fixed 
point and 
we eliminated $\delta \Omega _V$ by using (\ref{eq:Friedmann}). 
They can be decoupled by singling out
\begin{equation}
\left( \delta \varpi + \frac{\alpha - 3\lambda }{2} \delta \Sigma _+ 
\right) ^{\prime } = \left( q -2 \right) \left( \delta \varpi + \frac{\alpha 
- 3\lambda }{2} \delta \Sigma _+ \right) \ .
\end{equation}
The eigenvalues for the rest of $\delta \Sigma _+, \delta\varpi$ and 
$\delta \Omega _E$ are the roots of the quadratic equation
\begin{equation}
 x^2- (q-2)x - 3\Omega _E \frac{\lambda \left( \alpha - \lambda \right) ^2 
\left( \alpha - 3\lambda \right) - 4\left( \alpha - \lambda \right) 
\left( \alpha - 5\lambda \right) -32}{\left( \alpha - \lambda \right) 
\left( \alpha - 3\lambda \right) +8}   = 0 \ . 
\end{equation}
We obtain the following eigenvalues.
\begin{align}
\begin{split}
& \omega _{\Sigma _+ \mathchar`- \varpi \mathchar`- \Omega _E} = \frac{q-2 \pm \sqrt{\left( q-2 
\right) ^2 + 12\Omega _E \left( \lambda \left( \alpha - \lambda \right) 
-4\right)}}{2} \ , \\
& \omega _{\Sigma _+ \mathchar`- \varpi } = \omega _{\Sigma _-} = q-2 \ , \quad 
\omega _{\mathcal{N}_1} = -6 \frac{\left( \alpha - \lambda \right) ^2 
-4}{\left( \alpha - \lambda \right) \left( \alpha -3\lambda \right) +8} \ , 
\\
& \omega _{\mathcal{N}_2} = \omega _{\mathcal{N}_3} = \omega _{\mathcal{A}_1}
 = 6 \frac{\alpha ^2 -\lambda ^2 -4}{\left( \alpha -\lambda \right) 
\left( \alpha -3\lambda \right) + 8 } \ , \quad \omega _{\mathcal{E}_2} 
= \omega _{\mathcal{E}_3} = 3\Sigma _+ \ , \\
& \omega _{\mathcal{A}_2} = \omega _{\mathcal{A}_3} = 6 \frac{\lambda 
\left( \alpha - \lambda \right) }{\left( \alpha - \lambda \right) 
\left( \alpha - 3\lambda \right) + 8 } \ , \quad \omega _{\Gamma } 
= \frac{3\left( \alpha - \lambda \right) ^2 }{\left( \alpha - \lambda \right) 
\left( \alpha -3\lambda \right) + 8 } \ .
\end{split}
\end{align}
We find 
 the unstable orthogonal electric components 
$\omega _{\mathcal{E}_2},
\omega _{\mathcal{E}_3}$, which represents the tendency to preferring 
isotropic configuration.
Notice that the magnitude of these positive eigenvalues is of order 
$\Sigma _+$.
This means that once a trajectory approaches an anisotropic inflation, 
it generically
takes at least $\sim \Sigma _+^{-1}$ e-foldings before it leaves
 the fixed point. This estimate
is roughly in agreement with the numerical calculations for Abelian case
(see, however, the excepional case discussed in section
\ref{Quasi-single-component}).
$\Gamma $ also becomes unstable for $\lambda <0$. 
This growing effective YM coupling leads to a complicated
dynamical behaviour which cannot be understood in terms of fixed points.

%%%%%%%%%%%%%%%%%%%%%%%%%%%%%%%%%%%%%%%%%%%%%%%%%%%%%%%%%%%
%%%%%%%%%%%%%%%%%%%%%%%%%%%%%%%%%%%%%%%%%%%%%%%%%%%%%%%%%%%
%%%%%%%%%%%%%%%%%%%%%%%%%%%%%%%%%%%%%%%%%%%%%%%%%%%%%%%%%%%
\subsubsection{Anisotropic electric inflation {\rm (}double components{\rm )}}
%%%%%%%%%%%%%%%%%%%%%%%%%%%%%%%%%%%%%%%%%%%%%%%%%%%%%%%%%%%
%%%%%%%%%%%%%%%%%%%%%%%%%%%%%%%%%%%%%%%%%%%%%%%%%%%%%%%%%%%
%%%%%%%%%%%%%%%%%%%%%%%%%%%%%%%%%%%%%%%%%%%%%%%%%%%%%%%%%%%
There is another anisotropic inflationary fixed point with
$\mathcal{E}_1 = 0 , \mathcal{E}_2 \neq 0 , \mathcal{E}_3 \neq 0 $, which 
is given by
\begin{align}
\begin{split}
& \varpi = \frac{6\left( \alpha -2\lambda \right)}{\left( \alpha - \lambda 
\right) \left( \alpha -3\lambda \right) +2} \ , \quad q = \frac{2\alpha ^2 
-2\alpha \lambda -3\lambda ^2 -2}{\left( \alpha - \lambda \right) 
\left( \alpha -3\lambda \right) +2} \ , \\
& \Sigma _+ = - \frac{\alpha ^2 -\alpha \lambda - 4}{\left( \alpha - \lambda 
\right) \left( \alpha - 3\lambda \right) +2 } \ , \quad \Sigma _- 
=\mathcal{E}_1 = \Omega _B = \Gamma = 0 \ , \\
& \mathcal{N}_1 = \mathcal{N}_2 = \mathcal{N}_3 = 0 \ , \quad \Omega _V 
= 3 \frac{\left( \lambda ^2 - \alpha \lambda +2 \right) \left( 3\lambda ^2 
-2\alpha \lambda +2 \right)}{\left[ \left( \alpha - \lambda \right) 
\left( \alpha - 3\lambda \right) +2 \right] ^2} \ ,  \\ 
& \Omega _E = \frac{1}{3} \mathcal{E}_2^2 = \frac{1}{3}\mathcal{E}_3^2 
= 3 \frac{\left( \alpha ^2 -\alpha \lambda -4 \right) \left( 3\lambda ^2 
-2\alpha \lambda +2\right) }{\left[ \left( \alpha -\lambda \right) 
\left( \alpha - 3\lambda \right) +2 \right] ^2 } \ .
\end{split} \label{eq:electric_double}
\end{align}
The power exponent $p$ of the volume expansion is given by 
\begin{align}
p=
{(\alpha-\lambda)(\alpha-3\lambda)+2\over 3\alpha(\alpha-2\lambda)}
\,.
\end{align}

The existence condition for $\lambda <0$ is 
\begin{equation}
3\lambda ^2 - 2\alpha \lambda +2 > 0 \ , \quad \alpha ^2 -\alpha \lambda 
-4 > 0 \ ,
\end{equation}
which then implies $\Sigma _+ < 0$. The same type of symmetry as
 (\ref{shear_rot})
exists and there are other physically identical fixed points with two 
non-vanishing
electric components. 

The linearized equations take a similar form as the single component fixed 
point.
The coupled part is given by
\begin{align}
\begin{split}
 \delta \Sigma _+^{\prime } =& \left( q-2 + 6\Sigma _+^2  \right) \delta 
\Sigma _+ + \varpi \Sigma _+ \delta \varpi + \left( 2\Sigma _+ -1 \right) 
\delta \Omega _E \ , \\
 \delta \varpi ^{\prime } = &\left( q-2+ \varpi ^2 -\alpha \varpi \right) 
\delta \varpi + 6\Sigma _+ \left( \varpi - \alpha \right) \delta \Sigma _+ 
 + \left[ 2\varpi - 3\left( \alpha - \lambda \right) \right] \delta \Omega _E
 \ , \\
 \delta \Omega _E^{\prime } = & 2\Omega _E \left[ \left( 6\Sigma _+ + 1 
\right) \delta \Sigma _+ + \left( \varpi - \frac{\lambda }{2} \right) \delta 
\varpi + 2\delta \Omega _E \right] \ . 
 \end{split}
\end{align}
Again, one can decouple the equations with
\begin{equation}
\left( \delta \varpi - \left( \alpha -3\lambda \right) \delta \Sigma _+ 
\right) ^{\prime } = \left( q-2 \right) \left( \delta \varpi - \left( \alpha 
- 3\lambda \right) \delta \Sigma _+ \right) \ , 
\end{equation}
and the rest reduces to the quadratic equation
\begin{equation}
x^2 -(q-2)x  -3\Omega _E \frac{\lambda \left( \alpha - \lambda \right) ^2 
\left( \alpha -3\lambda \right) -2\left( \alpha - \lambda \right) 
\left( \alpha -4\lambda \right) -4}{\left( \alpha - \lambda \right) 
\left( \alpha - 3\lambda \right) +2 }=0 \ .
\end{equation}
The eigenvalues are as follows.
\begin{align}
\begin{split}
& \omega _{\Sigma _+ \mathchar`-\varpi \mathchar`- \Omega _E} = \frac{q-2 \pm \sqrt{\left( 
q-2 \right) ^2 + 12 \Omega _E \left( \lambda \left( \alpha - \lambda \right) 
-2 \right)}}{2} \ , \\
& \omega _{\Sigma _+ \mathchar`- \varpi } = q-2 \ , \quad \omega _{\Sigma _- 
\mathchar`- \mathcal{E}_- } = \frac{q-2 \pm \sqrt{\left( q-2 \right) ^2 
-24 \Omega _E}}{2} \ , \\
& \omega _{\mathcal{N}_1} = \frac{3}{2} \frac{\left( 3\alpha - 4\lambda 
\right) \left( \alpha + \lambda \right) -12}{\left( \alpha - \lambda \right)
 \left( \alpha - 3\lambda \right) +2} \ , \quad \omega _{\mathcal{E}_1} 
= -3\Sigma _+ \ , \\
& \omega _{\mathcal{N}_2} = \omega _{\mathcal{N}_3} = \omega _{\mathcal{A}_1} 
= -\frac{3}{2} \frac{\alpha ^2 -3\alpha \lambda +4\lambda ^2 -4}{\left( 
\alpha - \lambda \right) \left( \alpha -3\lambda \right) +2} \ , \\
& \omega _{\mathcal{A}_2} = \omega _{\mathcal{A}_3} = \frac{3}{2} 
\frac{\alpha ^2 + \alpha \lambda -4\lambda ^2 -4}{\left( \alpha - \lambda 
\right) \left( \alpha - 3\lambda \right) +2} \ , \quad \omega _{\Gamma } 
= \frac{3}{2} \frac{\left( \alpha - \lambda \right) \left( \alpha -2\lambda 
\right) }{\left( \alpha - \lambda \right) \left( \alpha - 3\lambda \right) 
+2} \ . 
\end{split}
\end{align}
Similarly to the single component case, the orthogonal electric field 
$\mathcal{E}_1$
and $\Gamma $ are unstable for $\lambda < 0 $. 
The typical time scale to leave this fixed point is again given by 
$\omega_{\mathcal{E}_1}^{-1}\sim |\Sigma_+|^{-1} $, 
which becomes long enough in the limit of 
$\lambda\rightarrow (\alpha^2-4)/\alpha$.
All the others are stable 
as long as
$\alpha < -\lambda $. 

%%%%%%%%%%%%%%%%%%%%%%%%%%%%%%%%%%%%%%%%%%%%%%%%%%%%%%%%%%%
%%%%%%%%%%%%%%%%%%%%%%%%%%%%%%%%%%%%%%%%%%%%%%%%%%%%%%%%%%%
%%%%%%%%%%%%%%%%%%%%%%%%%%%%%%%%%%%%%%%%%%%%%%%%%%%%%%%%%%%
\subsubsection{Isotropic electric inflation}
%%%%%%%%%%%%%%%%%%%%%%%%%%%%%%%%%%%%%%%%%%%%%%%%%%%%%%%%%%%
%%%%%%%%%%%%%%%%%%%%%%%%%%%%%%%%%%%%%%%%%%%%%%%%%%%%%%%%%%%
%%%%%%%%%%%%%%%%%%%%%%%%%%%%%%%%%%%%%%%%%%%%%%%%%%%%%%%%%%%
Assuming $\mathcal{E}_1^2 = \mathcal{E}_2^2 = \mathcal{E}_3^2 \neq 0$ 
characterizes
the isotropic inflationary solution, which was studied in 
\cite{Maeda2012}.
\begin{align}
\begin{split}
& \varpi = \frac{4}{\alpha -\lambda } \ , \quad q = \frac{\alpha + 
\lambda }{\alpha -\lambda } \ , \quad \Sigma _{\pm } = \Omega _E = \Gamma = 0 
\ , \\
& \Omega _V = -\frac{3\lambda (\alpha -\lambda )-4}{3(\alpha -\lambda )^2} \ ,
 \quad \Omega _E = \frac{\alpha (\alpha -\lambda ) -4}{(\alpha -\lambda )^2} 
\ .
\end{split} \label{eq:electric3}
\end{align}
The power exponent $p$ of the volume expansion is given by 
\begin{align}
p=
{\alpha-\lambda\over 2\alpha}
\,.
\end{align}

The existence condition under $\lambda <0$ is 
\begin{equation}
\alpha ^2 - \alpha \lambda -4 >0 \ .
\end{equation}

The linearization for perturbed variables
is straightforward and one obtains the eigenvalues as
\begin{align}
\begin{split}
& \omega _{\varpi \mathchar`- \Omega _E} 
= \frac{q-2 \pm \sqrt{\left( q-2 \right) ^2 - 4
\Omega _E \left( 3\lambda \left( \lambda - \alpha \right) +4 \right)}}{2} \ ,
\\
& \omega _{\Sigma _+ \mathchar`- \mathcal{E}_+ } 
= \omega _{\Sigma _- \mathchar`- \mathcal{E}_-} = 
\frac{q-2 \pm \sqrt{\left( q-2 \right) ^2 - 16 \Omega _E }}{2} \ , \\
& \omega _{\mathcal{N}_1} = \omega _{\mathcal{N}_2} = \omega _{\mathcal{N}_3} 
= \omega _{\mathcal{A}_1} = \omega _{\mathcal{A}_2} = \omega _{\mathcal{A}_3} 
= \frac{2\lambda }{\alpha -\lambda } \ , \quad \omega _{\Gamma } = 2 \ . 
\end{split}
\end{align}
Regardless of the parameters, $\Gamma $ is always unstable, while all 
the other
eigenvalues have negative real part, which is consistent with the previous 
studies.
A consequence
of this instability is oscillation of the gauge field, which was observed 
and studied
in the isotropic setup \cite{Maeda2012}. 

%%%%%%%%%%%%%%%%%%%%%%%%%%%%%%%%%%%%%%%%%%%%%%%%%%%%%%%%%%%
%%%%%%%%%%%%%%%%%%%%%%%%%%%%%%%%%%%%%%%%%%%%%%%%%%%%%%%%%%%
%%%%%%%%%%%%%%%%%%%%%%%%%%%%%%%%%%%%%%%%%%%%%%%%%%%%%%%%%%%
\subsection{Inflation with the magnetic components (Magnetic Inflation)}
%%%%%%%%%%%%%%%%%%%%%%%%%%%%%%%%%%%%%%%%%%%%%%%%%%%%%%%%%%%
%%%%%%%%%%%%%%%%%%%%%%%%%%%%%%%%%%%%%%%%%%%%%%%%%%%%%%%%%%%
%%%%%%%%%%%%%%%%%%%%%%%%%%%%%%%%%%%%%%%%%%%%%%%%%%%%%%%%%%%
Now we shadow the previous subsection for non-vanishing magnetic field.
While three types of fixed points are expected from the duality, some of them
do not appear as "fixed" when we express them in terms of $\mathcal{A}_A$.
Even when they are constant, there is arbitrariness which does not exist for
electric cases. All these features are related to the fact that they are 
essentially Abelian solutions while we have to deal with 
non-Abelian dynamics. 

%%%%%%%%%%%%%%%%%%%%%%%%%%%%%%%%%%%%%%%%%%%%%%%%%%%%%%%%%%%
%%%%%%%%%%%%%%%%%%%%%%%%%%%%%%%%%%%%%%%%%%%%%%%%%%%%%%%%%%%
%%%%%%%%%%%%%%%%%%%%%%%%%%%%%%%%%%%%%%%%%%%%%%%%%%%%%%%%%%%
\subsubsection{Anisotropic magnetic inflation {\rm (}single component{\rm )}} \label{sec:singleMag}
%%%%%%%%%%%%%%%%%%%%%%%%%%%%%%%%%%%%%%%%%%%%%%%%%%%%%%%%%%%
%%%%%%%%%%%%%%%%%%%%%%%%%%%%%%%%%%%%%%%%%%%%%%%%%%%%%%%%%%%
%%%%%%%%%%%%%%%%%%%%%%%%%%%%%%%%%%%%%%%%%%%%%%%%%%%%%%%%%%%
Analogous to the electric case, assuming $\mathcal{A}_1 = 0 , \mathcal{A}_2 
\neq 0,
\mathcal{A}_3 \neq 0$ leads to an inflationary solution with a background 
magnetic
field and non-vanishing $\Sigma _+$. Since the fixed points are located on 
the Abelian
boundary $\Gamma =0$, the electro-magnetic duality guarantees that one can 
obtain
the magnetic fixed point by replacing $\lambda $ with $-\lambda $ in 
(\ref{eq:singleE})
and taking
\begin{equation}
\Omega _E \rightarrow \Omega _B = \frac{1}{6} \mathcal{A}_2^2 \mathcal{A}_3^2 
\ ,  \quad \mathcal{E}_A = 0  \quad (A = 1,2,3) \ .
\end{equation}
It is also clear that there are other physically identical fixed points 
obtained by picking
$\mathcal{A}_2$ or $\mathcal{A}_3$ to be zero and applying the rotation 
(\ref{shear_rot}).
A difference is that the values of $\mathcal{A}_2$ and $\mathcal{A}_3$ are
 not determined
by the requirement of being a fixed point so that it is a one-parameter family 
of fixed points with a free parameter $\mathcal{A}_2 / \mathcal{A}_3$.

As for the perturbations of these fixed points,
 the decoupling of the linearized equations can be done in a similar fashion 
as in the electric case. 
We obtain the following eigenvalues for the perturbed variables.
\begin{align}
\begin{split}
& \omega _{\Sigma _+ \mathchar`- \varpi \mathchar`- \Omega _B } = \frac{q-2 
\pm \sqrt{\left( 
q-2 \right) ^2 -12\Omega _B \left( \lambda \left( \alpha + \lambda \right) 
+4 \right) }}{2} \ , \\
& \omega _{\Sigma _+ \mathchar`- \varpi } = \omega _{\Sigma _-} = q-2 \ , 
\quad 
\omega _{\mathcal{N}_1}=-3\Sigma _+ \ , \quad \omega _{\mathcal{N}_2} 
= \omega _{\mathcal{N}_3} = 3\Sigma _+ \ , \\
& \omega _{\mathcal{A}_1} = 3\Sigma _{+} \ , \quad \omega _{\mathcal{A}_2 
-\mathcal{A}_3} = 0 \ , \quad \omega _{\mathcal{E}_1} = -\lambda \varpi \ , 
\\
& \omega _{\mathcal{E}_2} =\omega _{\mathcal{E}_3} = 6\frac{\left( \alpha +
 \lambda \right) \left( \alpha -2\lambda \right) -4}{\left( \alpha + \lambda 
\right) \left( \alpha + 3\lambda \right) + 8}  \ , \quad \omega _{\Gamma } 
= \frac{3\left( \alpha ^2 -\lambda ^2 \right)}{\left( \alpha + \lambda \right)
\left( \alpha + 3\lambda \right) + 8 } \ . 
\end{split}
\end{align}
Note that the shear at the fixed point is given by
\begin{equation}
\Sigma _+ = 2\frac{\alpha ^2 + \alpha \lambda -4}{(\alpha + \lambda )(\alpha + 3\lambda ) + 8} >0 \label{eq:shearSingle}
\end{equation}
in this magnetic case.
It is clearly unstable against $\delta \mathcal{N}_2 , \delta \mathcal{N}_3$ 
and $\delta \mathcal{A}_1$,
namely the perturbation of magnetic fields $\mathcal{B}_2 , \mathcal{B}_3$ 
orthogonal
to the background $\mathcal{B}_1$ for $\lambda >0$, in parallel to the 
electric case. 
The zero eigenvalue
$\omega _{\mathcal{A}_2 -\mathcal{A}_3}$ reflects the fact that it is a
one-parameter family of fixed points and corresponds to the direction of
physical insignificant shift in the ratio $\mathcal{A}_2 / \mathcal{A}_3$. 
All the other eigenvalues can be made to have negative real parts by
imposing $\alpha < \lambda $. 

In the discussion above, we were implicitly assuming that the spatial
curvature vanishes in the background, which was necessary in the case of
electric inflation. The magnetic solutions are more complicated, however,
and $\mathcal{N}_A$ can take non-zero values while maintaining constant
$\mathcal{B}_A$. When this happens, the dynamics cannot be understood
in terms of   this
fixed point even though the behaviour  of shear and magnetic field
appears almost identical.
We will come back to this problem later in the analysis of our numerical
results.

%%%%%%%%%%%%%%%%%%%%%%%%%%%%%%%%%%%%%%%%%%%%%%%%%%%%%%%%%%%
%%%%%%%%%%%%%%%%%%%%%%%%%%%%%%%%%%%%%%%%%%%%%%%%%%%%%%%%%%%
%%%%%%%%%%%%%%%%%%%%%%%%%%%%%%%%%%%%%%%%%%%%%%%%%%%%%%%%%%%
\subsubsection{Anisotropic magnetic inflation {\rm (}double components{\rm )}}
%%%%%%%%%%%%%%%%%%%%%%%%%%%%%%%%%%%%%%%%%%%%%%%%%%%%%%%%%%%
%%%%%%%%%%%%%%%%%%%%%%%%%%%%%%%%%%%%%%%%%%%%%%%%%%%%%%%%%%%
%%%%%%%%%%%%%%%%%%%%%%%%%%%%%%%%%%%%%%%%%%%%%%%%%%%%%%%%%%%
It is easy to notice that any constant $\mathcal{A}_A$ cannot give constant
magnetic field with $\mathcal{B}_1 = 0 , \mathcal{B}_2 \neq 0 , \mathcal{B}_3 
\neq 0$
as long as one assumes $\mathcal{N}_A =0$. Nevertheless, using the equations
(\ref{eq:mag1}) - (\ref{eq:mag3}) with $\Gamma =0$ instead of (\ref{eq:a1}) -
(\ref{eq:a3}), it is clear that there is an inflationary solution dual to
(\ref{eq:electric_double}) with two non-vanishing constant magnetic components.
In fact, time-dependent vector potential can asymptotically realize that dyad 
configuration of magnetic field. To see how it is possible, let us evaluate 
equations 
(\ref{eq:a1}) - (\ref{eq:a3}) on this double magnetic field solution. 
They read
\begin{equation}
\mathcal{A}_1^{\prime } = \frac{3}{2} \Sigma _+ \mathcal{A}_1 \ , \quad 
\mathcal{A}_2^{\prime } = -\frac{3}{2} \Sigma _+ \mathcal{A}_2 \ , \quad 
\mathcal{A}_3^{\prime } = - \frac{3}{2} \Sigma _+ \mathcal{A}_3 \ ,
\end{equation}
where $\Sigma _+$ is a constant given by
\begin{equation}
\Sigma _+ = - \frac{\alpha ^2 +\alpha \lambda - 4}{\left( \alpha + \lambda 
\right) \left( \alpha + 3\lambda \right) +2 } <0 \ . \label{eq:shearDouble}
\end{equation}
Hence they are indeed consistent with constant $\mathcal{B}_2 $ and
$\mathcal{B}_3$ since
\begin{equation}
\mathcal{B}_2 \sim \mathcal{A}_1 \mathcal{A}_3 \ , \quad \mathcal{B}_3 \sim 
\mathcal{A}_1 \mathcal{A}_2 \ ,
\end{equation}
where $\mathcal{A}_2, \mathcal{A}_3$ grow exponentially while 
$\mathcal{A}_1$ decreases exponentially.
We then find 
 the exponentially growing $\mathcal{B}_1$, which represents the linear
instability against the orthogonal components discussed for the other
fixed points. 

This result is extended to include spatial curvature as
\begin{equation}
\mathcal{N}_1^{\prime } = - \frac{9}{2}\Sigma _+ \mathcal{N}_1 \ , \quad 
\mathcal{N}_2^{\prime } = \frac{3}{2} \Sigma _+ \mathcal{N}_2 \ , \quad 
\mathcal{N}_3^{\prime } = \frac{3}{2} \Sigma _+ \mathcal{N}_3 \ ,
\end{equation}
which implies
\begin{equation}
\mathcal{B}_2 \sim -\mathcal{A}_2 \mathcal{N}_2 \ , \quad \mathcal{B}_3 \sim 
- \mathcal{A}_3 \mathcal{N}_3 
\end{equation}
are asymptotically constant. Although the full stability analysis is 
complicated
because of the time-dependent background, the above analysis already indicates
the instability against $\mathcal{B}_1$. We also note that 
\begin{equation}
\omega _{\Gamma } = \frac{3}{2} \frac{(\alpha -\lambda )(\alpha 
+ 2\lambda )}{(\alpha + \lambda )(\alpha +3\lambda )+2} \  ,
\end{equation}
which is negative for $\alpha < \lambda $.
All the other eigenvalues are expected to have negative real part from the 
duality.
This is indeed confirmed   for the Abelian system  in the appendix.

%%%%%%%%%%%%%%%%%%%%%%%%%%%%%%%%%%%%%%%%%%%%%%%%%%%%%%%%%%%
%%%%%%%%%%%%%%%%%%%%%%%%%%%%%%%%%%%%%%%%%%%%%%%%%%%%%%%%%%%
%%%%%%%%%%%%%%%%%%%%%%%%%%%%%%%%%%%%%%%%%%%%%%%%%%%%%%%%%%%
\subsubsection{Isotropic magnetic inflation}
%%%%%%%%%%%%%%%%%%%%%%%%%%%%%%%%%%%%%%%%%%%%%%%%%%%%%%%%%%%
%%%%%%%%%%%%%%%%%%%%%%%%%%%%%%%%%%%%%%%%%%%%%%%%%%%%%%%%%%%
%%%%%%%%%%%%%%%%%%%%%%%%%%%%%%%%%%%%%%%%%%%%%%%%%%%%%%%%%%%
The duality transformation brings the isotropic electric inflation into the
corresponding magnetic one with $\mathcal{B}_1^2 = \mathcal{B}_2^2 = 
\mathcal{B}_3^2 \neq 0$.
The solution is given by replacing $\lambda $ with $-\lambda $ and 
interchanging
$\Omega _E$ and $\Omega _B$ in (\ref{eq:electric3}). $\mathcal{A}_A$ are also 
constant
this time. However, the constancy of $\mathcal{N}_A$ on this fixed 
point also follows
so that they do not have to be zero. In fact, the values of $\mathcal{A}_A$ and
$\mathcal{N}_A$ are arbitrary as long as they satisfy
\begin{equation}
\frac{1}{2}\left( \mathcal{A}_2 \mathcal{A}_3 - \mathcal{A}_1 \mathcal{N}_1 
\right) ^4 = \frac{1}{2}\left( \mathcal{A}_3 \mathcal{A}_1 - \mathcal{A}_2 
\mathcal{N}_2 
\right) ^4 =\frac{1}{2}\left( \mathcal{A}_1 \mathcal{A}_2 - \mathcal{A}_3 
\mathcal{N}_3 
\right) ^4 =\Omega_{B}=\frac{\alpha (\alpha +\lambda ) -4}{(\alpha 
+ \lambda )^2} \ ,
 %\quad 2 \ {\rm perms} \ .
\end{equation}
The eigenvalues read
\begin{align}
\begin{split}
& \omega _{\varpi \mathchar`-  \Omega _B} = \frac{q-2 \pm \sqrt{\left( q-2 
\right) ^2 - 4
\Omega _B \left( 3\lambda \left( \alpha + \lambda \right) +4 \right)}}{2} \ , 
\\
& \omega _{\Sigma _+ \mathchar`-  \mathcal{B}_+} = \omega _{\Sigma _- 
\mathchar`-  \mathcal{B}_-} =
\frac{q-2 \pm \sqrt{\left( q-2 \right) ^2 -16 \Omega _B }}{2} \ ,  \\ 
& \omega _{\mathcal{N}_1} = \omega _{\mathcal{N}_2} = \omega _{\mathcal{N}_3} 
= 0 \ , \quad \omega _{\mathcal{E}_1} = \omega _{\mathcal{E}_2} 
= \omega _{\mathcal{E}_3} = -\frac{4\lambda }{\alpha + \lambda } \ , \\
& \omega _{\Gamma } = \frac{\alpha - \lambda }{\alpha + \lambda } \ . 
\end{split}
\end{align}
It is easy to see that all the eigenvalues except for the three zeros posses
negative real parts for $\alpha < \lambda $. The zero eigenvalues correspond
to the arbitrariness in $\mathcal{A}_A$ and $\mathcal{N}_A$ and imply the
fixed points form a three-dimensional subset in the phase space. The local 
stability
of the isotropic inflation holds for diagonal Bianchi class A models as well.
 The time-scale of convergence can be estimated by looking at the smallest
eigenvalue, which is $\omega _{\Sigma _+ \mathchar`- \mathcal{B}_+}$ in this case. 
Using the approximation $\Omega _B \ll 1$ for realistic models, we obtain
\begin{equation}
\omega _{\Sigma _+ \mathchar`- \mathcal{B}_+} \sim \frac{4\Omega _B}{q-2} =
 -\frac{\alpha ^2 + \alpha \lambda -4}{(\alpha + \lambda )(\alpha +3\lambda ) }
\end{equation}
where we focused on the smaller of the two eigenvalues. Notice the interesting
similarity of this expression with the shear for the anisotropic fixed points 
(\ref{eq:shearSingle}) and (\ref{eq:shearDouble}), which decides their strength
of instability. Hence, one can expect that the dynamics around all of those 
fixed points have roughly the same characteristic time-scale. As a consequence,
the less anisotropic are those anisotropic inflations, the slower is 
the convergence to the final isotropic state.

%%%%%%%%%%%%%%%%%%%%%%%%%%%%%%%%%%%%%%%%%%%%%%%%%%%%%%%%%%%
%%%%%%%%%%%%%%%%%%%%%%%%%%%%%%%%%%%%%%%%%%%%%%%%%%%%%%%%%%%
%%%%%%%%%%%%%%%%%%%%%%%%%%%%%%%%%%%%%%%%%%%%%%%%%%%%%%%%%%%
\section{Convergence to the isotropic attractor in Bianchi type I}
%%%%%%%%%%%%%%%%%%%%%%%%%%%%%%%%%%%%%%%%%%%%%%%%%%%%%%%%%%%
%%%%%%%%%%%%%%%%%%%%%%%%%%%%%%%%%%%%%%%%%%%%%%%%%%%%%%%%%%%
%%%%%%%%%%%%%%%%%%%%%%%%%%%%%%%%%%%%%%%%%%%%%%%%%%%%%%%%%%%
When the gauge-kinetic coupling favors magnetic components ($\lambda >0$), 
we have seen that the isotropic magnetic inflation is locally stable for 
$\lambda > \alpha $ 
while there are anisotropic inflationary fixed points whose unstable directions
are given by the orthogonal components of the magnetic fields. Concerning
the global dynamics of the system, there are several questions to be addressed.
\begin{enumerate}
\item First of all, is the isotropic inflation really the global attractor? 
The linear
analysis suggests that when $\lambda > \alpha $ and $\alpha (\alpha 
+\lambda )>4$
are satisfied, there is no other stable fixed point in the dynamical system. 
However,
it does not rule out other future asymptotic states such as periodic orbits 
(limit cycles) or deterministic chaos.
\item The anisotropic fixed points that appear as saddles can in principle give
rise to an intermediate phase of inflation with a preferred direction and may 
leave
an observable signature in the primordial density fluctuation. Since the final
attractor is isotropic, the anisotropy may well exist only for large scales. 
How
plausible is this scenario? Can generic initial conditions take the universe
to a transient anisotropic inflation?
\item Strong initial spatial curvature of Bianchi type IX is expected to halt 
the 
expansion of the universe as was quantified in single-scalar inflation 
\cite{Kitada1993}. 
What is the effect of the spatial curvature in the presence of gauge fields? 
How
often does recollapse occur?
\end{enumerate}
To answer these questions, we carry out numerical calculations for a
variety of initial conditions. We solve 14 equations (\ref{eq:shearPlus}) -
(\ref{eq:GammaEv}) while independently monitoring the total energy density
\begin{equation}
\Omega _T = \Sigma ^2 + \Omega _K + \Omega _V + \Omega _N + \Omega _E 
+ \Omega _B 
\end{equation}
where
\begin{equation}
\Sigma ^2 = \Sigma ^2_+ + \Sigma ^2_- \ , \quad \Omega _N = 
\frac{\Gamma ^2}{12} \left[ \mathcal{N}_1^2 + \mathcal{N}_2^2 
+ \mathcal{N}_3^2 - 2\left( \mathcal{N}_1 \mathcal{N}_2 + \mathcal{N}_2 
\mathcal{N}_3 + \mathcal{N}_3 \mathcal{N}_1 \right) \right] \ , 
\end{equation}
and make sure that the Hamiltonian constraint (\ref{eq:Friedmann}), namely 
$\Omega _T = 1$, is maintained for consistency check and detecting any 
numerical 
instability.   In this section, we fix the model parameters $\alpha =2, 
\lambda =5$,
even though the slow-roll parameter for this case $\epsilon =q+1 \sim 0.5$ is 
too large
to be a realistic model of inflation. The reason for this choice is to keep 
the values
of density parameters at the fixed points relatively large, whence the
characteristic time-scale is relatively small, so that the presentation
becomes clearer.   Several other sets of parameters will be examined 
in the later sections. Our purpose in the present article is to derive 
qualitative features
of convergence under the influence of Yang-Mills coupling and spatial 
curvature.
For quantitative predictions, it will be necessary to choose more realistic 
potential
and guage-kinetic function. Since the number of variables is large, we start
from Bianchi type I ($\mathcal{N}_A =0$) in this section. The effect of spatial
curvature will be investigated in the next section.

%%%%%%%%%%%%%%%%%%%%%%%%%%%%%%%%%%%%%%%%%%%%%%%%%%%%%%%%%%%
%%%%%%%%%%%%%%%%%%%%%%%%%%%%%%%%%%%%%%%%%%%%%%%%%%%%%%%%%%%
%%%%%%%%%%%%%%%%%%%%%%%%%%%%%%%%%%%%%%%%%%%%%%%%%%%%%%%%%%%
\subsection{Abelian Bianchi type I subset}
%%%%%%%%%%%%%%%%%%%%%%%%%%%%%%%%%%%%%%%%%%%%%%%%%%%%%%%%%%%
%%%%%%%%%%%%%%%%%%%%%%%%%%%%%%%%%%%%%%%%%%%%%%%%%%%%%%%%%%%
%%%%%%%%%%%%%%%%%%%%%%%%%%%%%%%%%%%%%%%%%%%%%%%%%%%%%%%%%%%
The aim of this subsection is to study how initial states of shear, the scalar 
field
and the field strength of gauge fields affect the intermediate dynamics before
becoming isotropic. For this purpose, we set
$\Gamma = 0 $ to avoid complexity of non-Abelian dynamics and 
$ \mathcal{N}_{1,2,3}=0$ to reduce the number of variables.
\begin{figure}[htbp]
\begin{center}
\includegraphics[height=0.4\linewidth,width=1\linewidth]{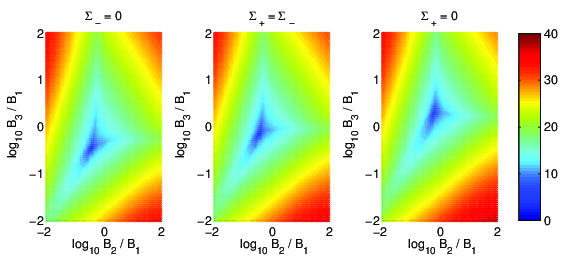}
\caption{The convergence time for different initial values of $\mathcal{B}_2 / 
\mathcal{B}_1$
and $\mathcal{B}_3 / \mathcal{B}_1$. From left to right, the initial shear 
variables are
taken to be $(\Sigma _+  , \Sigma _- ) = (\sqrt{0.2} , 0 ) ,(\sqrt{0.1},
\sqrt{0.1}) , 
(0 , \sqrt{0.2})$. The other conditions are explained in the body of the text. 
There is no significant dependence on initial $\Sigma _+ / \Sigma _-$. The 
typical convergence time appears to be between $10$ and $20$. While 
anisotropic 
phase lasts longer when the ratios between the magnetic components are extreme, note 
that the single component configuration $\mathcal{B}_2 \sim \mathcal{B}_3
\ll \mathcal{B}_1$ (left-bottom corners in the plots) is not efficient to 
generate anisotropy.}
\label{fig:magDirDep}
\end{center}
\end{figure}
Figure \ref{fig:magDirDep} shows the   colour-coded
 plot  of the time $\tau $ 
(e-folding number)
that elapsed before settling down to the isotropic magnetic inflation 
(convergence time). 
The initial ratios of $\mathcal{B}_2 , \mathcal{B}_3$ to $\mathcal{B}_1$
have been swept from $10^{-2}$ to $10^2$ while keeping fixed the other
initial conditions as
\begin{equation}
\Sigma ^2  = \frac{1}{6}\varpi ^2 = \Omega _V = \Omega _E = \Omega _B = 0.2 \ .
\label{eq:initial1}
\end{equation}
 The directions of electric field components are taken to be isotropic, namely
 \begin{equation}
 \mathcal{E}_1 = \mathcal{E}_2 = \mathcal{E}_3 \label{eq:initial2}
 \end{equation}
 and the three panels correspond to three representative choices of the 
direction of initial anisotropy
 \begin{equation}
 \Sigma _- = 0 \ , \quad \Sigma _+ = \Sigma _- \ , \quad \Sigma _+ = 0 \ .
 \end{equation}
In practice, one needs to specify the criterion for convergence since an orbit
never reaches the attractor exactly, but only approaches to it asymptotically.
Our choice is 
\begin{equation}
|\Sigma _+ |, | \Sigma _-  | < 10^{-3} \ , \quad \left| \varpi - \frac{4}{\alpha +\lambda } \right| , \left| \Omega _V - \frac{3\lambda ( \alpha + \lambda ) +4}{3(\alpha + \lambda )^2} \right| < 10^{-2} \ . \label{eq:criteria}
\end{equation}
 The latter two conditions are added to ensure that the isotropy is 
not accidental,
but due to the convergence to the isotropic magnetic inflation.
\begin{figure}[htbp]
\begin{center}
\includegraphics[height=0.26\linewidth]{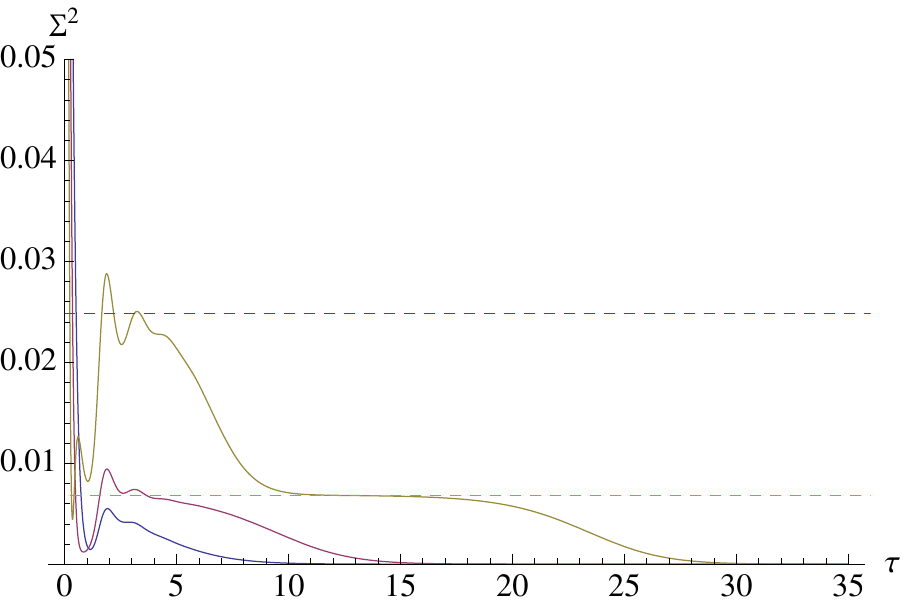}
\includegraphics[height=0.26\linewidth]{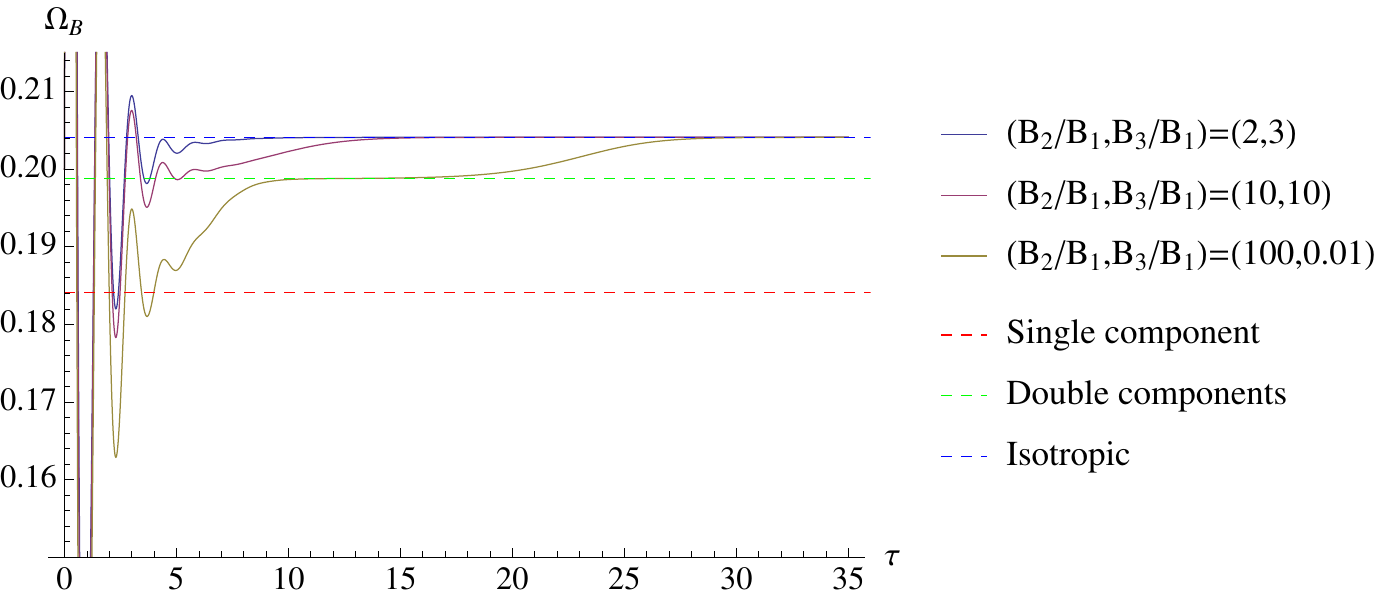}
\caption{Transient anisotropy and  convergence to the isotropic attractor. 
We plot the
time evolution of $\Sigma ^2$ and $\Omega _B$ for three initial magnetic 
configurations
$(\mathcal{B}_2 / \mathcal{B}_1 , \mathcal{B}_3 / \mathcal{B}_1 ) = (2,3) ,
(10,10),(100,0.01)$. The shear variabels satisfy $\Sigma _+ = \Sigma _-$ 
initially
and the conditions are the same as specified in (\ref{eq:initial1}) and 
(\ref{eq:initial2}).
The equilibrium values of $\Sigma ^2$ and $\Omega _B$ at each of the 
inflationary
fixed points are indicated by dotted lines. The maximum value of $\Sigma ^2$ 
for 
$(B_2 /B_1 , B_3 /B_1) =(10,10)$ roughly agrees with the equilibrium value of
double component inflation. The orbit $(B_2 /B_1 , B_3 /B_1) =(100,0.01)$ 
exhibits
an intermediate stationary period at double component inflation and also a hint
of initial attraction to single component inflation. While saddle behaviour 
is not so
clear in $(B_2 /B_1 , B_3 /B_1) =(2,3)$, it appears reasonable to attribute 
the initial
anisotropic period to a temporal approach to anisotropic fixed point.}
\label{fig:convergence}
\end{center}
\end{figure}
They show a clear tendency that a strong initial anisotropy in the 
configuration of
magnetic field results in a prolonged period of anisotropic phase before the 
universe
reaches the isotropic inflation (figure \ref{fig:convergence}). 
When the initial magnetic components are of the same order (blue regions in 
figure \ref{fig:magDirDep}), the typical time-scale of convergence 
to the attractor appears to be of order $\tau \sim 10$. This is much longer
than the case of single scalar inflation where the typical convergence time is
of order $\tau \sim 1$. Given that the extremum of anisotropy observed just 
before
convergence agrees with the value of shear at the double magnetic fixed 
point (figure \ref{fig:convergence}), it is reasonable to attribute the 
augmentation 
of convergence time to temporary attraction to those anisotropic solutions.
\begin{figure}[bhtp]
\begin{center}
\includegraphics[width=1.0\linewidth,height=0.4\linewidth]{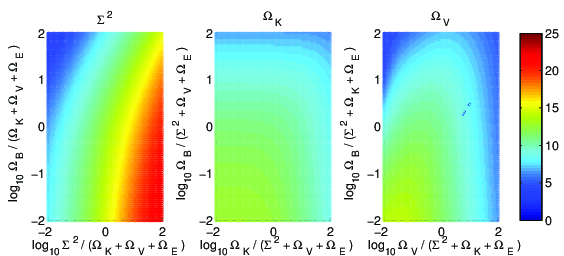}
\caption{Left: color-coded plot of convergence time for varying $\Sigma ^2 , 
\Omega _B$
and $\Omega _K = \Omega _V = \Omega _E$. The top-left corner corresponds to the
initial universe dominated by magnetic sector and the bottom-right to shear 
domination.
The dependence is rather weak (note the color scheme different from figure 
\ref{fig:magDirDep}). Centre and right: similar plots for varying $\Omega _K ,
 \Omega _B$
and $\Omega _V , \Omega _B$. The effect is even less significant.}
\label{fig:magRatioDep}
\end{center}
\end{figure}
 Once the attraction occurs, it requires another $\sim O(10)$ e-foldings 
in order
for becoming isotropic since   the eigenvalues for instability of anisotropic
inflation as well as stability of isotropic one are roughly proportional to
the value of shear during the anisotropic phases, 
 which is $\Sigma_{+}=-10/121$ 
for the double-component magnetic inflation with the current choice 
of the parameters.

Next, we study the dependence on the partition of the energy among
different sectors while respecting the Friedmann equation (\ref{eq:Friedmann}).
We fix the initial ratios between components of shear, electric 
and magnetic field as
\begin{equation}
\Sigma _+ = \Sigma _- \ , \quad \mathcal{E}_1 = \mathcal{E}_2 =\mathcal{E}_3 
\ , \quad \mathcal{B}_1 = \mathcal{B}_2 = \mathcal{B}_3 \ .
\end{equation}
In the left panel of figure \ref{fig:magRatioDep}, we vary 
$\Omega _B /(\Omega _K
+\Omega _V + \Omega _E )$ 
and $\Sigma ^2 / (\Omega _K + \Omega _V + \Omega _E )$ from
 $10^{-2}$ to $10^2$ while 
maintaining
$\Omega _K = \Omega _V = \Omega _E$. Similar prescriptions for
 $\Omega _B $ - $
\Omega _K$
and $\Omega _B$ - $\Omega _V$ have lead to the centre and right panels. It can 
be
seen that the convergence time is rather insensitive to the total energy 
densities
of each component (note the different color map from figure 
\ref{fig:magDirDep}). 
The convergence time can be large for exceedingly large shear density, 
but otherwise it is about 10 e-foldings.
While we have not presented any dependence on the directions
and strengths of electric field, we mention that we checked their irrelevance 
in 
deciding the convergence as is expected from the general tendency to 
suppression of electric fields for $\lambda >0$.

%%%%%%%%%%%%%%%%%%%%%%%%%%%%%%%%%%%%%%%%%%%%%%%%%%%%%%%%%%%
%%%%%%%%%%%%%%%%%%%%%%%%%%%%%%%%%%%%%%%%%%%%%%%%%%%%%%%%%%%
%%%%%%%%%%%%%%%%%%%%%%%%%%%%%%%%%%%%%%%%%%%%%%%%%%%%%%%%%%%
\subsection{Occurrence of oscillatory attractor in non-Abelian Bianchi I}
\begin{figure}[htbp]
\begin{center}
\includegraphics[width=1.0\linewidth,height=0.4\linewidth]{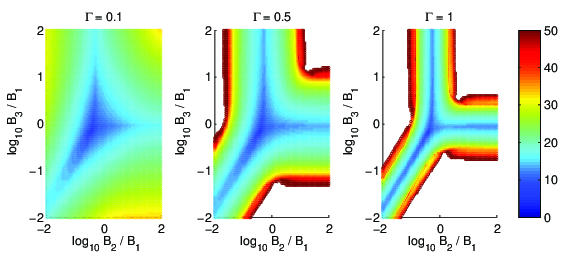}
\caption{The same type of plots as the centre panel of figure 
\ref{fig:magDirDep}
except that $\Gamma \neq 0$ initially here. The regions with long period of 
anisotropic
inflation are now replaced by the oscillatory final state (white). Near 
the edge
of the convergent area, the convergence time rises sharply. The larger is
the initial value of $\Gamma $, the more often the system ends up in the
oscillatory state.}
\label{fig:MagDirOsc}
\end{center}
\end{figure}

\begin{figure}[htbp]
\begin{center}
\includegraphics[width=0.48\linewidth]{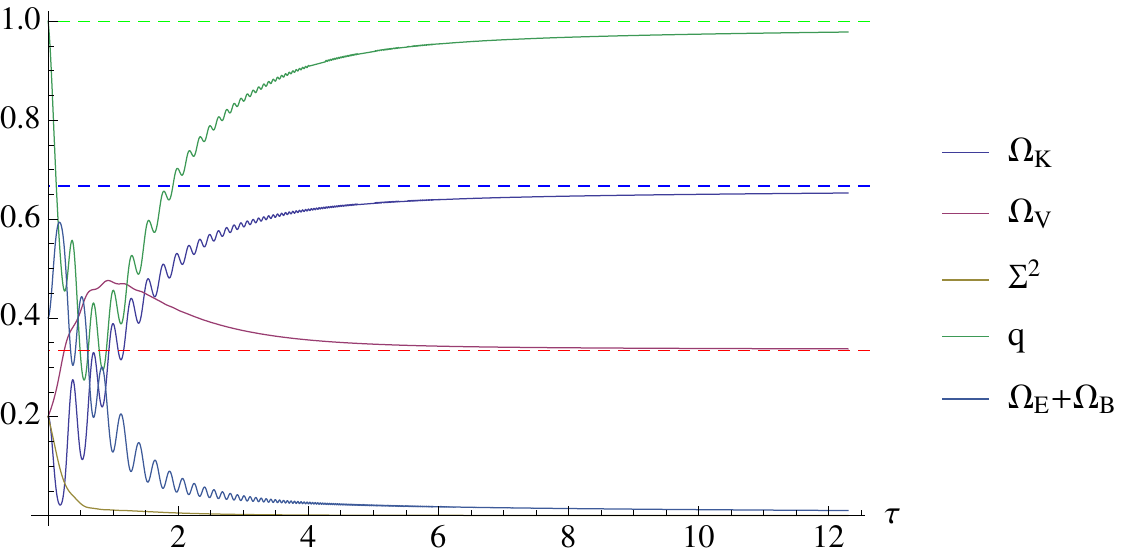}
\includegraphics[width=0.48\linewidth]{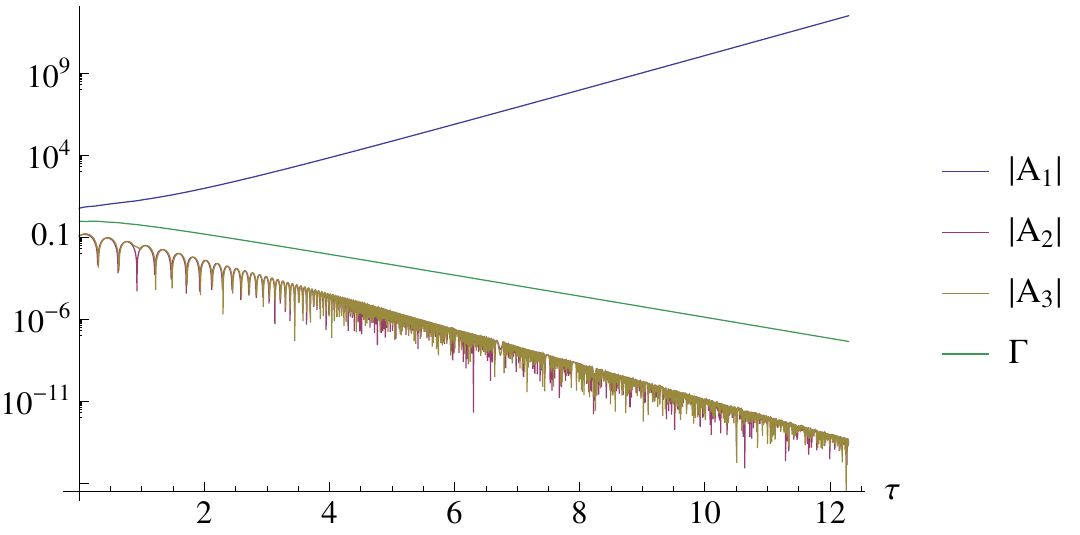}
\includegraphics[width=0.48\linewidth]{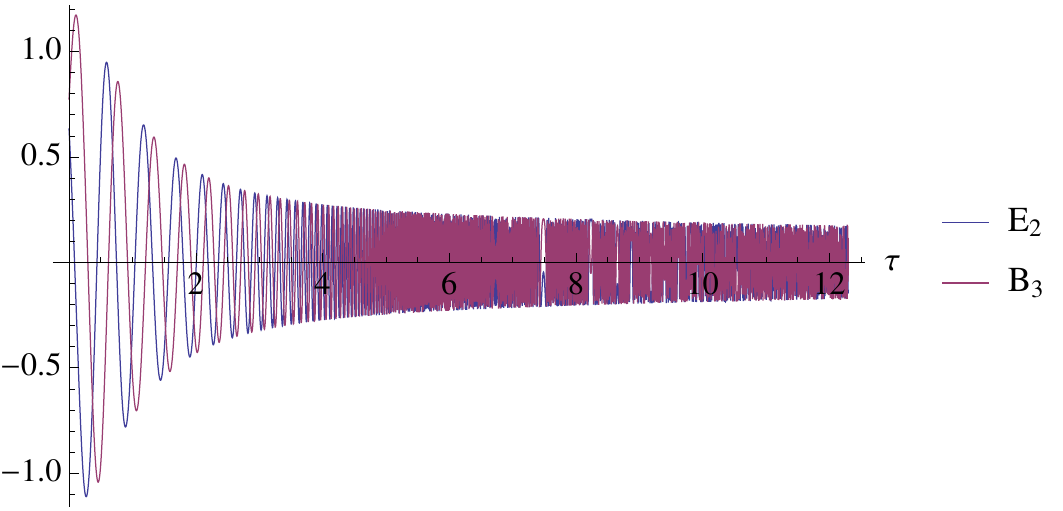}
\includegraphics[width=0.48\linewidth]{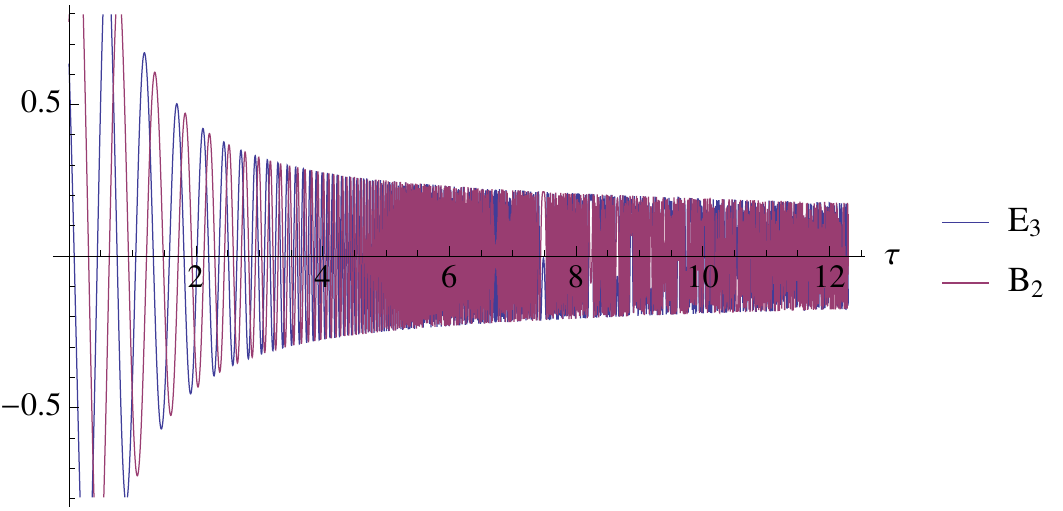}
\caption{A typical oscillatory behaviour for $\alpha =2 , \lambda =5$. 
The initial
conditions are $\Sigma ^2 = \Omega _K = \Omega _V = \Omega _E = \Omega _B 
=0.2, 
\Omega _N =0, \Sigma _+  = \Sigma _- , \mathcal{E}_3 = 2 \mathcal{E}_1 = 2
\mathcal{E}_2 ,
\mathcal{B}_2 = \mathcal{B}_3 = 100\mathcal{B}_1$. While not presented, 
$\mathcal{E}_1$
and $\mathcal{B}_1$ are exponentially suppressed. On the top-left panel,
the values of $\Omega _K , \Omega _V$ and $q$ for the power-law scalar
solution (\ref{eq:fix1}) are indicated by the dotted lines with the corresponding
colors.}
\label{fig:oscillation}
\end{center}
\end{figure}

%%%%%%%%%%%%%%%%%%%%%%%%%%%%%%%%%%%%%%%%%%%%%%%%%%%%%%%%%%%
%%%%%%%%%%%%%%%%%%%%%%%%%%%%%%%%%%%%%%%%%%%%%%%%%%%%%%%%%%%
%%%%%%%%%%%%%%%%%%%%%%%%%%%%%%%%%%%%%%%%%%%%%%%%%%%%%%%%%%%
Now we turn on the gauge coupling and repeat the same type of analysis
as in the previous subsection.  The figure \ref{fig:MagDirOsc} 
shows
the convergence time for $\Gamma = 0.1, 0.5,1$ initially when the ratios 
$\mathcal{B}_2 /\mathcal{B}_1 , \mathcal{B}_3 /\mathcal{B}_1$ are varied
under the same condition as in figure \ref{fig:magDirDep}.
First of all, one notices the sharp rise in convergence time when the initial 
magnetic
configuration is anisotropic. Anisotropic phase may last more than 50 e-folds
 for $\Gamma = 0.5 $ or $1$ compared to 30 e-folds for $\Gamma =0$. 
Nevertheless, the final state is still the isotropic magnetic inflation 
and the dynamics
 is qualitatively similar to Abelian Bianchi I. On the contrary, in the
 white regions that are separated by those boundaries of prolonged anisotropic
 period, we have been unable to observe the convergence to the 
isotropic magnetic
 inflation. The dynamics in those regions is characterized by 
rapid oscillations of gauge 
 field while the overall amplitudes of shear and 
gauge field energy density are decaying
 (figure \ref{fig:oscillation}). 
The period of oscillation decreases indefinitely so that 
the numerical calculations had to
be abandoned typically around $\tau = 10$.
 As one can see from figure \ref{fig:oscillation}, 
however, the system appears to settle down to a stationary state 
that is not a fixed
point and dominated by the scalar field.
The positive value of $q$ indicates that the expansion is no longer inflationary.

%{\bf Beyond these sharp boundaries (the white regions of 
%initial conditions), we expect an isotropic expansion only by 
%the scalar field, because the Yang-Mills energy and shear density 
%drop to negligiblly small values although the numerical calculations had 
%to be stopped before it reaches the isotropic attarctor.
%It is not an inflationary solution but a power law expansion with
%$p=2/\alpha^2<1$.
%}
Across those white regions, the dynamical behaviour shares several common features.
\begin{enumerate}
\item $\varpi, \Omega _V $ and $q$ appear to converge to constant values
which are consistent with the conventional power-law fixed point by the 
scalar field (solution (\ref{eq:fix1}) which is not inflationary for
 $\alpha =2$). While
$\Sigma _+^2 + \Sigma _-^2$ and $\Omega _E + \Omega _B$ are also
more or less constant, there contributions in the Friedmann equation and 
the other evolution equations for the isotropic variables (such as 
$\Omega _K ,\Omega _V $ and $q$) are subleading. 
Note, however, that
it does not mean the final state is the fixed point (\ref{eq:fix1}) 
since it was
already shown to be linearly unstable whence it cannot be an attractor.
\item There are always two components of electric as well as magnetic
field that survive. The remaining one component for each decays exponentially.
 The three separated white regions correspond to three possible choices of
two components out of three. The amplitudes of oscillation, and hence
the expansion normalized energy density of gauge field, approach constant 
in the expanding universe with $q = 1 \Leftrightarrow p = 1/2$, which means
the gauge field is effectively behaving as a radiation fluid. This may be expected
by observing that the structure of oscillation is similar to that of the familiar
electromagnetic plane-wave.
\item For the vector potential $\mathcal{A}_A$, one component grows
exponentially while the other two undergo oscillation damping. The growth
of the single component is faster than the decay of $\Gamma $ (top-right
panel in figure \ref{fig:oscillation}). 
\end{enumerate}
Given those observations, we look for an asymptotic oscillatory solution
as follows. First, we assume the back reaction of gauge field on the scalar
field dynamics 
is negligible and set $\varpi ^{\prime } = \Omega _V^{\prime } =0$ in
equations (\ref{scalar1}) and (\ref{scalar2}), obtaining 
\begin{equation}
\varpi = \frac{2\left( q+1 \right) }{\alpha } \ , \quad \Omega _V =
 -\frac{2(q+1)(q-2)}{3\alpha ^2} \ . \label{eq:oscillationV}
\end{equation}
 Further neglecting the effect of gauge field on the spacetime geometry,
one effectively arrives at the scalar power-law solution 
(discussed in section \ref{sec:scalar})
\begin{equation}
q = \frac{\alpha ^2 -2}{2} \ . 
\end{equation}
Recall that this solution is not inflationary (namely $q>0$) if $\alpha > \sqrt{2}$.
Second, assuming that $\mathcal{E}_1$ and $\mathcal{B}_1$ are decaying away,
we solve the evolution equation for $\Gamma \mathcal{A}_1$ to yield
\begin{equation}
\Gamma \mathcal{A}_1 = \Gamma (0) \mathcal{A}_1 (0) e^{q\tau } \ . \label{eq:oscillationG}
\end{equation}
Next, we use equations (\ref{eq:ele1}) - (\ref{eq:ele3}) and (\ref{eq:mag1}) - 
(\ref{eq:mag3}) to derive
\begin{align}
\begin{split}
& \mathcal{E}_2^{\prime } = \left[ q-1 -\frac{\lambda }{\alpha }(q+1) \right] 
\mathcal{E}_2 + \Gamma \mathcal{A}_1 \mathcal{B}_3 \ , \\
 & \mathcal{B}_3^{\prime } = \left[ q-1 + \frac{\lambda }{\alpha } (q+1) 
\right] \mathcal{B}_3 - \Gamma \mathcal{A}_1 \mathcal{E}_2 \ , 
\end{split} \label{eq:oscillationEv}
\end{align}
and an analogous set of equations for $\mathcal{E}_3$-$\mathcal{B}_2$.
Because of the exponential growth of $\Gamma \mathcal{A}_1$ for $q>0$, 
the contribution from the first terms soon become negligible. The approximated 
solution is then given by
\begin{equation}
 \mathcal{E}_2 = C \sin \left( \frac{\Gamma (0) 
\mathcal{A}_1 (0)}{q} e^{q\tau } + \phi _0 \right) \ , \quad  \mathcal{B}_3 =C 
\cos \left( \frac{\Gamma (0) \mathcal{A}_1 (0)}{q} e^{q\tau } + \phi _0 \right)
 \ . 
 \label{eq:oscSol}
\end{equation}
The appearance of exponential growth
inside the trigonometric functions explains
the increasingly rapid oscillations towards the end and the numerical difficulty
in that regime.

The fact that the spacetime evolution is effectively that of the 
power-law solution
in section \ref{sec:scalar} might appear puzzling since that solution 
is linearly
unstable for the current parameter set and thus cannot be an attractor. The
linear analysis does not apply here, however, due to the prominent effect of
$\Gamma \mathcal{A}_1$ terms in equations (\ref{eq:oscillationEv}), which
are higher order in perturbation. The net result of these terms is to force
the gauge field to behave as an isotropic radiation fluid by making its components
oscillate rapidly. Then, one might expect the fixed point (\ref{eq:fix1}) to effectively
``become stable" since its only unstable modes come from the gauge field that
now behaves as a radiation fluid. The power-law fixed point is stable against
radiation fluid (and shear) as long as $\alpha \leq 2$ (in the case of equality
it is marginally stable as we have seen here) so that this oscillatory 
regime can be an attractor.

\begin{figure}[htbp]
\begin{center}
\includegraphics[width=1.0\linewidth,height=0.4\linewidth]{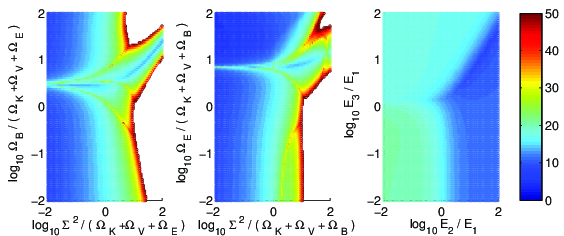}
\caption{Left and centre: the convergence time for varying $\Sigma ^2 $ and $\Omega _B$
or $\Omega _E$. We have $\Sigma _+ = \Sigma _- >0 , \mathcal{E}_1 = \mathcal{E}_2 
=\mathcal{E}_3, \mathcal{B}_1 = \mathcal{B}_2 = \mathcal{B}_3, \Gamma =1$ and 
$\Omega _K = \Omega _V = \Omega _E $ (or $=\Omega _B$). Again, $\Omega _B$
and $\Omega _E$ hardly influence the result while large $\Sigma ^2$ 
drives oscillation even in the isotropic initial magnetic field. Right: the dependence
on the initial direction of electric field. The initial condition is the equipartition 
(\ref{eq:initial1}) plus $\Sigma _+ = \Sigma _-  >0 , \mathcal{B}_1 = \mathcal{B}_2 
= \mathcal{B}_3, \Gamma = 1$. No significant dependence is seen.}
\label{fig:MagDirOsc2}
\end{center}
\end{figure}

It is observed that the condition for this oscillatory attractor to be reached
is similar to that for a long anisotropic period in the Abelian Bianchi type I. 
In other words,
the oscillation occurs when the components of magnetic fields have 
different magnitudes among them except for the configuration $\mathcal{B}_2
\sim \mathcal{B}_3 \ll \mathcal{B}_1 $ and its permutations. It is reasonable
considering the nature of the oscillation which requires two dominant
components with one decaying exponentially. 
 Although not presented
here, the initial direction of shear does not play an important role in deciding
the final state of the universe as in the Abelian case. The dependence on 
the initial value of $\Gamma $ is rather straightforward. The white regions
start to appear around $\Gamma =0.3$ and grow larger as $\Gamma $ 
increases. The effect of unequal density parameters is investigated
 in figure \ref{fig:MagDirOsc2}. The convergence time is again mostly
 insensitive to them except for $\Sigma ^2$, greater values of which 
 induce oscillations even for isotropic initial magnetic field.

%%%%%%%%%%%%%%%%%%%%%%%%%%%%%%%%%%%%%%%%%%%%%%%%%%%%%%%%%%%
%%%%%%%%%%%%%%%%%%%%%%%%%%%%%%%%%%%%%%%%%%%%%%%%%%%%%%%%%%%
%%%%%%%%%%%%%%%%%%%%%%%%%%%%%%%%%%%%%%%%%%%%%%%%%%%%%%%%%%%
\section{The effect of spatial curvature}
%%%%%%%%%%%%%%%%%%%%%%%%%%%%%%%%%%%%%%%%%%%%%%%%%%%%%%%%%%%
%%%%%%%%%%%%%%%%%%%%%%%%%%%%%%%%%%%%%%%%%%%%%%%%%%%%%%%%%%%
%%%%%%%%%%%%%%%%%%%%%%%%%%%%%%%%%%%%%%%%%%%%%%%%%%%%%%%%%%%
Now we are going to discuss how inclusion of spatial curvature changes
the results obtained in the previous section.  
 Again, $(\alpha , \lambda ) = (2,5)$ is
taken as the representative set of parameters.
 During inflation, spatial curvature
is widely believed to be irrelevant. Although the local stability analysis 
supports
this assumption, we will show that it does not apply to the global dynamics.
In general, they tend to increase the time for convergence to the isotropic 
inflationary state, if it is asymptotically realized,
and hence affect the physical predictions. 
It also drastically reduces the chance
to encounter the oscillatory attractor.

%%%%%%%%%%%%%%%%%%%%%%%%%%%%%%%%%%%%%%%%%%%%%%%%%%%%%%%%%%%
%%%%%%%%%%%%%%%%%%%%%%%%%%%%%%%%%%%%%%%%%%%%%%%%%%%%%%%%%%%
%%%%%%%%%%%%%%%%%%%%%%%%%%%%%%%%%%%%%%%%%%%%%%%%%%%%%%%%%%%
\subsection{Bianchi type II}
%%%%%%%%%%%%%%%%%%%%%%%%%%%%%%%%%%%%%%%%%%%%%%%%%%%%%%%%%%%
%%%%%%%%%%%%%%%%%%%%%%%%%%%%%%%%%%%%%%%%%%%%%%%%%%%%%%%%%%%
%%%%%%%%%%%%%%%%%%%%%%%%%%%%%%%%%%%%%%%%%%%%%%%%%%%%%%%%%%%
The simplest spatially curved homogeneous model is Bianchi type II for which
one of the $\mathcal{N}_A$'s is non-zero and the other two vanish. Without
loss of generality, one can assume that the non-zero component is positive.
\begin{figure}[htbp]
\begin{center}
\includegraphics[width=1.0\linewidth,height=0.4\linewidth]
{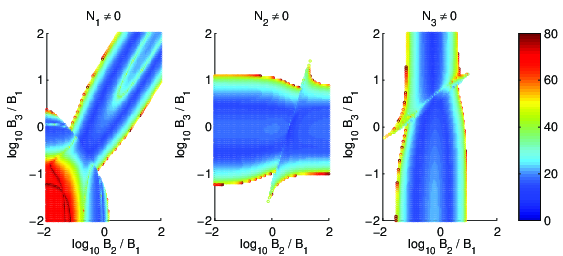}
\caption{Dependence of convergence time on anisotropy in
magnetic configuration for $\mathcal{N}_1 \neq 0$, $\mathcal{N}_2 \neq 0$
and $\mathcal{N}_3 \neq 0$ respectively.  
The shapes of the contour are significantly different from the 
corresponding plot for type I (figure \ref{fig:MagDirOsc}). When $\mathcal{N}_1
\neq 0$, oscillation occurs for $\mathcal{B}_2 \gg \mathcal{B}_3 $ or
$\mathcal{B}_3 \gg \mathcal{B}_2$ and for corresponding cyclic permutations
when $\mathcal{N}_2 \neq 0$ or $\mathcal{N}_3 \neq 0$. Also note that the
anomalously long convergence time in the bottom-left corner of the left panel.}
\label{fig:type2a}
\end{center}
\end{figure}
First of all, let us investigate the dependence on the anisotropy in the 
initial magnetic configuration, which turned out to be the deciding factor 
of the global dynamics for Bianchi type I. 
As for the initial data, we take all the density parameters to be equal
\begin{equation}
\Sigma ^2 = \Omega _K = \Omega _V = \Omega _N = \Omega _E = \Omega _B = 
\frac{1}{6} \label{eq:inicon2a}
\end{equation}
and set
\begin{equation}
\Sigma _+ = \Sigma _- \ , \quad \mathcal{E}_1 = \mathcal{E}_2 = \mathcal{E}_3 
\ , \quad \Gamma =1 \ . \label{eq:inicon2b}
\end{equation}
Figure \ref{fig:type2a} shows the convergence time
 for varying initial magnetic configuration with $\mathcal{N}_1 \neq 0
\,,  \mathcal{N}_2 \neq 0$  and $\mathcal{N}_3 \neq 0$. 
The initial conditions for which the 
attractor  is non-inflationary oscillatory solution
 are indicated by white as before. We see a significant change
in the
shape of white region compared to the corresponding plot for type I 
(right panel of figure 
 \ref{fig:MagDirOsc}). 
The three white regions in the Bianchi type I are reduced 
to two in the type II.  A speculative interpretation is
that the disappeared 
region corresponds to the single non-zero curvature component ($\mathcal{N}_1$
for the left panel and $\mathcal{N}_2$ for the centre). This means that 
the existence of spatial curvature  
prevents the system from evolving into the non-inflationary oscillatory solutions.
This hypothesis is   supported by the results for the other Bianchi types
(See below for types VIII and IX, and Appendix for 
types VI$_0$ and VII$_0$.)
\begin{figure}[htbp]
\begin{center}
\includegraphics[width=1.0\linewidth,height=0.4\linewidth]{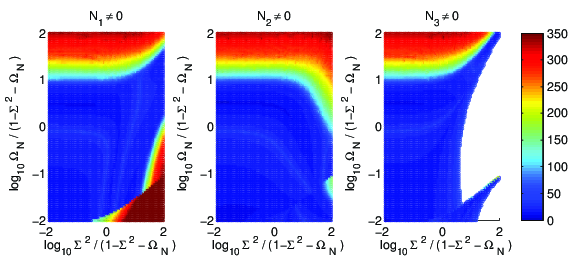}
\caption{Dependence of the convergence time on initial amplitudes of $\Sigma ^2$
and $\Omega _N$. The initial conditions satisfy $\Omega _K = \Omega _V = \Omega _E = \Omega _B$
and $\Sigma _+ = \Sigma _-  >0 , \mathcal{E}_1 = \mathcal{E}_2 =\mathcal{E}_3 ,
\mathcal{B}_1 = \mathcal{B}_2 = \mathcal{B}_3$. The tendency towards a long 
period of anisotropic inflation is 
observed for large $\Omega _N$. The appearance of oscillation in $\mathcal{N}_3 
\neq 0$ for large $\Sigma ^2$ is not totally surprising since the similar behaviour has been
noted in type I, although the reason for its absence in $\mathcal{N}_1 \neq 0$ and 
$\mathcal{N}_2 \neq 0$ is unknown.}
\label{fig:type2b}
\end{center}
\end{figure}

\begin{figure}[htbp]
\begin{center}
\includegraphics[height=0.24\linewidth]{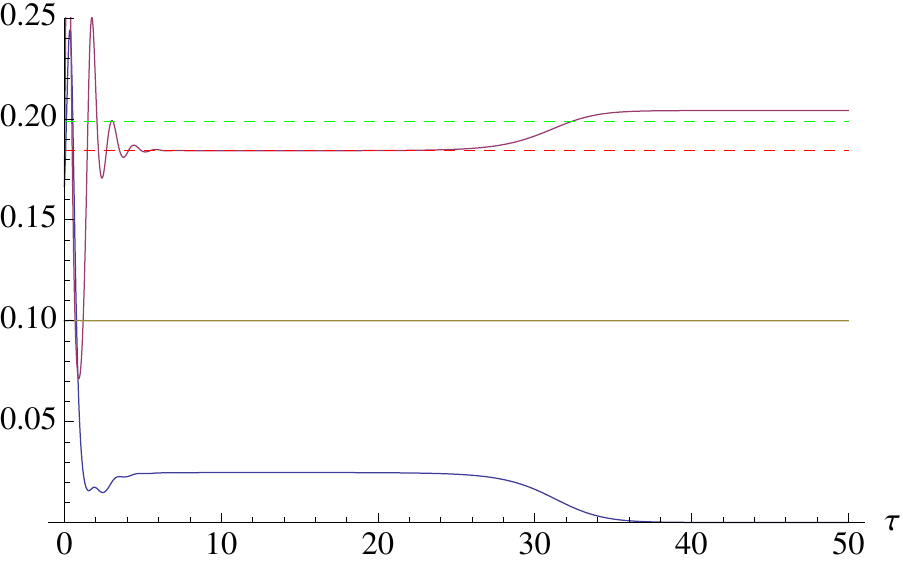}
\includegraphics[height=0.24\linewidth]{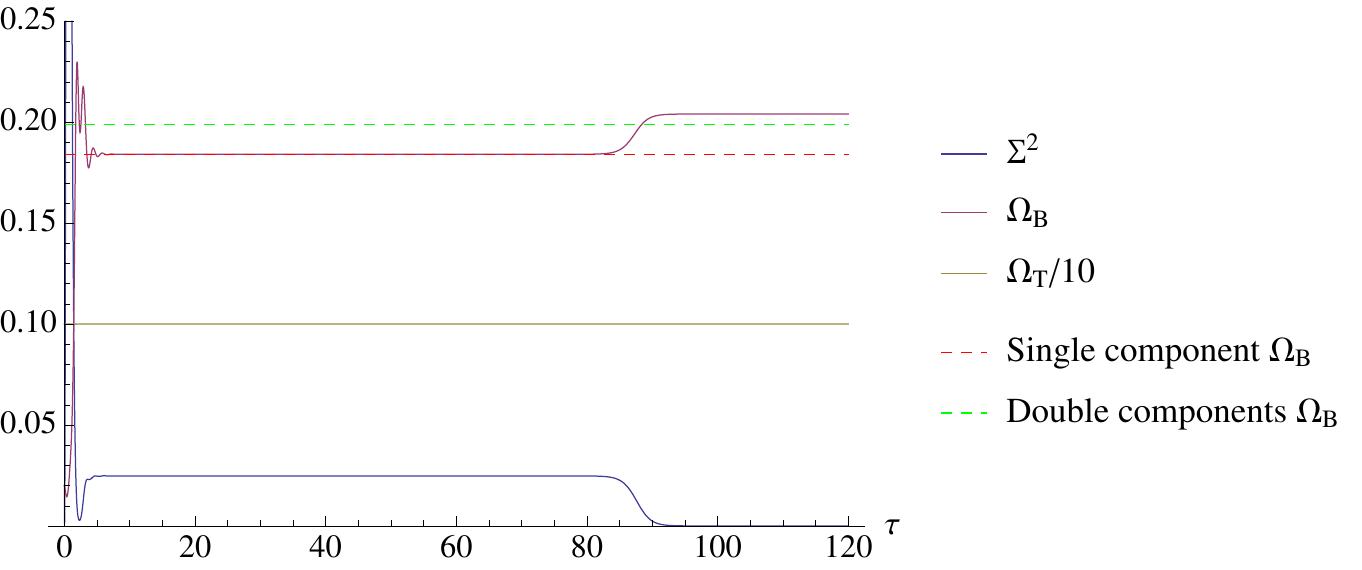}
\caption{Transient anisotropy and convergence to the isotropic attractor 
for Bianchi
type II. The initial conditions are; $\mathcal{N}_1 \neq 0 , \Sigma ^2 = 
\Omega _K = \Omega _V
= \Omega _N = \Omega _E = \Omega _B = 1/6, \mathcal{B}_1 = 10\mathcal{B}_2 =10
\mathcal{B}_3$
for left, $\mathcal{N}_2 \neq 0 , \Omega _N = 9/10, \Sigma ^2 = \Omega _K = 
\Omega _V =
\Omega _E = \Omega _B = 1/50, \mathcal{B}_1 =\mathcal{B}_2 = \mathcal{B}_3$. 
The
other conditions are as explained in the text. We also plot $\Omega _T $ to 
confirm that
the Hamiltonian constraint is maintained. Both saddles are clearly seen in the
 right
while the double component inflation is skipped in the left.}
\label{fig:type2ev}
\end{center}
\end{figure}

Figure \ref{fig:type2b} examines the effect of varying $\Omega _N$. 
Clearly, curvature-dominated universes undergo a particularly long period
of anisotropic inflation, which may last more than $300$ e-foldings. Along
with the left-bottom corner in the left panel of figure \ref{fig:type2a}, such
a prolonged anisotropic phase is not expected from the linear stability
analysis. In any of those regions, the single-component phase appears
to be responsible for the anisotropy as can be seen 
in figure \ref{fig:type2ev}. This is the topic of the next subsection.

%%%%%%%%%%%%%%%%%%%%%%%%%%%%%%%%%%%%%%%%%%%%%%%%%
%%%%%%%%%%%%%%%%%%%%%%%%%%%%%%%%%%%%%%%%%%%%%%%%%
%%%%%%%%%%%%%%%%%%%%%%%%%%%%%%%%%%%%%%%%%%%%%%%%%
\subsection{Quasi-single-component magnetic inflation}
\label{Quasi-single-component}
%%%%%%%%%%%%%%%%%%%%%%%%%%%%%%%%%%%%%%%%%%%%%%%%%
%%%%%%%%%%%%%%%%%%%%%%%%%%%%%%%%%%%%%%%%%%%%%%%%%
%%%%%%%%%%%%%%%%%%%%%%%%%%%%%%%%%%%%%%%%%%%%%%%%%
One may wonder why the anisotropic phases in type II (figure \ref{fig:type2ev})
could be as long as 100 e-folds while the stability analysis suggests that
the linear instability kicks in after $\sim \Sigma ^{-1} \sim 10$ e-folds. 
Although
the values of $\Sigma ^2$ and $\Omega _B$ during the anisotropic periods
are almost exactly those for the single-component magnetic fixed point
discussed in section \ref{sec:singleMag}, it turns out that the internal 
mechanism
is very different, involving the spatial curvature.

\begin{figure}[htbp]
\begin{center}
\includegraphics[height=0.23\linewidth]{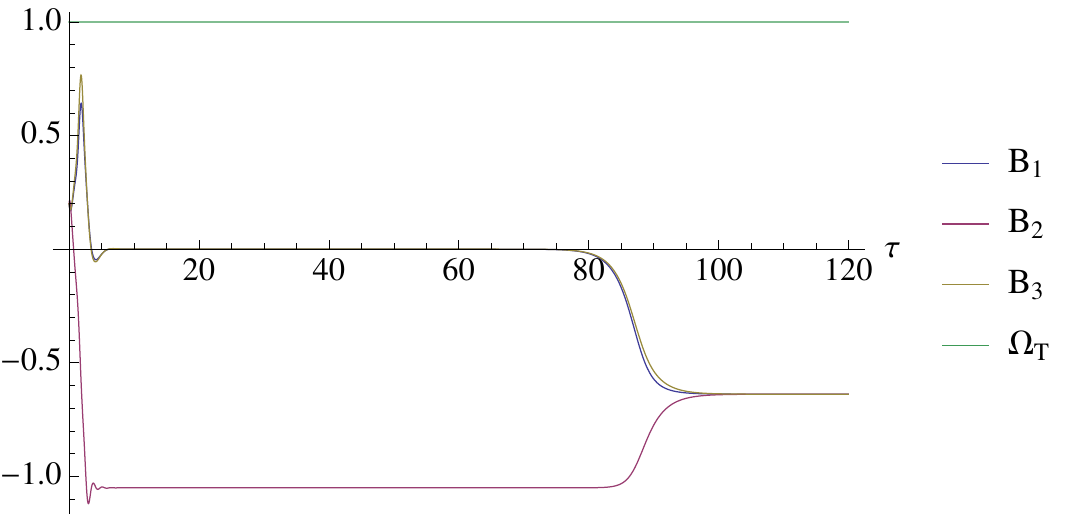}
\includegraphics[height=0.23\linewidth]{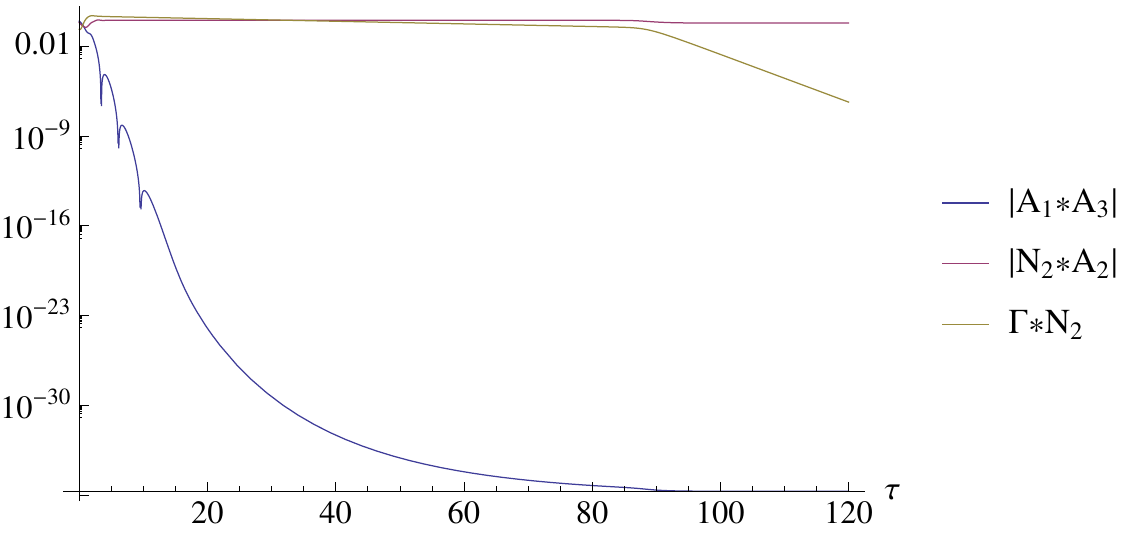}
\includegraphics[height=0.23\linewidth]{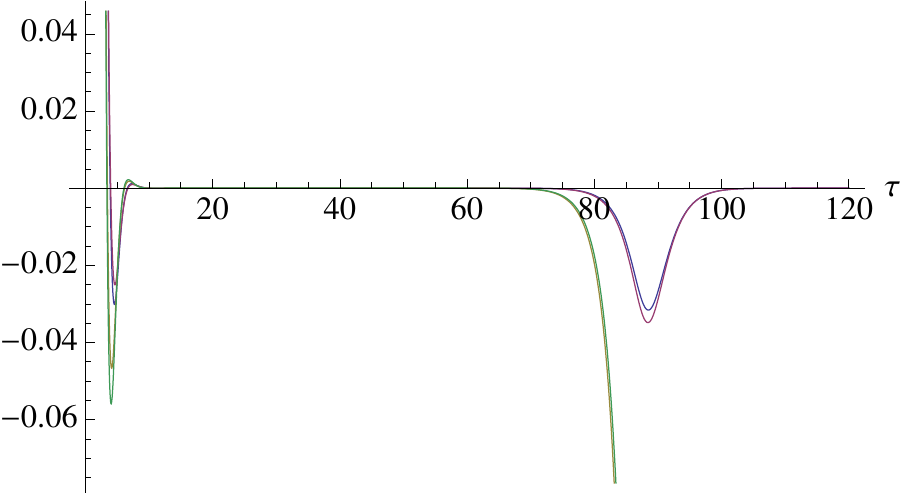}
\includegraphics[height=0.23\linewidth]{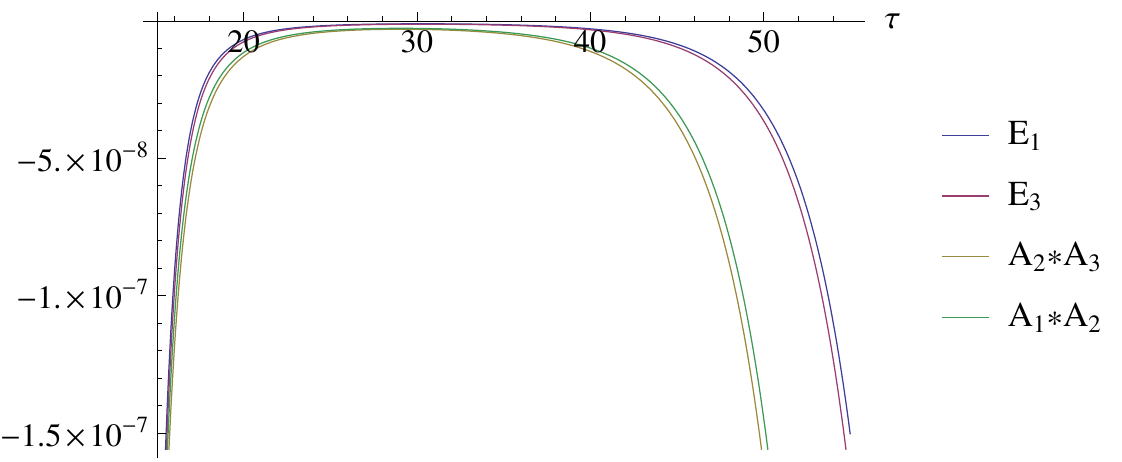}
\caption{Dynamics for initial conditions $\mathcal{N}_2 \neq 0, 
\Omega _N = 9/10,
\Sigma ^2 = \Omega _K = \Omega _V = \Omega _E = \Omega _B = 1/50, 
\mathcal{B}_1 = \mathcal{B}_2 = \mathcal{B}_3 $. Top-left: time evolution 
of magnetic components. Only $\mathcal{B}_2$ has a significant amplitude.
Top-right: comparison between the two contributions to $\mathcal{B}_2$. 
The vertical axis is logarithmic.
It is dominated by the curvature term. During the anisotropic phase, $\Gamma
\mathcal{A}_2$ decays only slowly. Bottom: plots of $\mathcal{E}_1 , 
\mathcal{E}_3 , \mathcal{B}_1 ,\mathcal{B}_3 $ for different time-scales.
Note that after $\tau \sim 90$, the electric components are exponentially
suppressed.}
\label{fig:type2ev2}
\end{center}
\end{figure}

Figure \ref{fig:type2ev2} shows time evolution of various quantities for the 
numerical solution presented in the right panel of figure \ref{fig:type2ev}.
From the top-left panel, the system experiences $\sim 80$ e-foldings of
anisotropic inflation with non-vanishing $\mathcal{B}_2$. If this was the
fixed point studied in section \ref{sec:singleMag}, $\mathcal{A}_1$ and
$\mathcal{A}_3$ should be roughly constant during that period. But the
top-right panel shows exponential decay of $\mathcal{A}_1 \mathcal{A}_3$.
Instead, $\mathcal{B}_2$ is dominated by the second term $-\mathcal{N}_2
\mathcal{A}_2$ which involves the spatial curvature. According to this
observation, one might suspect that there is another fixed point with 
non-vanishing $\mathcal{N}_2$ and constant $\mathcal{A}_2 \mathcal{N}_2$
with exactly the same apparent behaviour. However, such a solution does
not exist even asymptotically. In fact, one can see on the bottom-right that
the other components of magnetic field ($\mathcal{B}_1 =
 \mathcal{A}_2 \mathcal{A}_3$
and $\mathcal{B}_3 = \mathcal{A}_1 \mathcal{A}_2$) as well as electric
 components
do not behave as expected. If this could be understood by simple linear
perturbation around a fixed point, the negligibly small $\mathcal{E}_1, 
\mathcal{E}_3 ,
\mathcal{B}_1$ and $\mathcal{B}_3$ should have evolved monotonically 
(or oscillatory in case of imaginary eigenvalue) according to their
 stability.
Here they all show both exponential decay and growth deep inside the period
of the anisotropic inflation (between $\tau = 10$ and $\tau = 50$).
 In particular,
the growth of electric components around $\tau =50$ is completely against 
the principle of magnetic inflation discussed so far. 

This anomalous situation can be partially understood by taking a heuristic 
approach. First of all, let us assume that all the geometric and scalar-field
variables take their equilibrium values for the single-component magnetic
inflation. Further, from the numerical evidence, assume 
\begin{equation}
\begin{split}
& | \mathcal{B}_1 | \gg |\mathcal{B}_2 | \ , \ |\mathcal{B}_3 | \ , \
 |\mathcal{E}_1 | \ , \ | \mathcal{E}_2 | \ , \ |\mathcal{E}_3 | \ ,  \\
& |\mathcal{N}_1 \mathcal{A}_1 | \gg |\mathcal{A}_2 \mathcal{A}_3 | \ .
\end{split}
\end{equation}
Under these assumptions, we can derive the leading order behaviours
\begin{equation}
\mathcal{A}_1 \propto e^{3\Sigma _+ \tau } \ , \quad \mathcal{N}_1 
\propto e^{-3\Sigma _+ \tau } \ , \quad \Gamma \propto e^{(q-\Sigma _+)\tau }
 \ .
\end{equation}
These indeed give the constant magnetic field
$\mathcal{B}_1 \sim - \mathcal{N}_1 \mathcal{A}_1 $.
To see what happens to the other components of gauge field, we first note
that
\begin{equation}
\Gamma \mathcal{A}_1 \propto e^{(q+2\Sigma _+ ) \tau } \ ,
\end{equation}
with $0<\Sigma _+ , 1+q \ll 1$ for realistic inflationary models. Hence this
combination is always decaying with the exponent 
$|q + 2\Sigma _+ | \lesssim 1$ 
during the regime under consideration. The very slow decline as can be seen 
in figure \ref{fig:type2ev2} is expected to be rather peculiar for this set
of parameters ($q+2\Sigma_+=-1/37$ for $\alpha = 2, \lambda =5$). 
When it does happen, 
$\Gamma \mathcal{A}_1$
plays an important role in the evolution of the electromagnetic components
perpendicular to $\mathcal{B}_1$. Their evolution equations can be evaluated
as
\begin{equation}
\begin{split}
\mathcal{E}_2^{\prime } \sim & \left( 2q - 2 - \Sigma _+ \right) \mathcal{E}_2 + \Gamma \mathcal{A}_1 \mathcal{B}_3 \ , \\
\mathcal{E}_3^{\prime } \sim & \left( 2q - 2 -\Sigma _+ \right) \mathcal{E}_3 + \Gamma \mathcal{A}_1 \mathcal{B}_2 \ , \\
\mathcal{B}_2^{\prime } \sim & 3\Sigma _+ \mathcal{B}_2 -\Gamma \mathcal{A}_1 \mathcal{E}_3 \ , \\
\mathcal{B}_3^{\prime } \sim & 3\Sigma _+ \mathcal{B}_3 - \Gamma \mathcal{A}_1 \mathcal{E}_2 \ .
\end{split}
\end{equation} 
Note that the mixing terms involving $\Gamma \mathcal{A}_1$ would have been by
definition second order in perturbation for linear analysis in section \ref{sec:singleMag}. 
The dynamics of this coupled system crucially depends on the evolution of
$\Gamma \mathcal{A}_1$, but the generic effect of the interaction terms is 
again oscillatory. This does not easily allow the magnetic components to grow
indefinitely with the linear instability with respect to $\mathcal{B}_1$.
At some stage, the present approximation breaks down ($\Gamma \mathcal{A}_1
$ eventually dies away when $|q| > 2\Sigma _+$) and the system leaves this regime.
Numerical experiments suggest that this mechanism is in action whenever
there appears an anisotropic inflationary period of much longer than $30$ e-folds.
In more realistic models of inflation, however, since the decay of $\Gamma \mathcal{A}_1$ 
is expected to be much faster, the frequency of this event as well as 
the duration of anisotropic phase when it does occur should be much less than
in our setup.  This will be partially confirmed in section \ref{sec:param}.

%%%%%%%%%%%%%%%%%%%%%%%%%%%%%%%%%%%%%%%%%%%%%%%%%%%%%%%%%%%
%%%%%%%%%%%%%%%%%%%%%%%%%%%%%%%%%%%%%%%%%%%%%%%%%%%%%%%%%%%
%%%%%%%%%%%%%%%%%%%%%%%%%%%%%%%%%%%%%%%%%%%%%%%%%%%%%%%%%%%
\subsection{Generic Bianchi types}
%%%%%%%%%%%%%%%%%%%%%%%%%%%%%%%%%%%%%%%%%%%%%%%%%%%%%%%%%%%
%%%%%%%%%%%%%%%%%%%%%%%%%%%%%%%%%%%%%%%%%%%%%%%%%%%%%%%%%%%
%%%%%%%%%%%%%%%%%%%%%%%%%%%%%%%%%%%%%%%%%%%%%%%%%%%%%%%%%%%
The most general anisotropic cosmologies are Bianchi type VIII and IX for which
none of $\mathcal{N}_A$ vanishes. All the other types in class A can be 
considered as the boundary sets of these two  (where behaviours can be
very much different, see the Appendix for VI$_0$ and VII$_0$). 

\subsubsection{Bianchi type VIII}
For type VIII, one can take one of 
$\mathcal{N}_A$ to be negative with the other two being positive, 
which implies positive definite $\Omega _N$. 
\begin{figure}[htbp]
\begin{center}
\includegraphics[width=1.0\linewidth,height=0.4\linewidth]{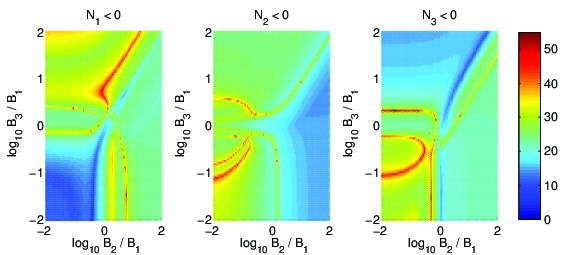}
\caption{Plots of convergence time for varying magnetic field with different 
choices
of negative $\mathcal{N}_A$. All the panels show qualitatively similar 
features. The
convergence is fast for the region where $\mathcal{N}_A <0$ and $\mathcal{B}_A$
is greater compared to the other two magnetic components. As far as these blue
regions are concerned, $\mathcal{N}_2 <0$ and $\mathcal{N}_3 <0$ are roughly
symmetric in reflection with respect to the diagonal $\mathcal{B}_2 = 
\mathcal{B}_3$.
The slight skew is expected as $\Sigma _- \neq 0$ initially. Overall, 
the convergence
appears to take longer than type I, probably because of the anisotropic nature 
of
type VIII spatial curvature. No oscillation is observed in type VIII.}
\label{fig:type8a}
\end{center}
\end{figure}
We repeat the same calculations as the type II with the initial conditions 
(\ref{eq:inicon2a}), (\ref{eq:inicon2b}) and the spatial curvature variables satisfying
\begin{equation}
|\mathcal{N}_1 | = |\mathcal{N}_2 | = |\mathcal{N}_3 | \ . \label{eq:inicon2c}
\end{equation}
Note that this configuration does not mean the spatial curvature is isotropic
because of the differing signatures.
The figure \ref{fig:type8a} shows plots for different choices of the
negative component in spatial curvature. It is observed that the anisotropy 
is suppressed in the region where $\mathcal{B}_A$ is dominant over 
the other magnetic components for $\mathcal{N}_A <0$. 
For example, for the initial data with $\mathcal{N}_1 <0$, 
the spacetime with $\mathcal{B}_1 \gg \mathcal{B}_2, 
\mathcal{B}_3$ will evolve rapidly into an isotropic one. This might
be partially explained by the structure of the evolution equations of
 $\mathcal{E}_1$
and $\mathcal{B}_1$. The coupling between negative $\mathcal{N}_1$ and 
$\mathcal{B}_1$ drives $\mathcal{E}_1$ initially, which in turn accelerates
the decay of $\mathcal{B}_1$ through $-\Gamma \mathcal{N}_1 \mathcal{E}_1$.
%It is because $\mathcal{E}_1$ increases and then 
%$\mathcal{B}_1$ decreases, resulting in an isotropic spacetime.

In other areas, the anisotropic phase is typically 
longer than that in type I. No obvious connection can be
seen between the initial anisotropy of magnetic components and
the convergence time.
Moreover, there is no oscillatory phase seen for type 
VIII even though $\Gamma =1$ initially. 

%%%%%%%%%%%%%%%%%%%%%%%%%%%%%%%%%%%%%%%%%%%%%%%%%%%%%%%%%%%
%%%%%%%%%%%%%%%%%%%%%%%%%%%%%%%%%%%%%%%%%%%%%%%%%%%%%%%%%%%
%%%%%%%%%%%%%%%%%%%%%%%%%%%%%%%%%%%%%%%%%%%%%%%%%%%%%%%%%%%
\subsubsection{Bianchi type IX}
\begin{figure}[htbp]
\begin{center}
\includegraphics[width=1.0\linewidth,height=0.4\linewidth]{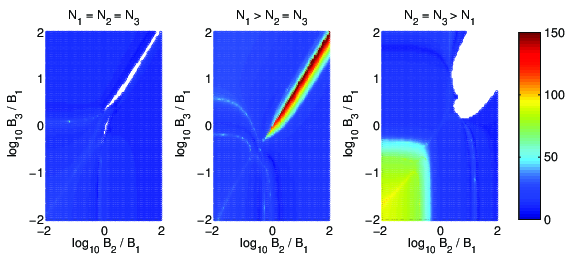}
\caption{Plots of convergence time for varying magnetic field in type IX. 
Left: isotropic initial spatial curvature. The white strips represent the 
initial conditions that resulted in the recollapse of the universe. 
Unless it takes place, the final isotropic state is achieved within $15$-$20$ e-folds. 
Centre: $\mathcal{N}_1= 10\mathcal{N}_2 = 10\mathcal{N}_3$ initially. Note that
$\Omega _N >0$ in this case so that the equipartition condition is
$\Sigma ^2 = \Omega _K = \Omega _V = \Omega _N = \Omega _E = \Omega _B = 1/6$.
In the red region, the intermediate evolution is the quasi-single-component magnetic inflation.
Right: $10\mathcal{N}_1 = \mathcal{N}_2 = \mathcal{N}_3$. The white region is
where the recollapse takes place.}
\label{fig:type9a}
\end{center}
\end{figure}

\begin{figure}[htbp]
\begin{center}
\includegraphics[height=0.26\linewidth]{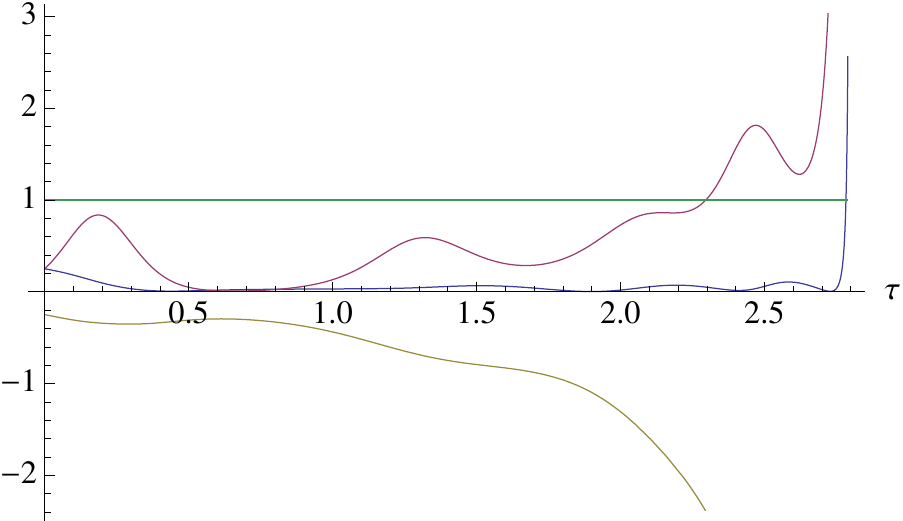}
\includegraphics[height=0.26\linewidth]{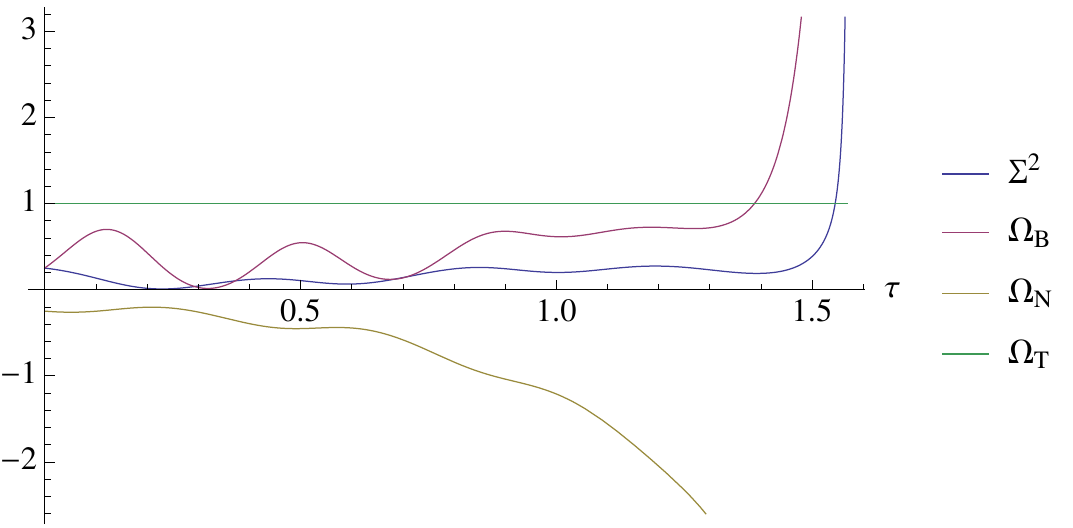}
\caption{Two examples of recollapse taking place for type IX. The initial 
conditions
are $\mathcal{N}_1 = \mathcal{N}_2 =\mathcal{N}_3 ,  \mathcal{B}_2 = 
\mathcal{B}_3 = 10 \mathcal{B}_1 $
for left panel and $\mathcal{N}_2 = \mathcal{N}_3 = 10\mathcal{N}_1 ,
 \mathcal{B}_2 = \mathcal{B}_3 = 10 \mathcal{B}_1 $ for right panel
 (each corresponding to a point in the white region of left and right panels
 in figure \ref{fig:type9a} respectively). 
 One can observe that the density parameters blow up, in particular
 $\Omega _N$ goes negative,
which indicates $H \rightarrow 0$. $\Omega _T$ is plotted to confirm that the 
calculation
is not ruined by the rapidly changing variables.}
\label{fig:recollapse}
\end{center}
\end{figure}

%%%%%%%%%%%%%%%%%%%%%%%%%%%%%%%%%%%%%%%%%%%%%%%%%%%%%%%%%%%
%%%%%%%%%%%%%%%%%%%%%%%%%%%%%%%%%%%%%%%%%%%%%%%%%%%%%%%%%%%
%%%%%%%%%%%%%%%%%%%%%%%%%%%%%%%%%%%%%%%%%%%%%%%%%%%%%%%%%%%
This subsection presents similar analysis for type IX for
which $\mathcal{N}_A >0 , A =1,2,3$. Since $\Omega _N$ is not necessarily
positive here, the initial conditions have to be modified
from (\ref{eq:inicon2a}). For left and right panels in figure
\ref{fig:type9a}, $\Omega _N$ is initially negative so that we take, for equipartition,
\begin{equation}
\Sigma ^2 = \Omega _K = \Omega _V = -\Omega _N = \Omega _E = \Omega _B = 
\frac{1}{4} \ . \label{eq:equip9}
\end{equation}
The initial shear and electric field are fixed as before
\begin{equation}
\Sigma _+ = \Sigma _- \ , \quad \mathcal{E}_1 = \mathcal{E}_2 =\mathcal{E}_3 
\ .
\end{equation}
Three different configurations of initial spatial curvature are studied:
\begin{equation}
\begin{split}
{\rm Left} &: \mathcal{N}_1 = \mathcal{N}_2 = \mathcal{N}_3 \ , \\
{\rm Centre} &: \mathcal{N}_1 = 10 \mathcal{N}_2 = 10 \mathcal{N}_3 \ , \\
{\rm Right} &: 10\mathcal{N}_1 = \mathcal{N}_2 = \mathcal{N}_3 \ .
\end{split}
\end{equation}
$\Gamma =1$ is common to all of them. They appear qualitatively similar. 
The white regions in this case do not represent oscillation but the recollapse
caused by growing negative $\Omega _N$  (figure \ref{fig:recollapse}). 
As far as we have checked, there is 
no oscillation observed for type IX either. 
For the isotropic spatial curvature, the convergence time is significantly shorter
than type VIII. Notice that the spatial curvature for type VIII is anisotropic
 in nature. When curvature is anisotropic in type IX, the order of magnitude
 of convergence time is similar to type VIII indeed. In the strip along the line
 $\mathcal{B}_2 = \mathcal{B}_3$ in the centre panel and the bottom-left
 region in the right panel, where convergence time is anomalously elongated,
  we confirmed that the anisotropic phase is given by the 
quasi-single-component   magnetic inflation. 
 It appears reasonable to conclude that the appearance of this phase is 
sensitive to the 
initial anisotropy in the spatial curvature but not the initial shear.

%%%%%%%%%%%%%%%%%%%%%%%%%%%%%%%%%%%%%%%%%%%%%%%%%%%%%%%%%%%
%%%%%%%%%%%%%%%%%%%%%%%%%%%%%%%%%%%%%%%%%%%%%%%%%%%%%%%%%%%
\subsection{Relation between oscillation and spatial curvature}
%%%%%%%%%%%%%%%%%%%%%%%%%%%%%%%%%%%%%%%%%%%%%%%%%%%%%%%%%%%
%%%%%%%%%%%%%%%%%%%%%%%%%%%%%%%%%%%%%%%%%%%%%%%%%%%%%%%%%%%
%%%%%%%%%%%%%%%%%%%%%%%%%%%%%%%%%%%%%%%%%%%%%%%%%%%%%%%%%%%

The numerical calculations suggest that spatial curvature generally suppresses
occurrence of the oscillatory attractor. As an effort to identify the 
mechanism, let us
look at the behaviour of the curvature variables $\mathcal{N}_A$ in the 
asymptotic
oscillatory solution derived in the previous section. From equations 
(\ref{eq:curv1}) -
(\ref{eq:curv3}), one can see
\begin{equation}
\mathcal{N}_A^{\prime } = \frac{1}{2} \left( q-1 + \frac{\lambda }{2}\varpi 
\right) \mathcal{N}_A \rightarrow \mathcal{N}_A \sim \mathcal{N}_A (0) \exp 
\frac{(\alpha + \lambda )q + (\lambda -\alpha )}{2\alpha } \tau \ ,
\end{equation}
where we used the condition of the power-law
``fixed point" (\ref{eq:oscillationV}). 
Given $q>0$ and $\lambda 
>\alpha >0$,
the exponent is positive so that $\mathcal{N}_A$ are all growing. 
The physically
relevant variables $N_A = \Gamma \mathcal{N}_A$ go as $\propto e^{q\tau}$ as 
well
so that the oscillatory regime is unstable against perturbations of spatial 
curvature. 
\begin{figure}[htbp]
\begin{center}
\includegraphics[height=0.25\linewidth]{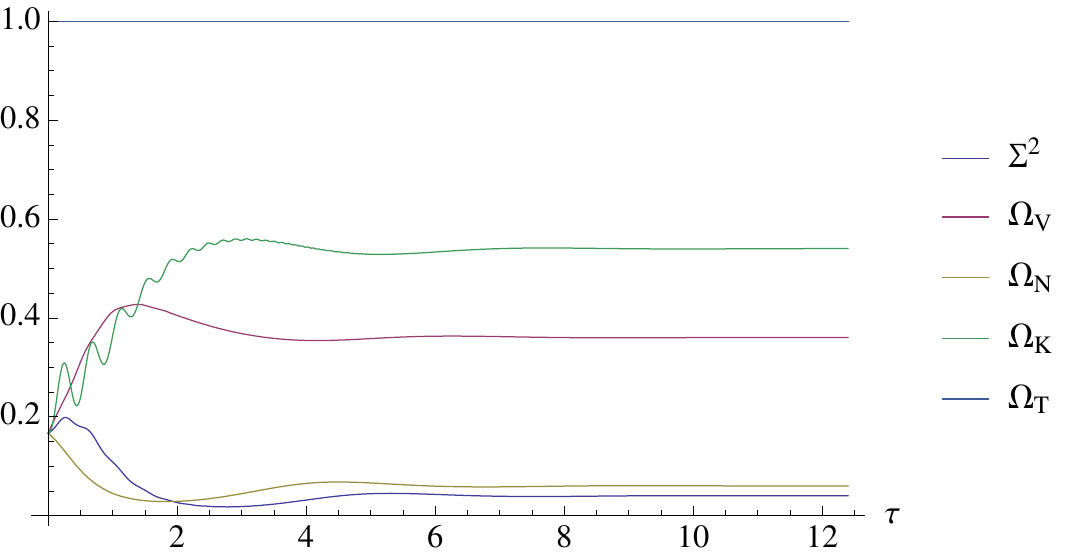}
\includegraphics[height=0.25\linewidth]{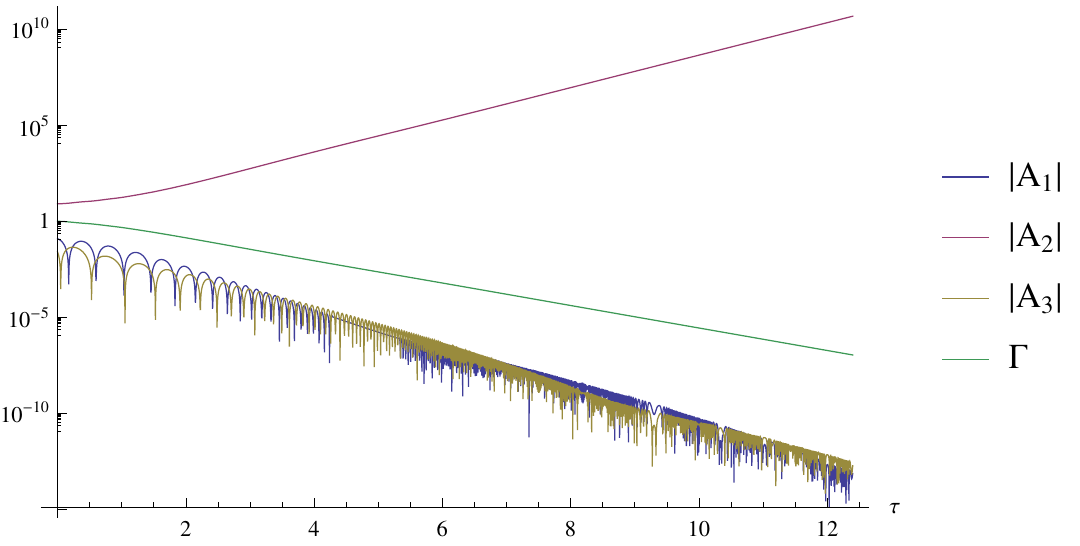}
\caption{Left: behaviour of the density parameters during type II oscillation. 
Note the 
significant contribution of $\Sigma ^2 $ and $\Omega _N$ in contrast to the 
type I case.
Right: time evolution of $\Gamma $ and components of vector potential for the 
same
numerical solution.  
The apparent discontinuity in the oscillating vector potential
is due to the logarithmic scale.
 The non-zero curvature component is $\mathcal{N}_1$ and 
the initial conditions are $\mathcal{B}_2 = 0.1 \mathcal{B}_1 ,
 \mathcal{B}_3 
= 10^{3/2}\mathcal{B}_1$.}
\label{fig:type2osc}
\end{center}
\end{figure}
 As the result, the oscillatory phases observed in type II (also IV$_0$ and
VII$_0$) are in fact of a different nature from those in type I.  Figure
\ref{fig:type2osc} shows what is going on when oscillation takes place in
Bianchi type II. We see that the system appears to settle down to stationary
attractor state before the Hamiltonian constraint breaks down due to numerical
difficulty caused by rapid oscillation. The difference from type I (figure 
\ref{fig:oscillation})
is the non-vanishing energy density of shear and curvature. The vector 
potential
appears to have the same feature of one growing and two damped oscillation 
components. To find the asymptotic type II solution similar to 
(\ref{eq:oscSol}), 
we assume $\mathcal{N}_1 \neq 0 , \Sigma _- =0$ and set $\Sigma _+ , \varpi , 
\Omega _V,
\Omega _N = \Gamma ^2 \mathcal{N}_1^2 /12$ all to be constant. From the 
equilibrium 
conditions, we derive 
\begin{align}
\begin{split}
& q = 4\Sigma _+ \ , \quad \Gamma ^2 \mathcal{N}_1^2 = 6\Sigma _+ 
(2\Sigma _+ -1) \ , \quad \varpi = \frac{2}{\alpha } \left( 4\Sigma _+ 
+1\right)  \ , \\
& \Omega _V = -\frac{4}{3\alpha ^2} \left( 4\Sigma _+ +1 \right) 
\left( 2\Sigma _+ -1\right) \ .
\end{split} \label{eq:sol2a}
\end{align} 
Note that $\Sigma _+$ has to be positive in order for $\Omega _V >0$. Now the
solution for equation (\ref{eq:GammaEv}) is 
\begin{equation}
\Gamma \sim \Gamma (0) \exp \left( \frac{\alpha - \lambda }{2\alpha } 
(4\Sigma _+ +1 ) \tau \right) 
\end{equation}
which is consistent with exponential decay seen in figure \ref{fig:type2osc} 
for
$\lambda > \alpha $. Let us assume that the oscillation occurs within 1- and
 3-components
of electric and magnetic fields and the 2-components die away. One can then 
solve the 
evolution equation for $\mathcal{A}_2$ (\ref{eq:a2}) to obtain
\begin{equation}
\Gamma \mathcal{A}_2 \sim \Gamma (0) \mathcal{A}_2 (0) e^{3\Sigma _+ \tau } \ ,
\end{equation}
which indicates the exponential growth of $\mathcal{A}_2$.
Analogous to type I, the evolution equations for the oscillatory components 
are 
\begin{align}
\begin{split}
\mathcal{E}_1^{\prime } =& \left[ 2\Sigma _+ - 1 -\frac{\lambda }{\alpha }
\left( 4\Sigma _+ +1\right) \right] \mathcal{E}_1 - \Gamma \mathcal{N}_1 
\mathcal{B}_1 + \Gamma (0)\mathcal{A}_2 (0) e^{3\Sigma _+ \tau}\mathcal{B}_3 
\ , \\
\mathcal{B}_3^{\prime } =& \left[ 5\Sigma _+ -1 +\frac{\lambda }{\alpha }
\left( 4\Sigma _+ +1 \right) \right] \mathcal{B}_3 -\Gamma (0) \mathcal{A}_2 
(0) e^{3\Sigma _+ \tau } \mathcal{E}_1 \ ,
\end{split}
\end{align}
with a corresponding set of equations for $\mathcal{E}_3$ and $\mathcal{B}_1$. 
Recalling $\Gamma \mathcal{N}_1$ is constant, the exponential
electromagnetic mixing terms soon 
dominate
the dynamics and we obtain the asymptotic solution
\begin{equation}
\mathcal{E}_1 = C \sin \left( \frac{\Gamma (0) \mathcal{A}_2 (0)}{3\Sigma _+} 
e^{3\Sigma _+ \tau } + \phi _0 \right) \ , \quad \mathcal{B}_3 = C \cos 
\left( \frac{\Gamma (0)\mathcal{A}_2 (0)}{3\Sigma _+} e^{3\Sigma _+ \tau } 
+ \phi _0 \right) \ . \label{eq:sol2b}
\end{equation}
The same construction works for assumption of growing $\mathcal{A}_3$, 
leading
to oscillation of $\mathcal{E}_1 , \mathcal{B}_2 $ and $\mathcal{E}_2 , \mathcal{B}_1$. 
However, growing $\mathcal{A}_1$ and decaying $\mathcal{A}_2 , \mathcal{A}_3$ 
are not compatible with decaying $\mathcal{E}_1$ and $\mathcal{B}_1$. It is due
to the definition of magnetic field
\begin{equation}
\mathcal{B}_1 =\mathcal{A}_2 \mathcal{A}_3 - \mathcal{N}_1 \mathcal{A}_1 
\rightarrow -\mathcal{N}_1 (0)\Gamma (0) \frac{\mathcal{A}_1}{\Gamma } \ ,
\end{equation}
which implies $\mathcal{B}_1 $ should be rapidly growing instead of 
oscillating. 
Hence, the spatial curvature in 1-direction $\mathcal{N}_1$ kills the 
possibility of 
oscillation in 2- and 3-components of electromagnetic field. This explains the 
disappearance of the white region for regime $\mathcal{B}_2 \sim \mathcal{B}_3 
\gg 
\mathcal{B}_1$ in the left panel of figure \ref{fig:type2a} that existed 
in type I
(figure \ref{fig:MagDirOsc}) since the oscillation in the region is precisely 
that
among 2- and 3-components as observed in figure \ref{fig:oscillation}. Note 
that the asymptotic solution given by (\ref{eq:sol2a}) and (\ref{eq:sol2b}) is again 
unstable
against perturbation of $N_2$ and $N_3$. Therefore, one can further speculate
that the absence of oscillation in type VIII and type IX is indeed attributed 
to the
presence of spatial curvature. To consolidate
this conclusion, it would be ideal to perform similar numerical calculations 
for
models with two non-vanishing components of spatial curvature, namely
type VI$_0$ and VII$_0$. Unfortunately, it turns out to be impossible 
for the initial data with the equipartition condition such as (\ref{eq:inicon2a})
because of a technical problem, which is explained and alleviated 
in the Appendix. The modified analysis there indeed suggests that
our conclusion is plausible.

\section{Dependence on the parameters $\alpha $ and $\lambda $}\label{sec:param}

So far, all the numerical results presented have been obtained for a particular
parameter set $(\alpha , \lambda ) =(2,5)$. As already explained, the reason for
the choice is mostly the convenience and clarity. In reality, one would have liked 
to have a smaller slow-roll parameter $\epsilon = q+1$. In this section, we demonstrate that the
qualitative features are more or less invariant against changes in $\alpha $ and
$\lambda $ while the quantitative ones can vary according to $\epsilon $. 
In particular, we find the general tendency that smaller $\epsilon $ results in
longer anisotropic period before the convergence to the final attractor.

\subsection{Strength of the stability of the isotropic magnetic inflation}
In the previous section, we saw that the typical convergence time for
$(\alpha , \lambda ) = (2,5)$ is of order $10$ e-foldings, except when
the curvature-driven single component magnetic inflation stretches the
anisotropic period. We argued that this order is determined by the magnitude
of eigenvalues that determine the instability of anisotropic inflations as 
well as the stability of isotropic solution. To test our hypothesis,
further numerical calculations have been carried out for six different
parameter sets: $(\alpha , \lambda ) = (2,5), (2,10),(2, 50),(3,10),(1,5)$,
 and $(1,3)$.
$(2,10)$ and $(2,50)$ have been chosen to demonstrate the dependence
on the slow-roll parameter ($\epsilon = 1/3, 1/13$ respectively). $(3,10)$
is expected to exhibit a similar behaviour as $(2,5)$ and $(1,5)$ is supposed
to provide a representative result for the situation where the scalar field
by itself is capable of accelerated expansion ($\alpha < \sqrt{2}$). 
The last set $(1,3)$ is a critical value at which $\Omega _B $ evaluated
at the isotropic magnetic inflation becomes zero and it merges with the
conventional power-law fixed point. When $\alpha < 1, \lambda <3$, the
usual cosmic-no-hair holds and the convergence should be very rapid. 
This merginal case is included here to give a flavor of that transition.
We have taken the initial conditions
\begin{equation}
\begin{split}
&\Sigma _+ = \Sigma _- = \sqrt{0.1} \ , \quad \mathcal{N}_1 = \mathcal{N}_2 = \mathcal{N}_3 = 0 \ , \quad \Gamma = 0.01 \\
&\mathcal{E}_1 = \mathcal{E}_2 = \mathcal{E}_3 = \sqrt{0.4} \ , \quad \varpi = \sqrt{1.2} \ , \quad \Omega _V = 0.2 \ , 
\end{split}
\end{equation}
and varied the magnetic components as
\begin{equation}
\frac{\mathcal{B}_2}{\mathcal{B}_1} , \frac{\mathcal{B}_3}{\mathcal{B}_1} \in [ 10^{-2} , 10^2 ] \ , \quad \Omega _B = 0.2 \ 
\end{equation}
 (the same prescription as the figure \ref{fig:magDirDep}). 
 \begin{figure}[htbp]
\begin{center}
\includegraphics[width = 0.9\linewidth]{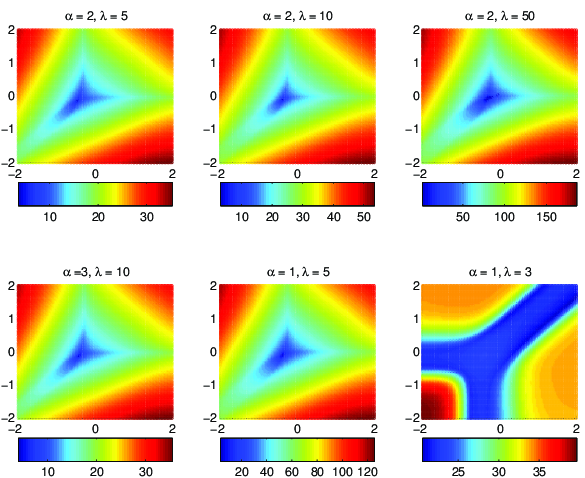}
\caption{Parameter dependence of the convergence time for Bianchi 
type I with $\Gamma = 0.01$ initially. The horizontal and vertical axes
represent ${\rm log} _{10} \mathcal{B}_2 / \mathcal{B}_1 $ and 
${\rm log}_{10} \mathcal{B}_3 / \mathcal{B}_1$ respectively. 
Note the different color codings for different sets of parameters. 
As expected, the convergence for $(1,3)$ is fast and insensitive
to the initial direction of magnetic field.}
\label{fig:compType1A}
\end{center}
\end{figure}
At a glance, the results
 (figure \ref{fig:compType1A}) look identical for all the parameter sets but $(\alpha , \lambda ) = (1,3)$ for which
 we should see a qualitatively different behaviour (the gauge field actually vanishes in the 
 final state). Even though the patterns are identical, however, the different color maps
 mean the time-scales of convergence vary significantly. 
 
According to the linear stability analysis in section \ref{sec:stability}, the convergence
time should be related to the values of $\Sigma _+$ for single- and double-component
magnetic inflations. 
\begin{figure}[htbp]
\begin{center}
\includegraphics[width=0.6\linewidth]{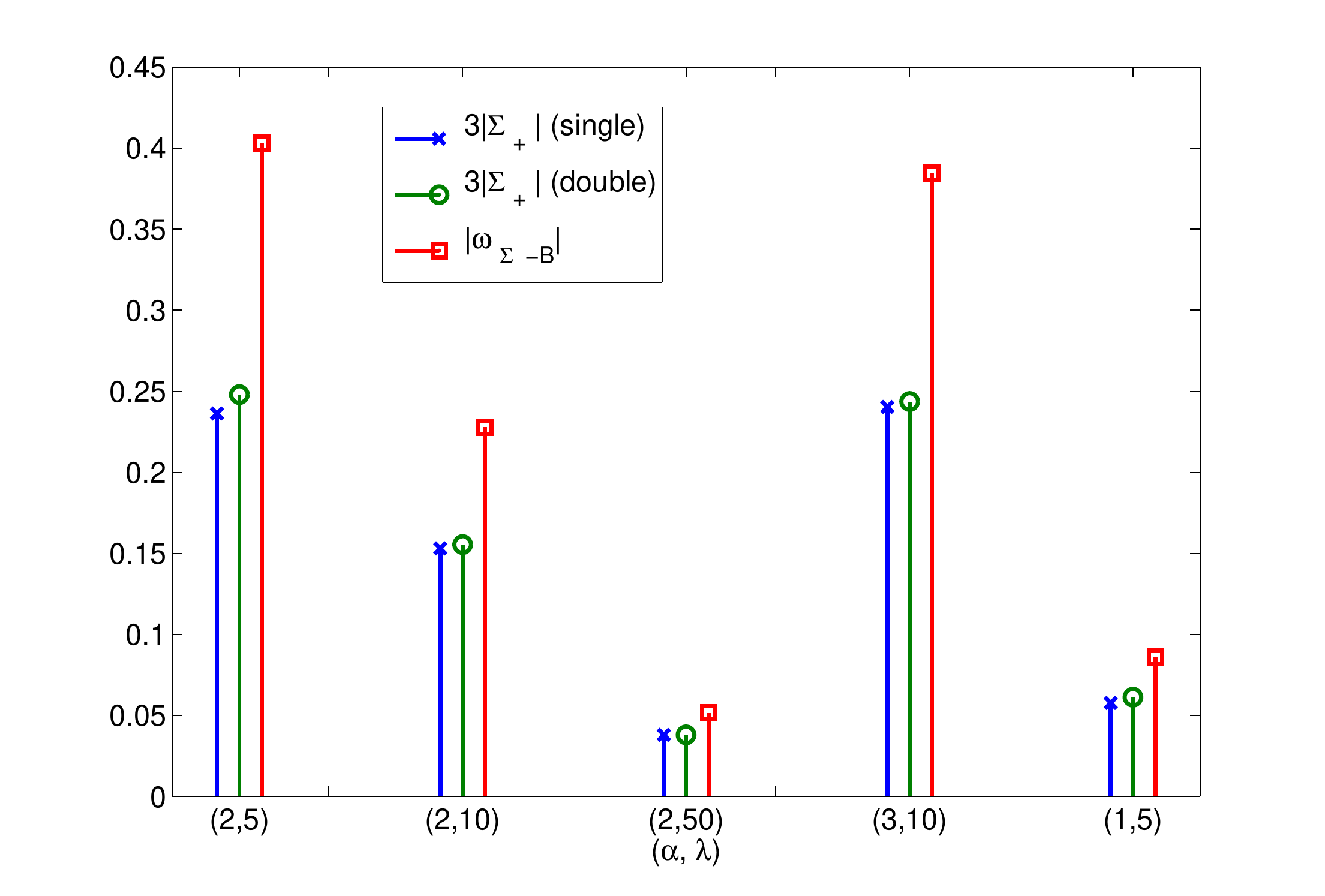}
\caption{Parameter dependence of the eigenvalues characterizing 
different magnetic inflations. The associated time-scales are determined
as the inverse of these values. They agree with the results in figure
\ref{fig:compType1A} well.}
\label{fig:paraTime}
\end{center}
\end{figure}
In figure \ref{fig:paraTime}, we plotted the characteristic eigenvalues for
single-component, double-component and isotropic magnetic inflations.
As already noted, the time-scales determined by these eigenvalues are 
all similar for each given parameter set. The values read off from the graph
explain the results of numerical calculations well. For instance, $(2,10)$
takes twice as much time as $(2,5)$, $(2,5)$ and $(3,10)$ should have
similar convergence time, and so on.

To see the relation between the convergence time and the slow-roll parameter $\epsilon $,
we note that for an acceptable inflationary model, 
 $\epsilon$ must be sufficiently small, i.e.,
\begin{equation}
\epsilon = q+1 = \frac{2\alpha }{\alpha + \lambda } \ll 1,
\end{equation}
which is roughy equivalent to $\alpha \ll \lambda $. Using this approximation, one
can see
\begin{equation}
\Sigma _+ \big| _{\rm single \mathchar`- comp} \sim 
\frac{\alpha \lambda -4}{3\lambda ^2 +8} \ , \quad \Sigma _+ 
\big| _{\rm double \mathchar`- comp} \sim 
\frac{\alpha \lambda -4}{3\lambda ^2 +2} \ .
\end{equation}
 Therefore, the characteristic time-scale should be at least
 \begin{equation}
 \tau _c \gtrsim \frac{3\lambda ^2 }{\alpha \lambda -4} > 
\frac{3\lambda }{\alpha } \sim 6\epsilon ^{-1} \ .
 \end{equation}
This implies that we should expect an anisotropic phase
of order $\epsilon ^{-1}$ at least. This will be an accurate estimate 
if $\alpha \lambda \gg 4$. 
On top of this generic lower bound, anisotropic phase should be further 
extended if 
$\alpha \lambda \sim 4$, which is exactly when $\Sigma _+$ become very small. 
 Given the observationally favored value $\epsilon \lesssim 10^{-2}$ 
\cite{PlanckCollaboration2013b}, it appears
 unlikely to see the convergence to the isotropic attractor before $100$ 
e-foldings.
 From another point of view, even when multiple gauge fields are present, the 
 period of anisotropic inflation may well be observable and tightly 
constrained. 
 
 \subsection{Effect on the oscillatory phase}
\begin{figure}[htbp]
\begin{center}
\includegraphics[width = 0.9\linewidth]{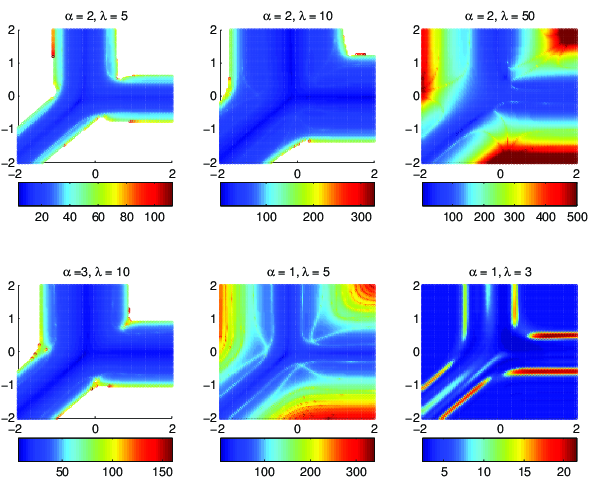}
\caption{Parameter dependence of non-Abelian ($\Gamma =1$) Bianchi I. The
initial conditions are the same as in figure \ref{fig:compType1A} 
except for $\Gamma $.
The axes are again directions of magnetic components.}
\label{fig:compType1N}
\end{center}
\end{figure}

 Figure \ref{fig:compType1N} shows the convergence time in non-Abelian
Bianchi I ($\Gamma =1$ initially) for the different parameter sets. 
The absence of the oscillatory phase for $\alpha =1$ is noted. 
It is expected since the exponential growth in equations 
(\ref{eq:oscillationG}) and (\ref{eq:oscillationEv}) happens only
 if $q$ is positive while it should be negative for $\alpha < \sqrt{2}$. 
This leads to  an interesting result that if the scalar field is capable 
of accelerated expansion  by itself, there will be no oscillation after 
introducing gauge fields. The area of 
 oscillation seems to be squeezed for greater $\lambda $ even for $\alpha =2$
 and almost disappears when $\lambda = 50$. Although the reason is not clear,
 we suspect that it is due to the back reaction of the gauge field onto the 
scalar field through equation (\ref{scalar1}) where its effect is amplified 
by the factor of $\lambda $. 
 
% A question arises: why does the oscillation still take place for $\alpha =3$ 
%for  which the power-law scalar solution does not exist ($\Omega _V$ would be
% negative for such a fixed point)?   

We also find the oscillation for $\alpha=3$. 
Since the isotropic fixed point (\ref{eq:fix1}) is physically admissible only for 
$\alpha \leq \sqrt{6}$, this oscillating solution does not correspond to 
the fixed point. In fact, the asymptotic state of the universe here is anisotropic.
 The nature of oscillation is rather akin to that of Bianchi type II. 
The energy density of shear as well as  gauge field cannot be 
ignored and settles down to a constant final value.

 \subsection{Suppression of the anomalously long anisotropic phase}
 
 In the discussion of the prolonged anisotropic phases encountered in the
 presence of spatial curvature, we mentioned that this would be a peculiar
 feature for the specific parameter values $(\alpha , \lambda ) = (2,5)$. 
\begin{figure}[htbp]
\begin{center}
\includegraphics[width = 0.9\linewidth]{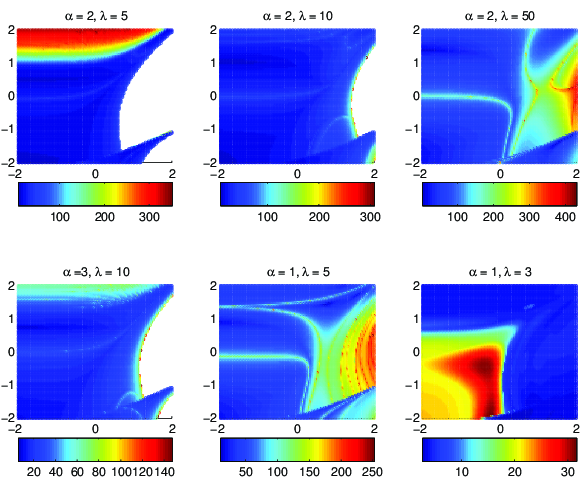}
\caption{Convergence time in type II ($\mathcal{N}_3 \neq 0$) for the 
different parameter sets under varying density parameters. The initial
conditions are $\Omega _K = \Omega _V = \Omega _E = \Omega _B , 
\Sigma _+ = \Sigma _- , \mathcal{E}_1 =\mathcal{E}_2 = \mathcal{E}_3,
\mathcal{B}_1 = \mathcal{B}_2 = \mathcal{B}_3$ and $\Gamma =1$. 
The x-axes; $\Sigma ^2 / \left( 1 - \Sigma ^2 - \Omega _N \right) $. The 
y-axes; $\Omega _N / \left( 1- \Sigma ^2 - \Omega _N \right) $.}
\label{fig:compType2N}
\end{center}
\end{figure}
Figure \ref{fig:compType2N} duly confirms our statement. Aside from the
top-left panel, the typical convergence time agrees with the one estimated
from the eigenvalues, and with the figure \ref{fig:compType1A}. In all the
cases from figure \ref{fig:compType1A} to \ref{fig:compType2N}, the contour 
patterns are very similar for different parameters. This implies that the
qualitative behaviours for $(\alpha , \lambda ) =(2,5)$ studied in detail
in the present paper are generic.

%%%%%%%%%%%%%%%%%%%%%%%%%%%%%%%%%%%%%%%%%%%%%%%%%%%%%%%%%%%
%%%%%%%%%%%%%%%%%%%%%%%%%%%%%%%%%%%%%%%%%%%%%%%%%%%%%%%%%%%
%%%%%%%%%%%%%%%%%%%%%%%%%%%%%%%%%%%%%%%%%%%%%%%%%%%%%%%%%%%
\section{Concluding remarks}
%%%%%%%%%%%%%%%%%%%%%%%%%%%%%%%%%%%%%%%%%%%%%%%%%%%%%%%%%%%
%%%%%%%%%%%%%%%%%%%%%%%%%%%%%%%%%%%%%%%%%%%%%%%%%%%%%%%%%%%
%%%%%%%%%%%%%%%%%%%%%%%%%%%%%%%%%%%%%%%%%%%%%%%%%%%%%%%%%%%

In this article, we studied a general class of anisotropic cosmological models
that contain a scalar field with exponential potential and an SU(2) gauge field
coupled to the scalar through exponential gauge-kinetic function. The governing
equations have been properly normalized so that all the important inflationary
solutions appear as fixed points of the dynamical system. We carried out the
detailed stability analysis for them and have explicitly confirmed that the 
only stable
attractor solution is the isotropic magnetic inflation. The peculiarity 
arising from
the Yang-Mills interaction and spatial curvature has also been pointed out, 
which
differentiates this model from the more regular U(1) triplet system. Then, 
extensive
numerical investigations have been made to survey a variety of initial 
conditions
and to see how convergence to the stable isotropic solution with magnetic 
field 
is achieved. We found several different types of oscillatory attractors that 
prevent
the universe from settling down to the isotropic inflation. The inclusion of 
fully generic
spatial curvature (type VIII and type IX) has been shown to restore the 
stability
of inflation globally. Nevertheless, the time it takes for the anisotropy 
to disappear
is rather significant, at the very least $10$ e-foldings. This estimate agrees
 with the 
characteristic eigenvalues of the isotropic as well as anisotropic inflationary
solutions and we have obtained the general lower bound for convergence 
time in terms of the slow-roll parameter, namely $\tau _c \gtrsim 
\epsilon ^{-1}$.
In summary, it is reasonable to conclude that the isotropic magnetic inflation
is stable for a large class of initial conditions in the general homogeneous 
cosmologies. From a physical point of view, however, it is very likely that 
we should see the signature of transient anisotropy, which may well last 
until the end of inflation. 

Although we have revealed new interesting 
properties of anisotropic inflation 
in the theory with gauge-kinetic coupling, this model may not be 
realistic due to the idealized and simplified choice of the action. 
The resent observational data rejects a power-law inflation model
\cite{PlanckCollaboration2013b}. It should also be pointed out that
the inflation is inherently eternal in the present model because the exponential
potential and gauge-kinetic coupling are scale-free so that there is
the problem of graceful exit.
An obvious next step will be to study the effects of the gauge-kinetic term
on more realistic inflationary models based on 
fundamental unified theories such as superstring theory, for example, 
brane inflation\cite{Dvali1999,Giddings2002,Kachru2003,Kachru2003a} 
or higher-order curvature inflation
\cite{Ishihara1986,Maeda1986,Ellis1999,Maeda2004,Akune2006,Bamba2007,Maeda2012}.
We should also find a way out of the accelerated expansion in those
models and provide a concrete example of successful reheating after
inflation.

From a phenomenological point of view, the relevance of inflaton-gauge
interaction in the context of statistical anisotropy in CMBR is made more
interesting by our finding that anisotropic
inflation is quite generic as far as our observational window of around 60
e-foldings is concerned even though the mathematically defined attractor
is isotropic. Quantitative estimates of CMBR anisotropy generated by the
transient anisotropic phase and its scale-dependence caused by the
transition from anisotropic phase to isotropic one may deserve more
serious attention. These studies are left for future projects.

%\begin{table}[htb]
%\begin{center}
%\small{
%\caption{}
%\vspace{2mm}\label{delta_table1}
%\begin{tabular}{l|c|c|c|c|c|c}
%\noalign{\global\arrayrulewidth1pt}
%\hline
%\noalign{\global\arrayrulewidth.4pt}
%Bianchi Type&I&II&VI$_0$&VII$_0$&VIII&IX\\
%\hline
%\hline
%anisotropic period
%&5-50&20-50&20-50&20-50&20-40&5-10\\
%\hline
%non-inflationary oscillation
%&{\small $\bigcirc$}&$\bigtriangleup$&$\bigtriangleup$&$\bigtriangleup$&
%$\times$&$\times$\\
%\noalign{\global\arrayrulewidth1pt}
%\hline
%\noalign{\global\arrayrulewidth.4pt}
%\end{tabular}}
%\end{center}
%\end{table}

%%%%%%%%%%%%%%%%%%%%%%%%%%%%%%%%%%%%%%%%%%%%%%%%%%%%%%%%%%%%%%%%%%%%%%
%%%%%%%%%%%%%%%%%%%%%%%%%%%%%%%%%%%%%%%%%%%%%%%%%%%%%%%%%%%%%%%%%%%%%%
\acknowledgments
%%%%%%%%%%%%%%%%%%%%%%%%%%%%%%%%%%%%%%%%%%%%%%%%%%%%%%%%%%%%%%%%%%%%%%
%%%%%%%%%%%%%%%%%%%%%%%%%%%%%%%%%%%%%%%%%%%%%%%%%%%%%%%%%%%%%%%%%%%%%%
We would like to thank John Barrow and Keiju Murata
 for valuable comments.
KY is also thankful to Hiroyuki Funakoshi and Shi Chun Su for
their very helpful advice on numerical calculations.
This work was partially supported by the Grant-in-Aid for 
Scientific Research
Fund of the JSPS (C)  (No.25400276).
KY would like to thank the Institute of Theoretical Astrophysics
in the University of Oslo, where a part of this work was conducted, 
for the support and hospitality.

\bibliographystyle{JHEP}
\bibliography{ref_YManiso,planck}

\providecommand{\href}[2]{#2}\begingroup\raggedright\begin{thebibliography}{10}

\bibitem{Starobinsky1980}
A.~Starobinsky, {\it {A new type of isotropic cosmological models without
  singularity}},  {\em Physics Letters B} {\bf 91} (Mar., 1980) 99--102.

\bibitem{Sato1981}
K.~Sato, {\it {First-order phase transition of a vacuum and the expansion of
  the Universe}},  {\em Monthly Notices of the Royal Astronomical Society} {\bf
  195} (May, 1981) 467--479.

\bibitem{Guth1981}
A.~H. Guth, {\it {Inflationary universe: A possible solution to the horizon and
  flatness problems}},  {\em Physical Review D} {\bf 23} (Jan., 1981) 347--356.

\bibitem{Albrecht1982}
A.~Albrecht and P.~Steinhardt, {\it {Cosmology for Grand Unified Theories with
  Radiatively Induced Symmetry Breaking}},  {\em Physical Review Letters} {\bf
  48} (Apr., 1982) 1220--1223.

\bibitem{Linde1982}
A.~Linde, {\it {A new inflationary universe scenario: A possible solution of
  the horizon, flatness, homogeneity, isotropy and primordial monopole
  problems}},  {\em Physics Letters B} {\bf 108} (Feb., 1982) 389--393.

\bibitem{Linde1983}
A.~Linde, {\it {Chaotic inflation}},  {\em Physics Letters B} {\bf 129} (Sept.,
  1983) 177--181.

\bibitem{Linde2005}
A.~Linde, {\it {Particle Physics and Inflationary Cosmology}},  {\em
  Contemporary Concepts in Physics} {\bf 5} (Mar., 2005) 270,
  [\href{http://xxx.lanl.gov/abs/0503203}{{\tt 0503203}}].

\bibitem{Linde2006}
A.~Linde, {\it {Inflation and String Cosmology}},  {\em Progress of Theoretical
  Physics Supplement} {\bf 163} (Mar., 2006) 295--322,
  [\href{http://xxx.lanl.gov/abs/0503195}{{\tt 0503195}}].

\bibitem{Linde2008}
A.~Linde, {\em {Inflationary Cosmology}}, vol.~738 of {\em Lecture Notes in
  Physics}.
\newblock Springer Berlin Heidelberg, Berlin, Heidelberg, May, 2008.

\bibitem{McAllister2008}
L.~McAllister and E.~Silverstein, {\it {String cosmology: a review}},  {\em
  General Relativity and Gravitation} {\bf 40} (Jan., 2008) 565--605,
  [\href{http://xxx.lanl.gov/abs/0710.2951}{{\tt arXiv:0710.2951}}].

\bibitem{Lyth2008}
D.~H. Lyth, {\em {Inflationary Cosmology}}, vol.~738 of {\em Lecture Notes in
  Physics}.
\newblock Springer Berlin Heidelberg, Berlin, Heidelberg, Feb., 2008.

\bibitem{Townsend2003}
P.~K. Townsend, {\it {Cosmic Acceleration and M-Theory}},  in {\em ICMP2003},
  (Lisbon), Aug., 2003.
\newblock \href{http://xxx.lanl.gov/abs/0308149}{{\tt 0308149}}.

\bibitem{Kachru2003}
S.~Kachru, R.~Kallosh, A.~Linde, and S.~Trivedi, {\it {de Sitter vacua in
  string theory}},  {\em Physical Review D} {\bf 68} (Aug., 2003) 046005,
  [\href{http://xxx.lanl.gov/abs/0301240}{{\tt 0301240}}].

\bibitem{Kachru2003a}
S.~Kachru, R.~Kallosh, A.~Linde, J.~Maldacena, L.~McAllister, and S.~P.
  Trivedi, {\it {Towards inflation in string theory}},  {\em Journal of
  Cosmology and Astroparticle Physics} {\bf 2003} (Oct., 2003) 013--013,
  [\href{http://xxx.lanl.gov/abs/0308055}{{\tt 0308055}}].

\bibitem{Maleknejad2013}
A.~Maleknejad, M.~M. Sheikh-Jabbari, and J.~Soda, {\it {Gauge Fields and
  Inflation}},  {\em Physics Reports} {\bf 528} (Dec., 2013) 161--261,
  [\href{http://xxx.lanl.gov/abs/1212.2921}{{\tt arXiv:1212.2921}}].

\bibitem{Yokoyama2008}
S.~Yokoyama and J.~Soda, {\it {Primordial statistical anisotropy generated at
  the end of inflation}},  {\em Journal of Cosmology and Astroparticle Physics}
  {\bf 2008} (Aug., 2008) 005, [\href{http://xxx.lanl.gov/abs/0805.4265}{{\tt
  arXiv:0805.4265}}].

\bibitem{Watanabe2009}
M.-a. Watanabe, S.~Kanno, and J.~Soda, {\it {Inflationary Universe with
  Anisotropic Hair}},  {\em Physical Review Letters} {\bf 102} (May, 2009)
  191302, [\href{http://xxx.lanl.gov/abs/0902.2833}{{\tt arXiv:0902.2833}}].

\bibitem{Bartolo2009a}
N.~Bartolo, E.~Dimastrogiovanni, S.~Matarrese, and A.~Riotto, {\it {Anisotropic
  trispectrum of curvature perturbations induced by primordial non-Abelian
  vector fields}},  {\em Journal of Cosmology and Astroparticle Physics} {\bf
  2009} (Nov., 2009) 028--028, [\href{http://xxx.lanl.gov/abs/0909.5621}{{\tt
  arXiv:0909.5621}}].

\bibitem{Bartolo2009}
N.~Bartolo, E.~Dimastrogiovanni, S.~Matarrese, and A.~Riotto, {\it {Anisotropic
  Bispectrum of Curvature Perturbations from Primordial Non-Abelian Vector
  Fields}},  {\em Journal of Cosmology and Astroparticle Physics} {\bf 2009}
  (Oct., 2009) 015--015, [\href{http://xxx.lanl.gov/abs/0906.4944}{{\tt
  arXiv:0906.4944}}].

\bibitem{Dimopoulos2009}
K.~Dimopoulos, M.~Karciauskas, D.~H. Lyth, and Y.~Rodr\'{\i}guez, {\it
  {Statistical anisotropy of the curvature perturbation from vector field
  perturbations}},  {\em Journal of Cosmology and Astroparticle Physics} {\bf
  2009} (May, 2009) 013--013, [\href{http://xxx.lanl.gov/abs/0809.1055}{{\tt
  arXiv:0809.1055}}].

\bibitem{Moniz2010}
P.~V. Moniz and J.~Ward, {\it {Gauge field back-reaction in Born–Infeld
  cosmologies}},  {\em Classical and Quantum Gravity} {\bf 27} (Dec., 2010)
  235009, [\href{http://xxx.lanl.gov/abs/1007.3299}{{\tt arXiv:1007.3299}}].

\bibitem{Dulaney2010}
T.~R. Dulaney and M.~I. Gresham, {\it {Primordial power spectra from
  anisotropic inflation}},  {\em Physical Review D} {\bf 81} (May, 2010)
  103532, [\href{http://xxx.lanl.gov/abs/1001.2301}{{\tt arXiv:1001.2301}}].

\bibitem{Gumrukcuoglu2010}
A.~E. G\"{u}mr\"{u}k\c{c}\"{u}oğlu, B.~Himmetoglu, and M.~Peloso, {\it
  {Scalar-scalar, scalar-tensor, and tensor-tensor correlators from anisotropic
  inflation}},  {\em Physical Review D} {\bf 81} (Mar., 2010) 063528,
  [\href{http://xxx.lanl.gov/abs/1001.4088}{{\tt arXiv:1001.4088}}].

\bibitem{Watanabe2010}
M.-a. Watanabe, S.~Kanno, and J.~Soda, {\it {The Nature of Primordial
  Fluctuations from Anisotropic Inflation}},  {\em Progress of Theoretical
  Physics} {\bf 123} (June, 2010) 1041--1068,
  [\href{http://xxx.lanl.gov/abs/1003.0056}{{\tt arXiv:1003.0056}}].

\bibitem{Emami2010}
R.~Emami, H.~Firouzjahi, S.~M.~S. Movahed, and M.~Zarei, {\it {Anisotropic
  Inflation from Charged Scalar Fields}},  {\em Journal of Cosmology and
  Astroparticle Physics} {\bf 2011} (Oct., 2010) 005--005,
  [\href{http://xxx.lanl.gov/abs/1010.5495}{{\tt arXiv:1010.5495}}].

\bibitem{Watanabe2011}
M.-a. Watanabe, S.~Kanno, and J.~Soda, {\it {Imprints of the anisotropic
  inflation on the cosmic microwave background}},  {\em Monthly Notices of the
  Royal Astronomical Society: Letters} {\bf 412} (Mar., 2011) L83--L87,
  [\href{http://xxx.lanl.gov/abs/1011.3604}{{\tt arXiv:1011.3604}}].

\bibitem{Murata2011}
K.~Murata and J.~Soda, {\it {Anisotropic inflation with non-abelian gauge
  kinetic function}},  {\em Journal of Cosmology and Astroparticle Physics}
  {\bf 2011} (June, 2011) 037--037,
  [\href{http://xxx.lanl.gov/abs/1103.6164}{{\tt arXiv:1103.6164}}].

\bibitem{Shiraishi2011}
M.~Shiraishi and S.~Yokoyama, {\it {Violation of the Rotational Invariance in
  the CMB Bispectrum}},  {\em Progress of Theoretical Physics} {\bf 126} (Nov.,
  2011) 923--935, [\href{http://xxx.lanl.gov/abs/1107.0682}{{\tt
  arXiv:1107.0682}}].

\bibitem{Namba2012}
R.~Namba, {\it {Curvature Perturbations from a Massive Vector Curvaton}},
  \href{http://xxx.lanl.gov/abs/1207.5547}{{\tt arXiv:1207.5547}}.

\bibitem{Bartolo2012}
N.~Bartolo, S.~Matarrese, M.~Peloso, and A.~Ricciardone, {\it {The anisotropic
  power spectrum and bispectrum in the f(phi) F\^{}2 mechanism}},
  \href{http://xxx.lanl.gov/abs/1210.3257}{{\tt arXiv:1210.3257}}.

\bibitem{Barnaby2011}
N.~Barnaby, R.~Namba, and M.~Peloso, {\it {Phenomenology of a pseudo-scalar
  inflaton: naturally large nongaussianity}},  {\em Journal of Cosmology and
  Astroparticle Physics} {\bf 2011} (Apr., 2011) 009--009,
  [\href{http://xxx.lanl.gov/abs/1102.4333}{{\tt arXiv:1102.4333}}].

\bibitem{Anber2012}
M.~M. Anber and L.~Sorbo, {\it {Non-Gaussianities and chiral gravitational
  waves in natural steep inflation}},  {\em Physical Review D} {\bf 85} (June,
  2012) 123537, [\href{http://xxx.lanl.gov/abs/1203.5849}{{\tt
  arXiv:1203.5849}}].

\bibitem{Kanno2009}
S.~Kanno, J.~Soda, and M.-a. Watanabe, {\it {Cosmological magnetic fields from
  inflation and backreaction}},  {\em Journal of Cosmology and Astroparticle
  Physics} {\bf 2009} (Dec., 2009) 009--009,
  [\href{http://xxx.lanl.gov/abs/0908.3509}{{\tt arXiv:0908.3509}}].

\bibitem{Barnaby2012}
N.~Barnaby, R.~Namba, and M.~Peloso, {\it {Observable non-Gaussianity from
  gauge field production in slow roll inflation, and a challenging connection
  with magnetogenesis}},  {\em Physical Review D} {\bf 85} (June, 2012) 123523,
  [\href{http://xxx.lanl.gov/abs/1202.1469}{{\tt arXiv:1202.1469}}].

\bibitem{Ferreira2013}
R.~J. Ferreira, R.~K. Jain, and M.~S. Sloth, {\it {Inflationary magnetogenesis
  without the strong coupling problem}},  {\em Journal of Cosmology and
  Astroparticle Physics} {\bf 2013} (Oct., 2013) 004--004.

\bibitem{Valenzuela-Toledo2009}
C.~A. Valenzuela-Toledo, Y.~Rodr\'{\i}guez, and D.~H. Lyth, {\it
  {Non-Gaussianity at tree and one-loop levels from vector field
  perturbations}},  {\em Physical Review D} {\bf 80} (Nov., 2009) 103519,
  [\href{http://xxx.lanl.gov/abs/0909.4064}{{\tt arXiv:0909.4064}}].

\bibitem{Valenzuela-Toledo2010}
C.~A. Valenzuela-Toledo and Y.~Rodr\'{\i}guez, {\it {Non-gaussianity from the
  trispectrum and vector field perturbations}},  {\em Physics Letters B} {\bf
  685} (Mar., 2010) 120--127, [\href{http://xxx.lanl.gov/abs/0910.4208}{{\tt
  arXiv:0910.4208}}].

\bibitem{Dimastrogiovanni2010}
E.~Dimastrogiovanni, N.~Bartolo, S.~Matarrese, and A.~Riotto, {\it
  {Non-Gaussianity and Statistical Anisotropy from Vector Field Populated
  Inflationary Models}},  {\em Advances in Astronomy} {\bf 2010} (Jan., 2010)
  1--21, [\href{http://xxx.lanl.gov/abs/1001.4049}{{\tt arXiv:1001.4049}}].

\bibitem{Karciauskas2011}
M.~Karciauskas, {\it {The Primordial Curvature Perturbation from Vector Fields
  of General non-Abelian Groups}},  {\em Journal of Cosmology and Astroparticle
  Physics} {\bf 2012} (Apr., 2011) 014--014,
  [\href{http://xxx.lanl.gov/abs/1104.3629}{{\tt arXiv:1104.3629}}].

\bibitem{Valenzuela-Toledo2011}
C.~A. Valenzuela-Toledo, Y.~Rodr\'{\i}guez, and J.~P.~B. Almeida, {\it
  {Feynman-like rules for calculating n -point correlators of the primordial
  curvature perturbation}},  {\em Journal of Cosmology and Astroparticle
  Physics} {\bf 2011} (Oct., 2011) 020--020,
  [\href{http://xxx.lanl.gov/abs/1107.3186}{{\tt arXiv:1107.3186}}].

\bibitem{Jain2012}
R.~K. Jain and M.~S. Sloth, {\it {On the non-Gaussian correlation of the
  primordial curvature perturbation with vector fields}},  {\em Journal of
  Cosmology and Astroparticle Physics} {\bf 2013} (Feb., 2013) 003--003,
  [\href{http://xxx.lanl.gov/abs/1210.3461}{{\tt arXiv:1210.3461}}].

\bibitem{Rodriguez2013}
Y.~Rodr\'{\i}guez, J.~P.~B. Almeida, and C.~A. Valenzuela-Toledo, {\it {The
  different varieties of the Suyama-Yamaguchi consistency relation and its
  violation as a signal of statistical inhomogeneity}},  {\em Journal of
  Cosmology and Astroparticle Physics} {\bf 2013} (Apr., 2013) 039--039,
  [\href{http://xxx.lanl.gov/abs/1301.5843}{{\tt arXiv:1301.5843}}].

\bibitem{Almeida2013}
J.~P.~B. Almeida, Y.~Rodr\'{\i}guez, and C.~A. Valenzuela-Toledo, {\it {The
  Suyama-Yamaguchi consistency relation in the presence of vector fields}},
  {\em Modern Physics Letters A} {\bf 28} (Feb., 2013) 1350012,
  [\href{http://xxx.lanl.gov/abs/1112.6149}{{\tt arXiv:1112.6149}}].

\bibitem{Anber2010}
M.~M. Anber and L.~Sorbo, {\it {Naturally inflating on steep potentials through
  electromagnetic dissipation}},  {\em Physical Review D} {\bf 81} (Feb., 2010)
  043534, [\href{http://xxx.lanl.gov/abs/0908.4089}{{\tt arXiv:0908.4089}}].

\bibitem{Dimopoulos2010a}
J.~M. Wagstaff and K.~Dimopoulos, {\it {Particle production of vector fields:
  Scale invariance is attractive}},  {\em Physical Review D} {\bf 83} (Jan.,
  2011) 023523, [\href{http://xxx.lanl.gov/abs/1011.2517}{{\tt
  arXiv:1011.2517}}].

\bibitem{Dimopoulos2011}
K.~Dimopoulos, G.~Lazarides, and J.~M. Wagstaff, {\it {Eliminating the
  $\eta$-problem in SUGRA hybrid inflation with vector backreaction}},  {\em
  Journal of Cosmology and Astroparticle Physics} {\bf 2012} (Feb., 2012)
  018--018, [\href{http://xxx.lanl.gov/abs/1111.1929}{{\tt arXiv:1111.1929}}].

\bibitem{Kanno2010}
S.~Kanno, J.~Soda, and M.-a. Watanabe, {\it {Anisotropic power-law inflation}},
   {\em Journal of Cosmology and Astroparticle Physics} {\bf 2010} (Dec., 2010)
  024--024, [\href{http://xxx.lanl.gov/abs/1010.5307}{{\tt arXiv:1010.5307}}].

\bibitem{Hervik2011}
S.~r. Hervik, D.~F. Mota, and M.~Thorsrud, {\it {Inflation with stable
  anisotropic hair: is it cosmologically viable?}},  {\em Journal of High
  Energy Physics} {\bf 2011} (Nov., 2011) 146,
  [\href{http://xxx.lanl.gov/abs/1109.3456}{{\tt arXiv:1109.3456}}].

\bibitem{Do2011}
T.~Q. Do and W.~F. Kao, {\it {Anisotropic power-law inflation for the
  Dirac-Born-Infeld theory}},  {\em Physical Review D} {\bf 84} (Dec., 2011)
  123009.

\bibitem{Do2011a}
T.~Q. Do, W.~F. Kao, and I.-C. Lin, {\it {Anisotropic power-law inflation for a
  two scalar fields model}},  {\em Physical Review D} {\bf 83} (June, 2011)
  123002.

\bibitem{Ohashi2013}
J.~Ohashi, J.~Soda, and S.~Tsujikawa, {\it {Anisotropic power-law
  k-inflation}},  \href{http://xxx.lanl.gov/abs/1310.3053}{{\tt
  arXiv:1310.3053}}.

\bibitem{Yamamoto2012}
K.~Yamamoto, {\it {Primordial fluctuations from inflation with a triad of
  background gauge fields}},  {\em Physical Review D} {\bf 85} (June, 2012)
  123504, [\href{http://xxx.lanl.gov/abs/1203.1071}{{\tt arXiv:1203.1071}}].

\bibitem{Maeda2012}
K.-i. Maeda and K.~Yamamoto, {\it {Inflationary dynamics with a non-Abelian
  gauge field}},  {\em Physical Review D} {\bf 87} (Jan., 2013) 023528,
  [\href{http://xxx.lanl.gov/abs/1210.4054}{{\tt arXiv:1210.4054}}].

\bibitem{Maleknejad2011}
A.~Maleknejad and M.~M. Sheikh-Jabbari, {\it {Non-Abelian gauge field
  inflation}},  {\em Physical Review D} {\bf 84} (Aug., 2011) 043515,
  [\href{http://xxx.lanl.gov/abs/1102.1932}{{\tt arXiv:1102.1932}}].

\bibitem{Maleknejad2012}
A.~Maleknejad, M.~Sheikh-Jabbari, and J.~Soda, {\it {Gauge-flation and cosmic
  no-hair conjecture}},  {\em Journal of Cosmology and Astroparticle Physics}
  {\bf 2012} (Jan., 2012) 016--016,
  [\href{http://xxx.lanl.gov/abs/1109.5573}{{\tt arXiv:1109.5573}}].

\bibitem{Adshead2012}
P.~Adshead and M.~Wyman, {\it {Natural Inflation on a Steep Potential with
  Classical Non-Abelian Gauge Fields}},  {\em Physical Review Letters} {\bf
  108} (June, 2012) 261302, [\href{http://xxx.lanl.gov/abs/1202.2366}{{\tt
  1202.2366}}].

\bibitem{Adshead2012a}
P.~Adshead and M.~Wyman, {\it {Gauge-flation trajectories in chromo-natural
  inflation}},  {\em Physical Review D} {\bf 86} (Aug., 2012) 043530,
  [\href{http://xxx.lanl.gov/abs/1203.2264}{{\tt arXiv:1203.2264}}].

\bibitem{Sheikh-Jabbari2012}
M.~Sheikh-Jabbari, {\it {Gauge-flation vs chromo-natural inflation}},  {\em
  Physics Letters B} {\bf 717} (Oct., 2012) 6--9,
  [\href{http://xxx.lanl.gov/abs/1203.2265}{{\tt arXiv:1203.2265}}].

\bibitem{Martinec2012}
E.~Martinec, P.~Adshead, and M.~Wyman, {\it {Chern-Simons EM-flation}},  {\em
  Journal of High Energy Physics} {\bf 2013} (Feb., 2013) 27,
  [\href{http://xxx.lanl.gov/abs/1206.2889}{{\tt arXiv:1206.2889}}].

\bibitem{Dimastrogiovanni2012}
E.~Dimastrogiovanni, M.~Fasiello, and A.~J. Tolley, {\it {Low-Energy Effective
  Field Theory for Chromo-Natural Inflation}},
  \href{http://xxx.lanl.gov/abs/1211.1396}{{\tt arXiv:1211.1396}}.

\bibitem{Adshead2013}
P.~Adshead, E.~Martinec, and M.~Wyman, {\it {Perturbations in Chromo-Natural
  Inflation}},  \href{http://xxx.lanl.gov/abs/1305.2930}{{\tt
  arXiv:1305.2930}}.

\bibitem{Adshead2013a}
P.~Adshead, E.~Martinec, and M.~Wyman, {\it {Gauge Fields and Inflation: Chiral
  Gravitational Waves, Fluctuations and the Lyth Bound}},
  \href{http://xxx.lanl.gov/abs/1301.2598}{{\tt arXiv:1301.2598}}.

\bibitem{Funakoshi2012}
H.~Funakoshi and S.~Renaux-Petel, {\it {A modal approach to the numerical
  calculation of primordial non-Gaussianities}},  {\em Journal of Cosmology and
  Astroparticle Physics} {\bf 2013} (Feb., 2013) 002--002,
  [\href{http://xxx.lanl.gov/abs/1211.3086}{{\tt arXiv:1211.3086}}].

\bibitem{PlanckCollaboration2013a}
{\bf Planck} Collaboration, P.~A.~R. Ade et~al., {\it {Planck 2013 Results.
  XXIV. Constraints on primordial non-Gaussianity}},
  \href{http://xxx.lanl.gov/abs/1303.5084}{{\tt arXiv:1303.5084}}.

\bibitem{Wainwright1997}
J.~Wainwright and G.~F.~R. Ellis, {\em {Dynamical Systems in Cosmology}},
  vol.~-1.
\newblock Cambridge University Press, 1997.

\bibitem{Wald1983}
R.~Wald, {\it {Asymptotic behavior of homogeneous cosmological models in the
  presence of a positive cosmological constant}},  {\em Physical Review D} {\bf
  28} (Oct., 1983) 2118--2120.

\bibitem{Kitada1992}
Y.~Kitada and K.-i. Maeda, {\it {Cosmic no-hair theorem in power-law
  inflation}},  {\em Physical Review D} {\bf 45} (Feb., 1992) 1416--1419.

\bibitem{Kitada1993}
Y.~Kitada and K.-i. Maeda, {\it {Cosmic no-hair theorem in homogeneous
  spacetimes}},  {\em Classical and Quantum Gravity} {\bf 10} (Jan., 1993)
  703--734.

\bibitem{Maleknejad2012a}
A.~Maleknejad and M.~M. Sheikh-Jabbari, {\it {Revisiting cosmic no-hair theorem
  for inflationary settings}},  {\em Physical Review D} {\bf 85} (June, 2012)
  123508, [\href{http://xxx.lanl.gov/abs/1203.0219}{{\tt arXiv:1203.0219}}].

\bibitem{PlanckCollaboration2013b}
{\bf Planck} Collaboration, P.~A.~R. Ade et~al., {\it {Planck 2013 results.
  XXII. Constraints on inflation}},
  \href{http://xxx.lanl.gov/abs/1303.5082}{{\tt arXiv:1303.5082}}.

\bibitem{Dvali1999}
G.~Dvali and S.-H. Tye, {\it {Brane inflation}},  {\em Physics Letters B} {\bf
  450} (1999), no.~1 72--82.

\bibitem{Giddings2002}
S.~B. Giddings, S.~Kachru, and J.~Polchinski, {\it {Hierarchies from fluxes in
  string compactifications}},  {\em Physical Review D} {\bf 66} (Nov., 2002)
  106006.

\bibitem{Ishihara1986}
H.~Ishihara, {\it {Cosmological solutions of the extended Einstein gravity with
  the Gauss-Bonnet term}},  {\em Physics Letters B} {\bf 179} (1986), no.~3
  217--222.

\bibitem{Maeda1986}
K.~Maeda, {\it {Cosmological solutions with Calabi-Yau compactification}},
  {\em Physics Letters B} {\bf 166} (1986), no.~1 59--64.

\bibitem{Ellis1999}
J.~Ellis, N.~Kaloper, K.~Olive, and J.~Yokoyama, {\it {Topological R4
  inflation}},  {\em Physical Review D} {\bf 59} (Apr., 1999) 103503.

\bibitem{Maeda2004}
K.-i. Maeda and N.~Ohta, {\it {Inflation from M-theory with fourth-order
  corrections and large extra dimensions}},  {\em Physics Letters B} {\bf 597}
  (2004), no.~3 400--407.

\bibitem{Akune2006}
K.~Akune, K.-i. Maeda, and N.~Ohta, {\it {Inflation from superstring and
  M-theory compactification with higher order corrections. - II. - Case of
  quartic Weyl terms}},  {\em Physical Review D} {\bf 73} (May, 2006) 103506.

\bibitem{Bamba2007}
K.~Bamba, Z.-K. Guo, and N.~Ohta, {\it {Accelerating Cosmologies in the
  Einstein-Gauss-Bonnet Theory with a Dilaton}},  {\em Progress of Theoretical
  Physics} {\bf 118} (Nov., 2007) 879--892.

\end{thebibliography}\endgroup
%%%%%%%%%%%%%%%%%%%%%%%%%%%%%%%%%%%%%%%%%%%%%%%%%%%%%%%%%%%%%%%%%%%%%%
%%%%%%%%%%%%%%%%%%%%%%%%%%%%%%%%%%%%%%%%%%%%%%%%%%%%%%%%%%%%%%%%%%%%%%
\appendix

%%%%%%%%%%%%%%%%%%%%%%%%%%%%%%%%%%%%%%%%%%%%%%%%%%%%%%%%%%%
%%%%%%%%%%%%%%%%%%%%%%%%%%%%%%%%%%%%%%%%%%%%%%%%%%%%%%%%%%%
%%%%%%%%%%%%%%%%%%%%%%%%%%%%%%%%%%%%%%%%%%%%%%%%%%%%%%%%%%%
\section{Abelian dynamics}
%%%%%%%%%%%%%%%%%%%%%%%%%%%%%%%%%%%%%%%%%%%%%%%%%%%%%%%%%%%
%%%%%%%%%%%%%%%%%%%%%%%%%%%%%%%%%%%%%%%%%%%%%%%%%%%%%%%%%%%
%%%%%%%%%%%%%%%%%%%%%%%%%%%%%%%%%%%%%%%%%%%%%%%%%%%%%%%%%%%
In this section, we discuss the formulation of the problem in the case
of three Abelian gauge fields and present some numerical results for the
purpose of comparison. Let us assume the condition (\ref{eq:diagCon}) and
fix the frame such that all the anisotropic variables are diagonal. 
First of all, 
we note that the vector potential completely
disappears from equations by setting $g=0$ in (\ref{eq:dynamicalB}), 
(\ref{eq:constraintB}), (\ref{eq:dynamicalE}) and (\ref{eq:constraintE}).
Hence, we do not need normalized variables $\mathcal{A}_A$ and 
$\Gamma $. Using the standard normalization (\ref{eq:normal1}) and
\begin{equation}
\mathcal{E}_A = \frac{e^{\lambda \varphi /2}E_A}{H} \ , \quad \mathcal{B}_A = 
\frac{e^{\lambda \varphi /2}B_A}{H} \ , 
\end{equation}
the Maxwell's equations are given as
\begin{align}
\begin{split}
\mathcal{E}_1^{\prime } = & \left( q -1 -\frac{\lambda }{2} \varpi -2\Sigma _+ 
\right) \mathcal{E}_1 - N_1 \mathcal{B}_1 \ , \\
\mathcal{E}_2^{\prime } = & \left( q-1-\frac{\lambda }{2} \varpi +\Sigma _+ + 
\sqrt{3}\Sigma _- \right) \mathcal{E}_2 -N_2 \mathcal{B}_2 \ , \\
\mathcal{E}_3^{\prime } =& \left( q-1 - \frac{\lambda }{2} \varpi +\Sigma _+ 
-\sqrt{3}\Sigma _- \right) \mathcal{E}_3 - N_3 \mathcal{B}_3 \ , \\
\mathcal{B}_1^{\prime } = & \left( q -1 +\frac{\lambda }{2} \varpi -2\Sigma _+
 \right) \mathcal{B}_1 + N_1 \mathcal{E}_1 \ , \\
\mathcal{B}_2^{\prime } = & \left( q-1+ \frac{\lambda }{2} \varpi +\Sigma _+ 
+ \sqrt{3}\Sigma _- \right) \mathcal{B}_2 +N_2 \mathcal{E}_2 \ , \\
\mathcal{B}_3^{\prime } =& \left( q-1 + \frac{\lambda }{2} \varpi +\Sigma _+
 -\sqrt{3}\Sigma _- \right) \mathcal{B}_3 + N_3 \mathcal{E}_3 \ ,
\end{split}
\end{align}
with the spatial curvature variables satisfying
\begin{align}
\begin{split}
N_1^{\prime } = & (q-4\Sigma _+) N_1 \ , \\
N_2^{\prime } = & (q+2\Sigma _+ + 2\sqrt{3}\Sigma _- )N_2 \ , \\
N_3^{\prime } = & (q+2\Sigma _+ -2\sqrt{3}\Sigma _- ) N_3 \ .
\end{split}
\end{align}
For the other equations, one only has to replace $\Gamma \mathcal{N}_A$
by $N_A$. Note that the magnetic fields $\mathcal{B}_A$
do not appear from the vector potential as long as it is assumed to
be homogeneous.

One can confirm that the type I invariant set $N_A =0$ coincides
with what we called Abelian boundary $\Gamma =0$ in the dynamical system
discussed in the main body of the article. 

The Abelian system behaves much better than the non-Abelian one. For
instance, the double-component magnetic inflation is a usual fixed point and
for the case of $\mathcal{B}_1=0$, the eigenvalues are given by
\begin{align}
\begin{split}
& \omega _{\Sigma _+ -\varpi -  \Omega _B} = \frac{q-2 \pm \sqrt{\left( q-2 
\right) ^2 + 12 \Omega _B \left( \lambda \left( \alpha + \lambda \right) -2
 \right)}}{2} \ , \\
& \omega _{\Sigma _+ - \varpi } = q-2 \ , \quad \omega _{\Sigma _- 
- \mathcal{B}_- } = \frac{q-2 \pm \sqrt{\left( q-2 \right) ^2 
-24 \Omega _B}}{2} \ , \\
& \omega _{N_1} = 3\frac{2\alpha ^2 +2\alpha \lambda -\lambda ^2 -6}{(\alpha 
+\lambda )(\alpha +3\lambda )+2} \ , \quad \omega _{N_2} = \omega _{N_3} 
= -3\frac{\lambda ^2 -2}{(\alpha +\lambda )(\alpha +3\lambda )+2} \ , \\
& \omega _{\mathcal{B}_1} = -3\Sigma _+ \  , \quad \omega _{\mathcal{E}_1} 
= 3 \frac{\alpha ^2 -\alpha \lambda -4\lambda ^2 -4}{(\alpha +\lambda )(\alpha 
+3\lambda )+2} \ , \quad \omega _{\mathcal{E}_2} = \omega _{\mathcal{E}_3} 
= \frac{-6\lambda (\alpha +2\lambda )}{(\alpha +\lambda )(\alpha +3\lambda )+2} \ ,
\end{split}
\end{align}
which confirms that all the eigenvalues except for $\omega _{\mathcal{B}_1}$ 
possess negative real part provided that $\lambda $ is sufficiently greater than unity.

\begin{figure}[htbp]
\begin{center}
\includegraphics[width=0.9\linewidth]{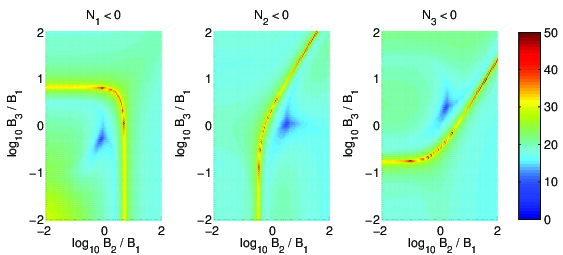}
\caption{Dependence of convergence time on initial magnetic configuration
for Abelian Bianchi type VIII. 
Each panel should be compared with the corresponding one
in figure \ref{fig:type8a}. While red contours are clearly identified between 
Abelian and non-Abelian plots, the latter exhibits more complicated patterns 
and longer anisotropic period.}
\label{fig:type8Abelian}
\end{center}
\end{figure}

\begin{figure}[htbp]
\begin{center}
\includegraphics[width=0.9\linewidth]{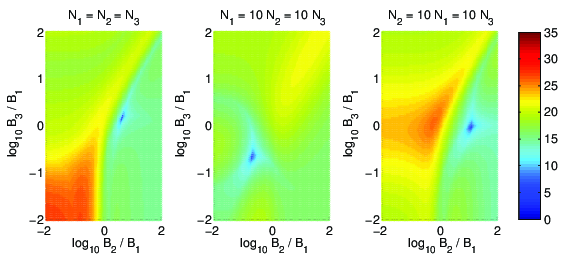}
\caption{Convergence time for Abelian Bianchi type IX. The left 
and center panels
have counterparts in the left and center of figure 
\ref{fig:type9a}
respectively. For Abelian models, anisotropic type IX curvature does not 
extend period of anisotropic inflation significantly.}
\label{fig:type9Abelian}
\end{center}
\end{figure}

Figures \ref{fig:type8Abelian} and \ref{fig:type9Abelian} show how initial
magnetic configuration affects the convergence time for Abelian case. 
As before, our initial conditions are taken to be equipartition of energy 
(either (\ref{eq:inicon2a}) or (\ref{eq:equip9}) depending on the signature
of $\Omega _N$) and $\Sigma _+ = \Sigma _- , \mathcal{E}_1 = \mathcal{E}_2 
=\mathcal{E}_3$. For type VIII, three different choices of the negative 
spatial 
curvature component with (\ref{eq:inicon2c}) are taken. For type IX, we have 
tried the isotropic initial curvature as well as two different preferred 
directions.
Comparing these with figures \ref{fig:type8a} and \ref{fig:type9a}, overall patterns 
look similar for type III while the absence of the recollapsing and anisotropic
strips in type IX is noted. From the discussion in section \ref{Quasi-single-component}, 
the enhancement of anisotropy should not happen in Abelian dynamics
and the result confirms it.

%%%%%%%%%%%%%%%%%%%%%%%%%%%%%%%%%%%%%%%%%%%%%%%%%%%%%%%%%%%
%%%%%%%%%%%%%%%%%%%%%%%%%%%%%%%%%%%%%%%%%%%%%%%%%%%%%%%%%%%
%%%%%%%%%%%%%%%%%%%%%%%%%%%%%%%%%%%%%%%%%%%%%%%%%%%%%%%%%%%
\section{Alternative numerical results tailored 
for Bianchi type VI$_0$ and  VII$_0$}
%%%%%%%%%%%%%%%%%%%%%%%%%%%%%%%%%%%%%%%%%%%%%%%%%%%%%%%%%%%
%%%%%%%%%%%%%%%%%%%%%%%%%%%%%%%%%%%%%%%%%%%%%%%%%%%%%%%%%%%
%%%%%%%%%%%%%%%%%%%%%%%%%%%%%%%%%%%%%%%%%%%%%%%%%%%%%%%%%%%
When the rank of 3 by 3 matrix $n_{AB}$ is 2, the homogeneous spatial geometry 
is classified into type VI$_0$ or VII$_0$ according to $\lambda _1 \lambda _2 
\gtrless 0$
where $\lambda _1$ and $\lambda _2$ are the two non-zero eigenvalues of 
$n_{AB}$. 
In the setup of the present article, one can assume $n_1 >0 , n_2 <0$ for 
VI$_0$
and $n_1 >0 , n_2 >0$ for VII$_0$ without loss of generality. 

For the purpose of our numerical calculations, 
it is desirable to specify the initial conditions of 
$\mathcal{B}_A$ instead of $\mathcal{A}_A$ since the
former give physical measure of dynamical contribution of 
magnetic components. 
However, $\mathcal{A}_A$'s being the fundamental
variables in terms of which the actual equations have been written down,
one has to solve (\ref{eq:magdef}) for $\mathcal{A}_A$'s to set the initial
conditions for each given run of numerical calculation. In an attempt to generate a 
figure analogous to figure \ref{fig:type2a}, let us assume that we are given
non-zero values for $\mathcal{B}_1, \mathcal{B}_2, \mathcal{B}_3 , 
\mathcal{N}_1 , \mathcal{N}_2$ while $\mathcal{N}_3$ vanishes. 
Eliminating $\mathcal{A}_2 $ and $\mathcal{A}_3$, one obtains
a quartic equation for $\mathcal{A}_1$
\begin{equation}
\mathcal{N}_1 \mathcal{A}_1^4 + \mathcal{B}_1 \mathcal{A}_1^3 - \mathcal{B}_2 
\mathcal{B}_3 \mathcal{A}_1 - \mathcal{N}_2 \mathcal{B}_3^2 = 0 \ .
\end{equation}
Since the highest power of $\mathcal{A}_1$ is even, this equation may
not have a real root depending on the values of magnetic field and
spatial curvature. At least a real root always exists for type II, VIII or IX
since the resulting algebraic equation is either odd or proportional to
$\mathcal{A}_1$. Since we are varying the ratios of magnetic components
by a factor of $10^4$, the parameter regions in which no real root exists 
are encountered during the calculations and it is impossible to produce
density plots that can be compared to figures \ref{fig:type2a}, 
\ref{fig:type8a}
and \ref{fig:type9a}.

To circumvent this problem, we abandon the initial equipartition of energy and
directly specify the components of vector potential $\mathcal{A}_1 ,\mathcal{A}_2 ,
\mathcal{A}_3$. The primary goal of studying type VI$_0$ and VII$_0$ is
to confirm that the oscillation of electric and magnetic fields occurs only on
 the 
plane perpendicular to the single direction for which the curvature is absent.
Since we already saw that the varying initial energy densities for magnetic 
field
or spatial curvature do not very much change the conditions for oscillation 
in
the case of type I and II, we expect to see the aforementioned feature. Our 
setup
for the numerical calculations is as follows. First of all, we fix the initial
 values of
shear, electric field, scalar kinetic energy and $\Gamma $ as
\begin{equation}
\Sigma _+ = \Sigma _- = \sqrt{0.1} \  , \quad \mathcal{E}_1 = \mathcal{E}_2 = 
\mathcal{E}_3 = \sqrt{0.4} \ , \quad \varpi = \sqrt{1.2} \ , \quad \Gamma = 1 
\ . \label{eq:iniMod1}
\end{equation}
Next, we choose the 2- and 3-components of magnetic field to be varied. 
The specific
prescription is given by
\begin{equation}
\mathcal{A}_2 \mathcal{A}_3 = \frac{\sqrt{0.6}}{100} \ , \quad 
\frac{\mathcal{A}_1}{\mathcal{A}_2} , \frac{\mathcal{A}_1}{\mathcal{A}_3} 
\in \left[ 10^{-2} , 10^2 \right] \ , \quad \mathcal{A}_1 >0 \ , \mathcal{A}_2 
> 0 \ , \mathcal{A}_3 > 0 \ .
\end{equation}
Note that this corresponds to setting
\begin{equation}
\mathcal{B}_1 = \frac{\sqrt{0.6}}{100} \ , \quad 
\frac{\mathcal{B}_2}{\mathcal{B}_1} , \frac{\mathcal{B}_3}{\mathcal{B}_1}
 \in \left[ 10^{-2} , 10^2 \right] 
\end{equation}
for Bianchi type I, in which case $\Omega _B$ would vary between 
$\sim 10^{-4}$ and $\sim 0.2$.
When the spatial curvature exists, the magnetic field receives extra 
contributions involving
$\mathcal{N}_A$ and the condition $\Omega _B < 0.4$ that is required by the 
Hamiltonian
constraint and positivity of $\Omega _V$ with (\ref{eq:iniMod1}) is not 
guaranteed. To reduce
the chance of breaching this upper bound, we fix the amplitude of each spatial
 curvature variable
to be $| \mathcal{N}_A | = 0.1$ whenever one is non-zero so that the 
modification to $\mathcal{B}_A$
is suppressed by a factor of $10$. This results in initial curvature density
\begin{equation}
\Omega _N = \begin{cases}
1/12  & ({\rm type \ II}) \ , \\
1/ 3  & ({\rm type \ VI}_0) \ , \\
0  & ({\rm type \ VII}_0) \ .
\end{cases}
\end{equation}
Finally, the scalar potential is determined by
\begin{equation}
\Omega _V = 0.4 - \Omega _B - \Omega _N 
\end{equation}
with its positivity being checked for each iteration. It turns out that the 
regions discarded
due to a negative value of $\Omega _V$ are indiscernible in the 
following plots. 
\begin{figure}[htbp]
\begin{center}
\includegraphics[width=0.9\linewidth]{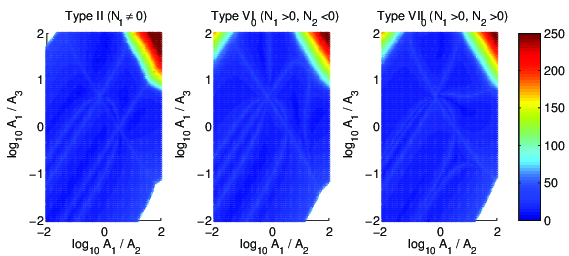}
\caption{Dependence of convergence time on initial vector potential. We chose
$\mathcal{N}_1 > 0$ for type II, $\mathcal{N}_1 >0 , \mathcal{N}_2 < 0$ for
 type
VI$_0$ and $\mathcal{N}_1 >0 , \mathcal{N}_2 >0$ for type VII$_0$. The
 disappearance
of oscillatory region at the top-left corner is clearly seen in type VI$_0$
 and VII$_0$.}
\label{fig:alt1}
\end{center}
\end{figure}

\begin{figure}[htbp]
\begin{center}
\includegraphics[width=0.9\linewidth]{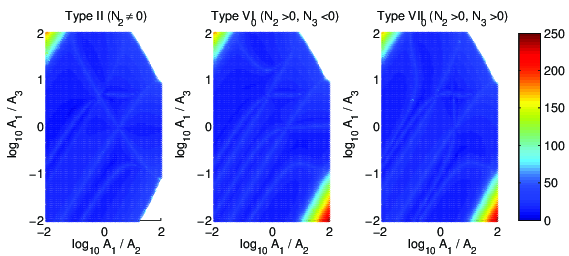}
\caption{Here the non-zero curvature components are taken to be $\mathcal{N}_2 
>0$,
$\mathcal{N}_2 >0 , \mathcal{N}_3 < 0$ and $\mathcal{N}_2 >0 , 
\mathcal{N}_3 >0$
for type II, VI$_0$ and VII$_0$ respectively. Since $\mathcal{N}_1 = 0$, 
oscillation
is allowed in top-right corner for which initially $\mathcal{A}_1 \gg 
\mathcal{A}_2 , \mathcal{A}_3$.}
\label{fig:alt2}
\end{center}
\end{figure}

Figures \ref{fig:alt1} and \ref{fig:alt2} show the results for type II, 
type VI$_0$
and type VII$_0$. The difference between the two is the choice of non-zero
components of $\mathcal{N}_A$. Note that now it does matter which ones to
be taken non-zero since the initial conditions for shear and magnetic field are
anisotropic. As in the main text, the white regions represent the initial 
conditions
which lead to the oscillatory final state. One can clearly see the absence of 
oscillation when the directions of initially large components of vector 
potential 
coincide with those of the non-zero spatial curvature components. 
Interestingly, in this scheme, the disappeared oscillation is almost completely
substituted by a prolonged period of anisotropic inflation. This observation 
agrees with the results for type I in section 4. 
%%%%%%%%%%%%%%%%%%%%%%%%%%%%%%%%%%%%%%%%%%%%%%%%%%%%%%%%%%%%%%%%%%%%%%%%%%%%%%
%%%%%%%%%%%%%%%%%%%%%%%%%%%%%%%%%%%%%%%%%%%%%%%%%%%%%%%%%%%%%%%%%%%%%%%%%%%%%%%
%%%%%%%%%%%%%%%%%%%%%%%%%%%%%%%%%%%%%%%%%%%%%%%%%%%%%%%%%%%
%%%%%%%%%%%%%%%%%%%%%%%%%%%%%%%%%%%%%%%%%%%%%%%%%%%%%%%%%%%
%%%%%%%%%%%%%%%%%%%%%%%%%%%%%%%%%%%%%%%%%%%%%%%%%%%%%%%%%%%
\end{document}